\documentclass[aps,prb,amsmath,amssymb,floatfix,twocolumn]{revtex4-1}
\usepackage[dvipsnames]{xcolor}
\usepackage{graphicx}
\usepackage{times}
\usepackage{bm}
\usepackage{hyperref}
\usepackage{subfigure}
\usepackage{color}
\usepackage{hyperref}
\usepackage{bbold}
\usepackage{amssymb}

\newcommand{\ncmd}{\newcommand}
\ncmd{\nn}{\nonumber}
\ncmd{\pg}[1]{\textcolor{red}{#1}}
\ncmd{\mbf}[1]{\bs{#1}}
\ncmd{\Lam}{\Lambda}
\ncmd{\lam}{\lambda}
\ncmd{\Gam}{\Gamma}
\ncmd{\gam}{\gamma}
\ncmd{\sig}{\sigma}
\ncmd{\Dl}{\Delta}
\ncmd{\dl}{\delta}
\ncmd{\kap}{\kappa}
\ncmd{\Om}{\Omega}
\ncmd{\om}{\omega}
\ncmd{\mc}{\mathcal}
\ncmd{\eps}{\epsilon}
\ncmd{\veps}{\varepsilon}
\ncmd{\vphi}{\varphi}
\ncmd{\vtheta}{\vartheta}
\ncmd{\note}[1]{{\color{red}{#1}}}
\ncmd{\new}[1]{{\texttt{#1}  } }
\ncmd{\eq}[1]{Eq. \eqref{#1}}
\ncmd{\bs}{\boldsymbol}
\ncmd{\pll}{\parallel}
\ncmd{\dsty}{\displaystyle}

\begin{document}

\title{Part II: Witten effect and $\mathbb{Z}$-classification of axion angle $\theta=n \pi$}

\author{Alexander C. Tyner$^{1}$ and Pallab Goswami$^{1,2}$}
\affiliation{$^{1}$ Graduate Program in Applied Physics, Northwestern University, Evanston, Illinois, 60208, USA}
\affiliation{$^{2}$ Department of Physics and Astronomy, Northwestern University, Evanston, Illinois 60208, USA}

\date{\today}

\begin{abstract} 
The non-trivial third homotopy class of three-dimensional topological insulators leads to quantized, magneto-electric coefficient or axion angle $\theta= n \pi$, with $n \in \mathbb{Z}$. In Part I, we developed tools for computing $n$ from a staggered symmetry-indicator $\kappa_{AF,j}$ and Wilson loops of non-Abelian, Berry connection in momentum-space, which clearly distinguished between magneto-electrically trivial ($n=0$), and non-trivial ($n=2s$) topological crystalline insulators. In this work, we perform $\mathbb{Z}$-classification of real-space, topological response or $\theta$ by carrying out thought experiments with magnetic, Dirac monopoles. We demonstrate this for non-magnetic and magnetic topological insulators by computing induced electric charge on monopoles or Witten effect. We show that both first- and higher- order topological insulators can exhibit quantized, magneto-electric response, irrespective of the presence of gapless surface-states, and corner-states. Special attention is paid to the response of octupolar higher-order topological insulator, which was originally predicted to be magneto-electrically trivial. The important roles of fermion zero-modes, $\mathcal{CP}$, and flavor symmetries are critically addressed. Our work outlines a unified theoretical framework for addressing dc topological response and topological quantum phase transitions, which cannot be reliably predicted by symmetry-based classification scheme. 
\end{abstract}

\maketitle

\section{Introduction} Topological properties of defects and textures of order parameters and gauge fields in real space are generally classified by homotopy groups~\cite{Thouless,Volovik,Shifman}. The homotopy analysis can also be applied for topological classification of quasi-particle band structures in reciprocal space, with Bloch Hamiltonian and Berry connection respectively assuming the roles of order parameter and gauge fields~\cite{Thouless,Volovik,FuKaneMele2007,FuKane,Moore2007,qi2008topological,RyuLudwigPRB,Roy20093D}. 
For three-dimensional (3D) insulators, preserving  parity ($\mathcal{P}$) and time-reversal ($\mathcal{T}$) symmetries, tunneling configurations of $SU(2)$ Berry connection $\bs{A}_i(\bs{k})$ of $i$-th Kramers degenerate band can be classified by the Chern-Simons invariant~\cite{qi2008topological,RyuLudwigPRB,ryu2010topological}
\begin{eqnarray}\label{A11}
\mathcal{CS}_i &=&\frac{1}{8\pi^2} \;  \int d^{3}k  \; \epsilon^{abc} \; \text{Tr}[A_{a,i}\partial_{b}A_{c,i}+\frac{2i}{3}A_{a,i}A_{b,i}A_{c,i}], \nn \\ \\
&=&\frac{n_i}{2}, \; \text{with} \; n_i \in \mathbb{Z}.
\end{eqnarray} 
When a proper Abelian gauge-fixing procedure is implemented, the winding number $n_i$ describes the third homotopy class of $i$-th Kramers-degenerate band, and the diagonal matrix 
$n_{ij} =n_i \; \delta_{ij}$
identifies the third homotopy class of $2N \times 2N $ Bloch Hamiltonian~\cite{tyner2021symmetry}. Similarly, the third homotopy class of magnetic systems can be described by a diagonal matrix of Abelian Chern-Simons invariants of non-degenerate bands~\cite{moore2008topological,Lapierre2021,tyner2021symmetry}. 

Field theory calculations suggest that the \emph{adiabatic, electrodynamic response} of $\mathcal{T}$-symmetric topological insulators (TIs) is described by the effective action~\cite{qi2008topological} 
 \begin{equation}\label{eq:1}
S_{eff}= \int dt d^3x \left[\frac{1}{2} (\epsilon  \bs{E}^2 - \frac{1}{\mu} \bs{B}^2) +  
 \frac{\theta e^2}{4 \pi^2 \hbar}  \mathbf{E}\cdot \mathbf{B} \right],
\end{equation} 
where $\epsilon$ ($\mu$) is the dielectric permittivity (magnetic permeability) of medium, and 
\begin{equation}\label{gentheta}
\theta= 2\pi \sum_{i}^{l} \mathcal{CS}_i =  \pi \sum_{i=1}^{l} n_i = n \pi,
\end{equation} 
is the quantized, magneto-electric (ME) coefficient or axion angle~\cite{PhysRevLett.38.1440,Wilczek1987}, and $l$ is the number of occupied bands. Depending on the detailed topological properties of occupied bands, one can realize (i) $\mathbb{Z}_2$ TIs with ME coefficient $\theta=(2s+1) \pi$, (ii) $\mathbb{Z}_2$-trivial, TIs with ME coefficient $\theta= 2s \pi$, and (iii) $\mathbb{Z}_2$-trivial, and magneto-electrically trivial, TIs with $\theta=0$, but $n_i \neq 0$~\cite{tyner2021symmetry}. 

Under periodic boundary conditions (PBC) in $(3+1)$-D space-time, the electromagnetic Berry phase
\begin{eqnarray}
e^{i \frac{S_\theta}{\hbar}} = e^{i  \frac{\theta e^2}{h^2}  \int dt d^3x  \mathbf{E}\cdot \mathbf{B}} 
=e^{i n \pi N_E N_B }
\end{eqnarray}
can distinguish between $n=(2s+1)$ and $n=2s$, when the instanton number of electromagnetic fields $N_E N_B$ is an \emph{odd integer}, with $N_E $ ($N_B $) being the number of electric (magnetic) flux quantum~\cite{QiZhangmonopole,Karch2009,Vazifeh2010,Chen2011}. In past fourteen years, many intriguing consequences of axion electrodynamics have been studied for weakly correlated, non-magnetic $\mathbb{Z}_2$ TIs~\cite{EssinMagnetoelectric,Essin2010,malashevich2010theory,QiZhangmonopole,Karch2009,Vazifeh2010,Chen2011,Maciejko1,TseMacDonald,rosenberg2010witten,RosenbergWormhole,Coh2011,zhao2012magnetic,wu2016quantized,zirnstein2017time,zirnstein2020topological}, anti-ferromagnetic TIs~\cite{li2010dynamical,Mong,Oshikawa,dziom2017observation,Zhang2019,zhu2021tunable}, chiral TIs~\cite{RyuMooreLudwig,NeupertRyuPRB}, and fractionalized TIs~\cite{Maciejko2,Swingle2011,Metlitski2013}.

However, the direct calculation of $\theta$ for realistic band structures has remained a formidable task, and limited progress has been made for $\mathbb{Z}_2$ TIs~\cite{EssinMagnetoelectric,Essin2010,malashevich2010theory,Coh2011,varnava2020}. 
In a companion paper (Part I), we have shown how to determine $n_{i}$ from tight-binding models and $|n_{i} |$ from \emph{ab initio} band structures by avoiding direct calculation of Chern-Simons invariant~\cite{tyner2021symmetry}. This development compelled us to raise the following questions. 
(i) Can $n=0$ and $n=2s$ be distinguished with electromagnetic probes? (ii) Can different odd integers be distinguished? (iii) How does topological quantum phase transition (TQPT) affect $\theta$? To answer these questions, 
we will go beyond the adiabatic theory of ME response, and perform thought experiments by embedding a test, Dirac monopole of strength $g=\frac{\hbar m}{2e}$ inside TIs, with $m \in \mathbb{Z}$~\cite{dirac1931quantised}. 

Both magnetic flux-tubes~\cite{Laughlin1,Laughlin2} and monopoles~\cite{Haldane} are extensively used as singular probes of Chern number, describing the second homotopy class of 2D quantum Hall systems~\cite{TKNN}. In this work, we will establish monopoles as singular probes of third homotopy class of 3D insulators. While the electric charge is odd under charge-conjugation ($\mathcal{C}$) and even under $\mathcal{P}$ and $\mathcal{T}$, the magnetic charge $g$ is odd under $\mathcal{C}$, $\mathcal{P}$, and $\mathcal{T}$. Therefore, $eg=\hbar m/2$ is a topologically non-trivial pseudo-scalar that directly couples to $\theta$, allowing us to determine dc ME coefficient. 

For an infinite system, described by Eq.~\ref{eq:1}, with a constant $\theta \neq 0$, monopoles can bind electric charge and turn into dyons~\cite{witten1979dyons}. This phenomenon is known as the Witten effect (WE). The $\theta$-term leads to modified constitutive relations 
\begin{eqnarray}
\bs{D}=\epsilon \bs{E} + \frac{\theta e^2}{4\pi^2 \hbar} \bs{B}, \; \bs{H}= \frac{\bs{B}}{\mu} - \frac{\theta e^2}{4\pi^2 \hbar} \bs{E}. 
\end{eqnarray}
In the absence of external electric fields, 
static monopoles give rise to the displacement field 
\begin{equation}
\bs{D}(\bs{r})=\bs{P}_{ME}(\bs{r}) = \frac{\theta e^2}{4\pi^2 \hbar} \bs{B}(\bs{r})= \frac{e m \theta}{8 \pi^2} \frac{\hat{\bs{r}}}{r^2},
\end{equation}
where $\bs{P}_{ME}(\bs{r})$ is the ME polarization, and the induced or bound electric charge density 
\begin{equation} 
\delta \rho(\bs{r})=-\bs{\nabla} \cdot \bs{P}_{ME} = - \frac{em \theta}{ 2 \pi } \delta^3(\bs{r}).  \label{rhoe}
\end{equation}
Due to the oversimplified nature of $S_{eff}$, the induced electric charge density turns out to be sharply localized on monopoles, and the total induced electric charge
\begin{equation}
\delta Q= \int d^3 r^\prime \delta \rho(\bs{r}^\prime) = -\frac{e m \theta}{2 \pi}
\end{equation} 
does not depend on the radius of Gaussian surface. For half-filled, insulators, preserving $\mathcal{C}$, $\mathcal{P}$ and $\mathcal{T}$ symmetries 
\begin{equation}\label{Wittencont}
\delta Q= - \frac{e m n }{2}
\end{equation}
 can only acquire integer and half-integer values. \emph{Therefore, $\delta Q/e$ for a unit monopole $m=+1$ can track 3D winding number $n$ or $\mathbb{Z}$-classification of third homotopy class}. Furthermore, two insulators with distinct winding numbers $n_1$ and $n_2$ can show identical WE, if $m_1 n_1 = m_2 n_2$. 

Non-perturbative, numerical calculations of WE for TIs were reported by Rosenberg and Franz~\cite{rosenberg2010witten}. They showed a minimal Dirac monopole ($m=+1$), embedded in half-filled, strong $\mathbb{Z}_2$ TIs could support $\delta Q=\pm e/2$, both in the presence and absence of Zeeman coupling between fermion spin and spatially varying magnetic field of monopole. No connection with the third homotopy classification was made. The validity of their results for half-filled systems, in the absence of Zeeman coupling was questioned in Ref.~\onlinecite{zhao2012magnetic}.

In this work, we will perform comprehensive analysis of fermion spectrum and WE for general $m$ and $n$, without introducing spatially varying Zeeman coupling between fermion spin and monopoles. Our manuscript is organized as follows. In Sec.~\ref{Continuum}, we describe analytical and numerical calculations for a continuum or spherical model of first-order TIs (FOTIs), which \emph{support gapless surface-states under all orientations}. In Sec.~\ref{FOTI}, we describe thought experiments on tight-binding models of FOTIs, exhibiting \emph{gapless surface-states along all high-symmetry axes}. In Sec.~\ref{SOTI}, Sec.~\ref{AppA}, and Sec.~\ref{TOTI1}, we respectively describe our analysis for magneto-electric/chiral higher-order topological insulators (HOTI), magnetic HOTIs, and octupolar HOTIs, which can possess \emph{gapless or gapped surface-states, depending on the orientation of surface}. The effects of spatially varying Zeeman coupling are briefly discussed in Appendix~\ref{AppC}. For conceptual clarity, we work with minimal four-band models in Secs.~\ref{Continuum}-\ref{AppA}. In Sec.~\ref{TOTI1} we deal with eight-band model. 
Our main results are as follows. 

\begin{figure*}[t]
\centering
\subfigure[]{
\includegraphics[scale=0.25]{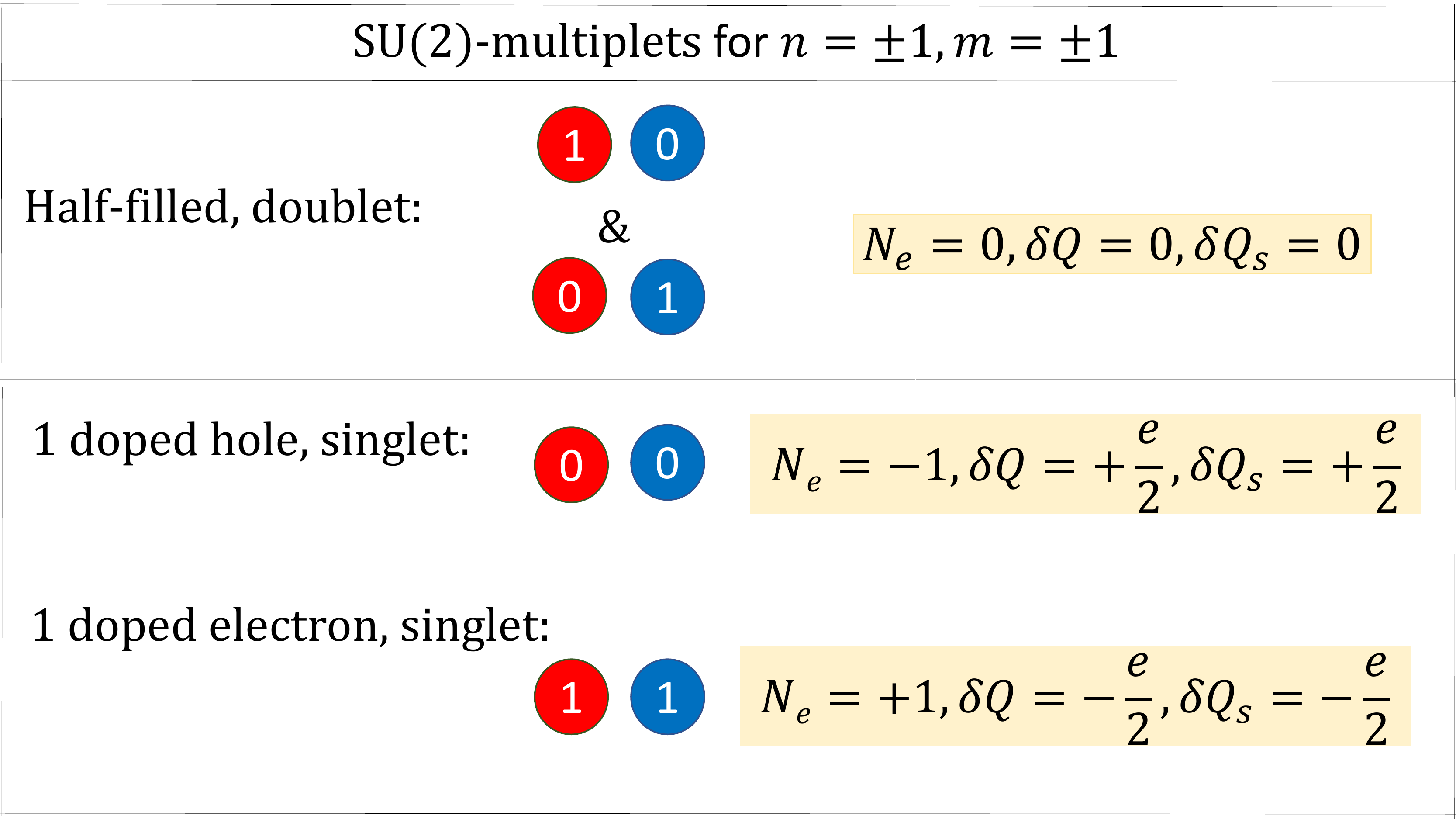}
\label{fig:presentation1}}
\subfigure[]{
\includegraphics[scale=0.25]{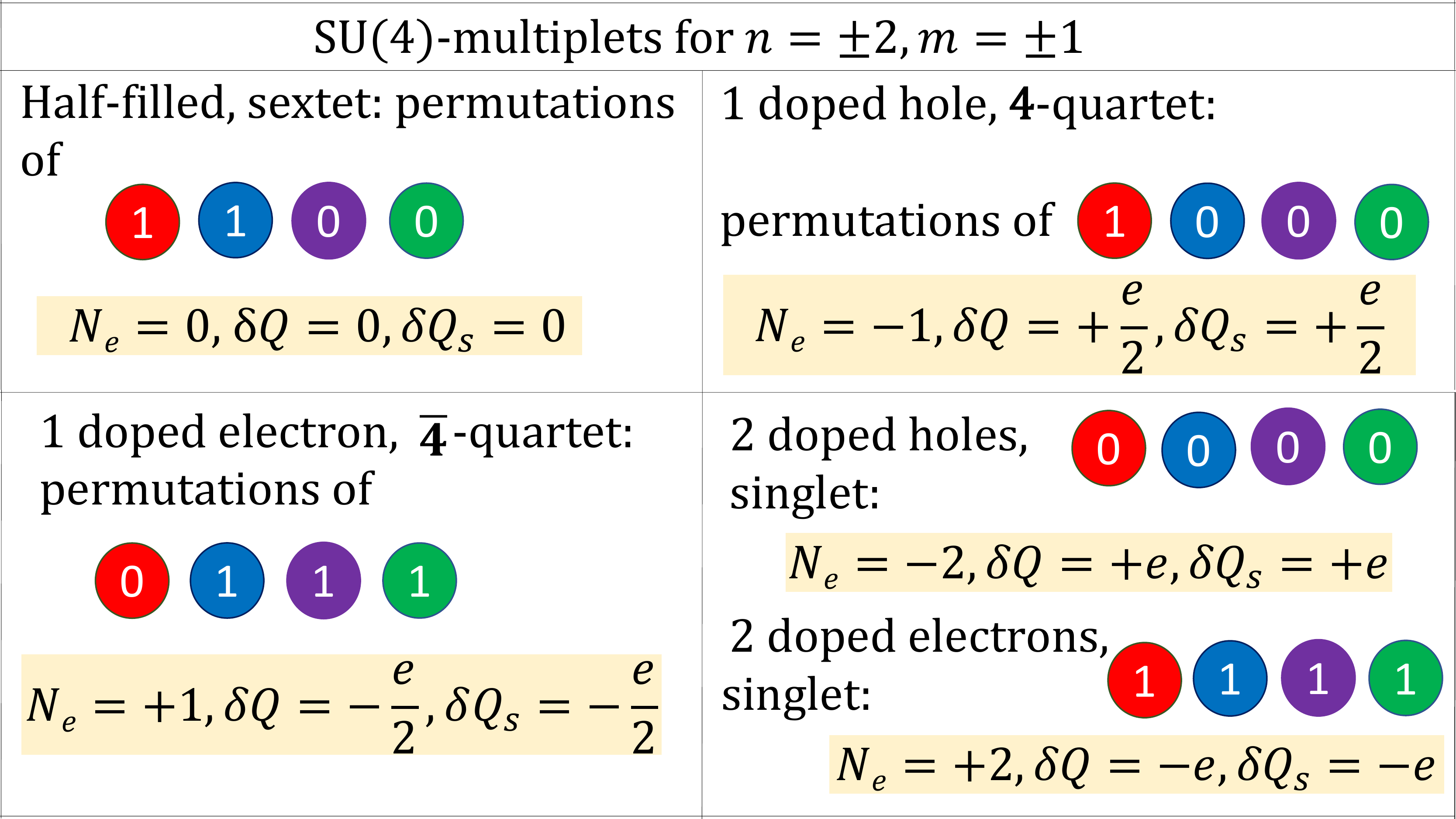}
\label{fig:presentation2}}
\caption{Schematics of spin-charge separation and $\mathbb{N}$-classification of bulk topology of $\mathcal{C}$-, $\mathcal{P}$-, and $\mathcal{T}$- symmetric first-order topological insulators under open boundary conditions. In the absence of a pseudo-scalar training field ($M^\prime=0$), the degenerate half-filled state cannot exhibit Witten effect. However, by doping electron or hole into insulators we can probe induced the electric charge on monopole for charged $SU(2|nm|)$-multiplets. The occupation numbers of fermion zero-modes are labeled by $0$ and $1$, and $N_e$ corresponds to the number of doped electrons. The induced charge on monopole (surface) is denoted by $\delta Q$ ($\delta Q_s$). Topological insulators with winding numbers $n= \pm 1$ also exhibit identical $SU(4)$-multiplets in the presence of double monopoles $m=\pm 2$. A trivial insulator does not support spin-charge separation.  }
\label{Presentation}
\end{figure*}

\begin{enumerate}

\item In Secs.~\ref{Continuum} and ~\ref{FOTI}, we show that a static monopole, embedded in an \emph{infinite}, $\mathcal{C}$-, $\mathcal{P}$- and $\mathcal{T}$- symmetric FOTI supports $|n m|$ number of exponentially localized, fermion zero-modes (FZMs). The FZMs give rise to the \emph{vacuum expectation value of pseudo-scalar mass operator}, which is proportional to the chirality of FZMs or $\text{sgn}(n m)$. Therefore, infinite FOTIs in the presence of an isolated, static monopole break $\mathcal{P}$, $\mathcal{T}$, $\mathcal{CP}$, and $\mathcal{CT}$ symmetries. Consequently, $\text{sgn}(n)$ and $\mathbb{Z}$-classification can be determined, either by computing the expectation value of pseudo-scalar mass or WE. 

\item However, under open boundary conditions (OBC), the gapless surface-states of \emph{finite-size} FOTIs also give rise to $|n m|$-number of FZMs of opposite chirality. The hybridization between monopole-localized and surface-localized FZMs leads to $2 |n m|$-number of exponentially split, near zero-modes (NZMs). As the chirality of FZMs is obscured by hybridization, \emph{finite-size, half-filled, $\mathcal{C}$-, $\mathcal{P}$- and $\mathcal{T}$- symmetric FOTIs cannot exhibit WE}. 

\item Therefore, finite-size effects must be subdued with \emph{a pseudo-scalar, training field ($M^\prime$) to observe WE}. We will calculate the induced charge enclosed by a Gaussian sphere of radius $R$, as a scaling function $\delta Q(L, R,\xi, M^\prime) \neq 0$, where $L$ is the system size, the correlation length $\xi \sim |M|^{-1}$, and $2|M|$ is the bulk band gap, when $M^\prime=0$. We show that maximum value of $\delta Q(L, R,\xi, M^\prime)$ in the adiabatic, scaling regime $\xi \ll |M^\prime|^{-1} \ll L$ approaches quantized results of Eq.~\ref{Wittencont} (see Subsec.~\ref{WEHF} ). As the overall half-filled system is charge neutral, compensating charge $\delta Q_s=-\delta Q$ would be localized around the boundary.

\item \emph{In the absence of pseudo-scalar training field}, we can only detect $\mathbb{N}$-classification of FOTIs by studying ``spin-charge separation" [see Fig.~\ref{Presentation} ]. In the thermodynamic limit ($L \to \infty$), the energy splitting due to hybridization can be ignored, and the charge-neutral, half-filled state follows the self-conjugate representation of $\mathfrak{su}(2|nm|)$ algebra, with degeneracy 
\begin{equation}\label{grounddeg}
N_G=\frac{(2|nm|)!}{[(|nm|)!]^2}.
\end{equation}

By doping one electron (or hole) at a time, we will probe $\mathcal{C}$-odd, charged multiplets that follow non-self-conjugate representations of $\mathfrak{su}(2|nm|)$ (see Subsec.~\ref{WED}). When the number of doped electrons ($N_e$) is varied between $-|n m|$ and $|n m|$, the maximum value of monopole-bound charge $\delta Q(L, R,\xi, M^\prime=0)$ saturates to $-e N_e/2$, and the surface-bound charge also shows saturation $\delta Q_s \to -e N_e/2$. Thus, maximum values of $\delta Q$ shows oscillations between half-integer and integer values, until NZMS become completely empty or occupied. The completely occupied and empty NZMs describe non-degenerate, many-body states that follow singlet representation of $\mathfrak{su}(2|nm|)$. This method is useful for identifying $|n|$ with unit monopoles $m=\pm 1$, and the equivalence between $|n_1 m_1|=|n_2 m_2|$. Moreover, TQPTs can be addressed by studying the finite-size scaling of $\delta Q (L , R, \xi, M^\prime=0)$, as a function of $L/\xi$.

\item It is often assumed that the induced electric charge $\delta Q= s e$, with $s \in \mathbb{Z}$ cannot be measured for crystalline systems, as the total charge inside a unit cell can be changed by an integer multiple of $e$. Such arguments do not apply to the results of thought experiments with monopoles, as the induced electric charge density is not described by a Dirac delta function [Eq.~\ref{rhoe}]. The maximum value of $\delta Q$ occurs for $R=R^\ast$, such that $\delta Q$ remains distributed over multiple unit cells. The compensating charge of opposite sign also stays distributed over multiple unit cells, close to the boundary for TIs, and over the entire sample for topological quantum critical points (TQCPs). 

\item In Sec.~\ref{SOTI}, we identify the third homotopy class of $\mathcal{C}$-preserving, but $\mathcal{P}$- and $\mathcal{T}$- breaking chiral, HOTIs~\cite{Schindlereaat0346}. This is a special class of $\mathcal{CP}$-odd, $\mathcal{PT}$-even ME insulators that can exhibit $\theta=\pm \pi$, protected by the combined $C^z_4\mathcal{T}$ symmetry, where $C^z_4$ denotes four-fold rotations about $z$-axis. Akin to FOTIs, chiral HOTIs also support FZMs on monopole and surface under OBC, and the half-filled state must be trained with a pseudo-scalar field ($M^\prime$) to observe WE.  In the absence of training field, the spin-charge separation of chiral HOTI under OBC follows Fig.~\ref{fig:presentation1}.

As the system supports direction-sensitive gapless surface states, the surface-localized FZMs can be entirely eliminated, with suitable boundary conditions. The non-degenerate ground state, under \emph{mixed boundary conditions} (MBC) can display WE, without any pseudo-scalar training field. 

To gain physical insights into tunneling configurations of $SU(2)$ Berry connection, we probe the second homotopy class of $C^z_4$-symmetric planes, with a magnetic $\pi$-flux tube. We show these planes, which are described by 2D quadrupolar HOTIs~\cite{Benalcazar61,BBHPrb} support quantized $SU(2)$ Berry flux~\cite{tyner2020topology,tyner2021quantized,sur2022mixed}. The presence (absence) of non-trivial third homotopy class is precisely related to the presence (absence) of tunneling of $SU(2)$ Berry flux~\cite{tyner2021symmetry}.

\item In Sec.~\ref{AppA}, we consider an example of $C^z_4$-symmetric, magnetic TIs, breaking $\mathcal{C}$, $\mathcal{P}$, $\mathcal{T}$, and the combined discrete symmetries $\mathcal{CP}$, $\mathcal{PT}$, and $\mathcal{CT}$. Even under OBC, the states localized on monopole and surface are not degenerate. \emph{Hence, the third homotopy class of such insulators can be detected by computing WE without any pseudo-scalar training field}. With a magnetic $\pi$-flux tube, we demonstrate the existence of  $C^z_4$-symmetry-protected, tunneling configuration of Berry connection. 

\item In Sec.~\ref{TOTI1}, we consider octupolar HOTIs that preserve $\mathcal{C}$, $\mathcal{P}$, and $\mathcal{T}$ symmetries, and display gapped surface-states, and corner-localized zero-energy-states, under cubic-symmetry-preserving OBC~\cite{Benalcazar61,BBHPrb}. This system combines the exotic physics of FZMs, $\mathcal{CP}$-violation, and $SU(2)$ flavor-symmetry-breaking. 

In Refs.~\onlinecite{Benalcazar61,BBHPrb}, Benalcazar \emph{et al.} have claimed that the octupolar HOTIs do not support Chern-Simons coefficient and dipolar ME response. \emph{Contrary to such claims, we show that the octupolar HOTIs can exhibit quantized ME effect, with $\theta= \pm 2 \pi$, in the presence of infinitesimal, pseudo-scalar training field}. This result is independent of the existence of corner-localized states, which we substantiate by studying fermion spectrum and WE under OBC and MBC. We also address spin-charge separation and TQPT by doping electrons and holes. Moreover, 
with magnetic $\pi$-flux tube, we demonstrate the $C_3$-symmetry-protected, tunneling configuration of $SO(5)$ Berry connection, along $[111]$ direction~\cite{sur2022mixed}.

\item Throughout this work, we emphasize that only the monopole-localized FZMs precisely describe universal aspects of bulk topology. The surface- and corner- localized modes are additional contributions, which depend on the details of model Hamiltonians (FOTIs vs. HOTIs), and non-universal aspects of boundary conditions. Indiscriminate use of bulk-boundary correspondence usually leads to incorrect conclusions about bulk topology. 
 \end{enumerate}

\begin{figure}
\centering
\subfigure[]{
\includegraphics[scale=0.8]{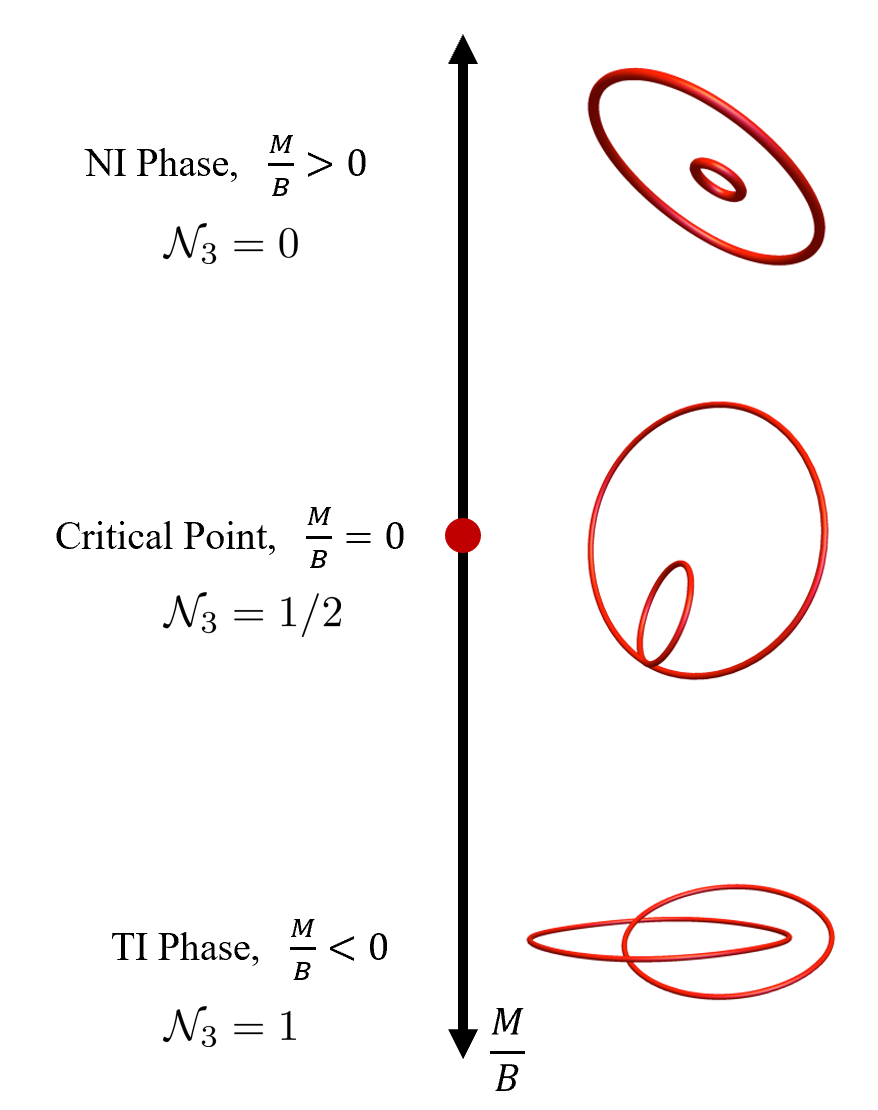}
\label{fig:linkingV}}
\subfigure[]{
\includegraphics[scale=0.32]{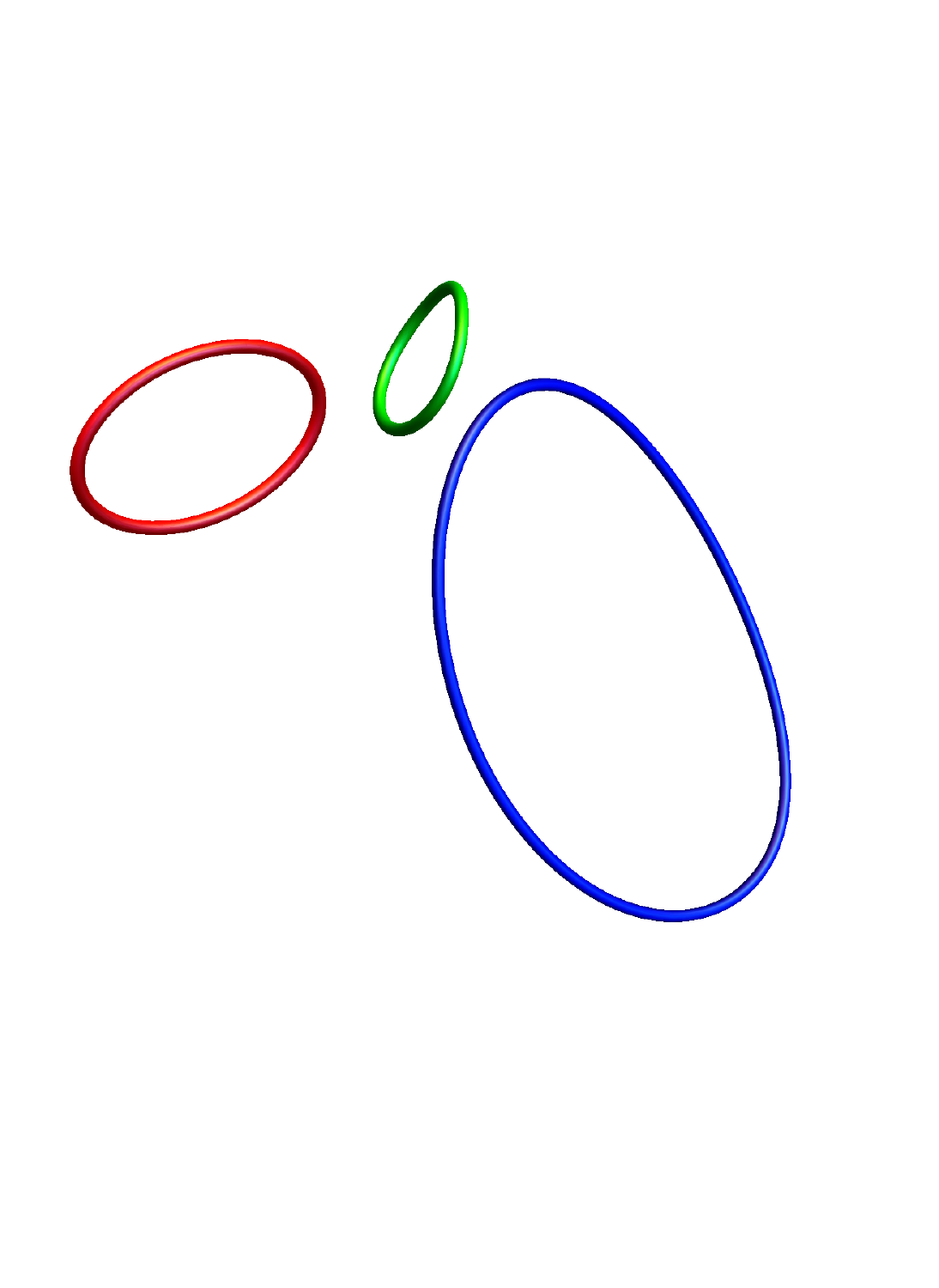}
\label{fig:unlinked}}
\subfigure[]{
\includegraphics[scale=0.32]{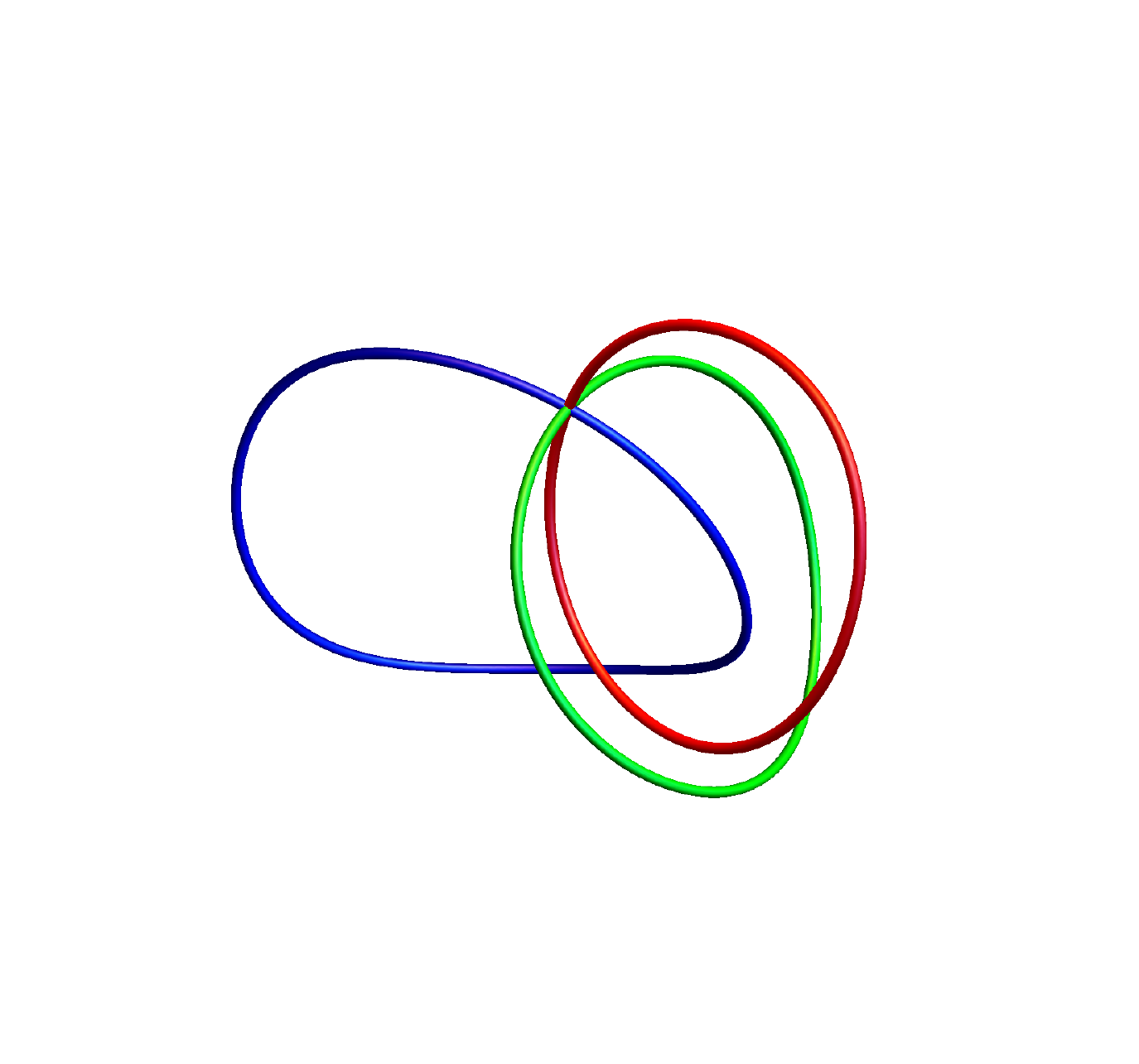}
\label{fig:criticallinked}}
\subfigure[]{
\includegraphics[scale=0.32]{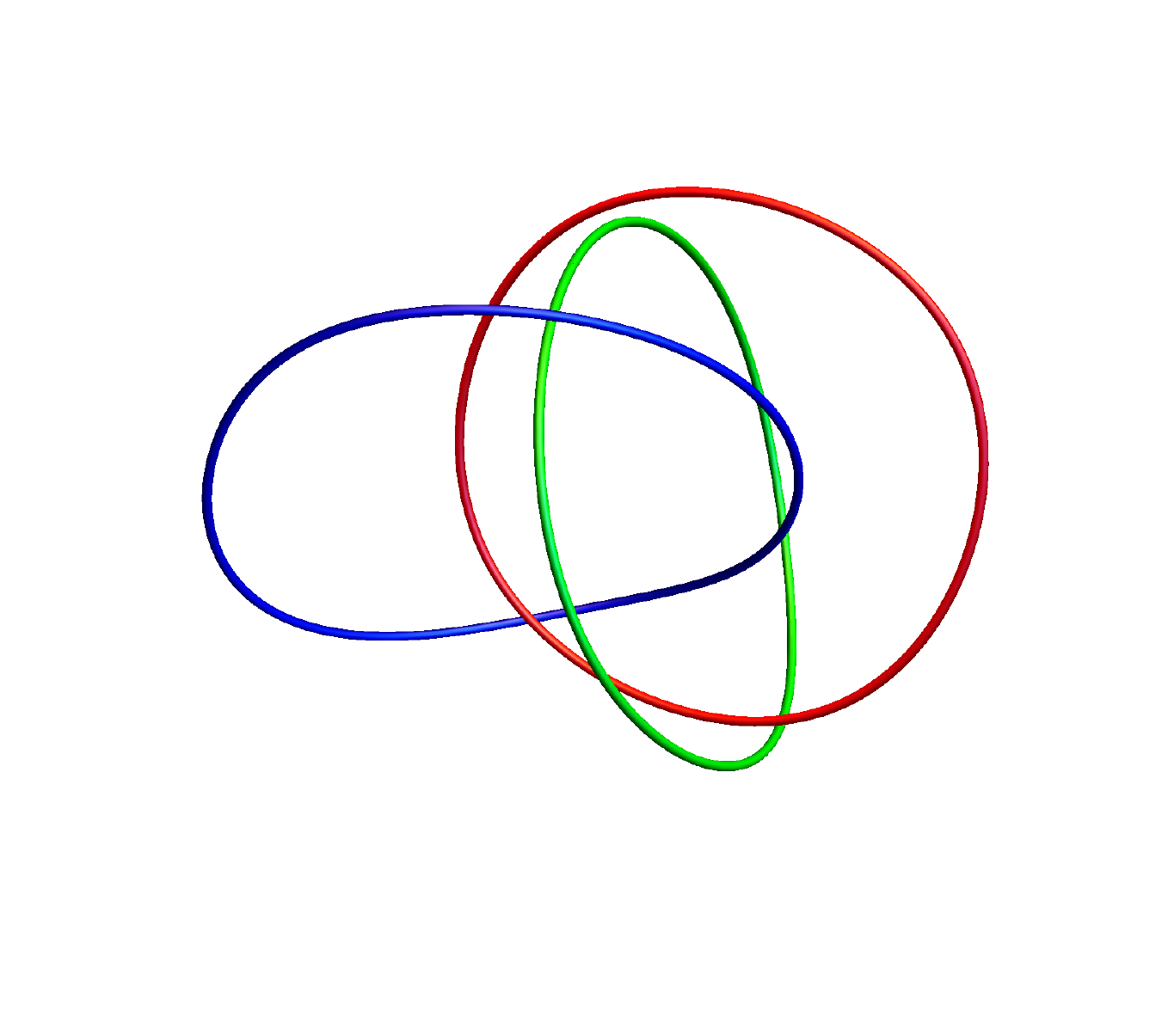}
\label{fig:Hopflinked}}
\caption{(a) Phase diagram of continuum model [Eq.~\ref{cont1} ]. The winding number of $O(4)$ unit vector [Eq.~\ref{winding1} ] controls the third homotopy class of $SU(2)$ Berry connection, which causes self-linking of lines of Berry curvatures for all three color components or spin projections. They are unlinked in the trivial (NI) phase, touching at the quantum critical point, and Hopf-linked in the topologically non-trivial (TI) phase. Mutual linking of three color components for (b) NI, (c) critical point, and (d) TI.}
\label{fig:MutualLink}
\end{figure}

\section{Continuum theory}\label{Continuum}
To gain analytical and numerical insights, we first work with the following continuum model of spherical TIs
\begin{eqnarray}\label{cont1}
H_{sph}(\bs{k})&=&  \bs{N}(\bs{k}) \cdot \boldsymbol \Gamma =\hbar v \sum_{j=1}^{3} \; k_j \Gamma_j + (M+B \bs{k}^2) \Gamma_{5}, \nn \\
\end{eqnarray}
which operates on a four-component spinor $\Psi(\bs{k})$, and $\Gamma_{j=1,2,3}=\tau_{1}\otimes \sigma_{j}$, $\Gamma_{4}=\tau_{2}\otimes \sigma_{0}$, and $\Gamma_5=\tau_3 \otimes \sigma_0$ are five, mutually anti-commuting $4 \times 4$ matrices. The $2\times 2$ identity and Pauli matrices $\tau_{i=0,1,2,3}(\sigma_{i=0,1,2,3})$ act on orbital/parity (spin) index. The $\mathcal{P}$, $\mathcal{T}$, and $\mathcal{C}$ symmetries are described by the operations $\Gamma_5 H (-\bs{k}) \Gamma_5= H (\bs{k})$, $ \Gamma_{31} H^\ast(-\bs{k}) \Gamma_{31}=H(\bs{k})$, and $ \Gamma_{25}H^\ast(-\bs{k})\Gamma_{25}=-H(\bs{k})$, respectively, and $\Gamma_{ab}=[\Gamma_a, \Gamma_b]/(2i)$. 

The $O(4)$ unit vector $\hat{\bs{N}}(\bs{k})$, describing maps from $\mathbb{R}^3 \to S^3$ can be written as
\begin{eqnarray}
\hat{\bs{N}}&=&[\mathrm{sgn}(v) \; \sin (\alpha(\bs{k})) \; \hat{\bs{k}}, \cos (\alpha(\bs{k}))], \\
\cos (\alpha(\bs{k})) &=& \frac{M+B \bs{k}^2}{\sqrt{\hbar^2 v^2 \bs{k}^2 + (M+B \bs{k}^2)^2}}.
\end{eqnarray}
where $0 \leq \alpha \leq \pi$ is the polar angle of $S^3$.
The topologically non-trivial, instanton configuration of $\hat{N}(\bs{k})$ is classified by the third spherical homotopy group $\pi_3(S^3)=\mathbb{Z}$ and the winding number is given by
\begin{eqnarray}
\mathcal{N}_3 &=& \frac{1}{2\pi^2} \;  \int d^3k \; \epsilon^{\mu \nu \rho \lambda} \hat{N}_{\mu}\partial_1 \hat{N}_{\nu} \partial_2 \hat{N}_{\rho} \partial_3  \hat{N}_{\lambda}, \label{winding} \\
&=&\frac{ 1}{2} \text{sgn}(vB) \; \left[1- (1- \delta_{M,0}) \; \mathrm{sgn}(M B) \right], \label{winding1}
\end{eqnarray}
where $\epsilon^{\mu \nu \rho \lambda}$ is the four-dimensional, Levi-Civita symbol, and the Greek indices take values from $1$ through $4$. At the TQCP with $M=0$, $\alpha(\bs{k}) $ can only scan the northern or southern hemisphere of $S^3$, leading to the \emph{sphaleron configuration, in reciprocal space, possessing half-integer winding number} $\mathcal{N}_3 = \mathrm{sgn}(vB)/2$. In Fig.~\ref{fig:MutualLink}, we illustrate topological properties of $SU(2)$ Berry connection.

\subsection{Monopoles and fermion zero-modes}\label{AppB}
Next we carry out analytical calculations of FZMs in the presence of a Dirac monopole at the origin $\bs{r}=(0,0,0)$, for an infinite system. For convenience, we first perform a global unitary rotation $\psi(\bs{r}) \to e^{i \frac{\pi}{4} \Gamma_{45} }\chi(\bs{r})$, such that mass operators $\psi^\dagger \Gamma_5 \psi \to - \chi^\dagger \Gamma_4 \chi$, and $\psi^\dagger \Gamma_4 \psi \to  \chi^\dagger \Gamma_5 \chi$, and obtain the rotated, and gauged Hamiltonian operator
\begin{eqnarray}\label{eq:contGauge}
H_{sph}[\bs{a}]&=&-i \hbar v \sum_{j=1}^{3} D_j \Gamma_j - (M - B \bs{D}^2) \Gamma_4 + M^\prime \Gamma_5, \nn \\
&=& \begin{bmatrix}
M^\prime \sigma_0 & \mathcal{D}^\dagger \\ 
\mathcal{D} & -M^\prime \sigma_0
\end{bmatrix} , 
\end{eqnarray}
where $D_j=(\partial_j - i \frac{e}{\hbar} a_j)$ is the covariant derivative, $\bs{a}$ is the vector potential, and
\begin{eqnarray}
\mathcal{D}=-i\hbar v \sum_{j=1}^{3} D_j \sigma_j - i(M - B \bs{D}^2)\sigma_0.
\end{eqnarray} 
In the absence of monopole, $\mathcal{D} \to - i |\bs{N}(k)| u(\bs{k})$, and the third homotopy class of $SU(2)$ matrix 
\begin{equation}
u(\bs{k})=\hat{N}_5(\bs{k}) \sigma_0 + i \hat{N}_j(\bs{k}) \sigma_j
\end{equation}
is given by $\mathcal{N}_3$. The third homotopy class of $u^\dagger(\bs{k})$ is given by $-\mathcal{N}_3$. Following Ref.~\onlinecite{wu1976dirac}, we work with the non-singular gauge,
\begin{equation}\label{eq:7}
    \bs{a}(\bs{r})=\begin{cases}
    \frac{g}{r\sin\theta}(1-\cos\theta) \; \hat{\phi},\;0\leq\theta<\frac{\pi}{2}+\delta\\
    \frac{-g}{r\sin\theta}(1+\cos\theta) \; \hat{\phi},\;\frac{\pi}{2}-\delta<\theta\leq\pi 
        \end{cases}.
\end{equation}
The wave sections for the northern and southern hemispheres are related by the gauge transformation $\psi = e^{i 2 m \phi} \psi$.

When $M^\prime=0$, the $\mathbb{Z}_2$ particle-hole symmetry follows from $\{H[\bs{a}], \Gamma_5 \}=0$. As a consequence of this symmetry, if $\chi_{n}(\bs{x})$ is an eigenstate of $H[\bs{a}]$ with energy $E_{n}$, $\chi^\prime_{n} = \Gamma_5 \chi_{n} (\bs{x})$ will be an eigenstate with energy $-E_{n} $. Hence, the finite-energy eigenstates $\chi_n$ and $\Gamma_5 \chi_n$ are orthogonal, and $\langle \chi^\dagger_n \Gamma_5 \chi_n \rangle =0$. If normalizable, FZMs exist, they must be eigenstates of $\Gamma_5$ with eigenvalue $s=+1$ (positive chirality) or $s=-1$ (negative chirality). Furthermore, these solutions must satisfy $\mathcal{D} \chi_{0,+} =\mathcal{D}_+ \chi_{0,+}=0$, and $\mathcal{D}^\dagger \chi_{0,-} =\mathcal{D}_-  \chi_{0,-}=0$, where $\mathcal{D}_+=\mathcal{D}^\dagger\mathcal{D}$, and $\mathcal{D}_-=\mathcal{D} \mathcal{D}^\dagger$. If there are $n_s$ FZMs with chirality $s$, 
\begin{equation}
\langle \chi^\dagger \Gamma_5 \chi \rangle=\sum_s \langle \chi^\dagger_{0,s} \Gamma_5 \chi_{0,s} \rangle = (n_+ -n_-).
\end{equation}
The goal of index theorem is to find a precise relationship between $(n_+ - n_-)$, the bulk invariant $\mathcal{N}_3$, and the monopole strength $m$. 

Due to spherical symmetry, $H[\bs{a}]$, $\bs{J}^2$, and $J_z$ can be simultaneously diagonalized~\cite{Kazama1977,shnir2006magnetic}, where the total angular momentum, and the orbital angular momentum operators are given by
\begin{eqnarray}
\bs{J}&=&\bs{L} + \frac{\hbar}{2} \tau_0 \otimes \boldsymbol \sigma, \\
\bs{L}&=&-i \hbar \; \bs{r} \times \bs{D} - \frac{\hbar}{2} m \hat{\bs{r}}.
\end{eqnarray}
The FZMs can be labeled by the following quantum numbers: (i) the total angular momentum $j=(|m|-1)/2$, (ii) the projection of total angular momentum $-j \leq j_z \leq +j$, and (iii) the orbital angular momentum $l = j+1/2= |m|/2$. 

Without the higher gradient term $-B \bs{D}^2 \Gamma_4$, the radial differential operator of linearized Dirac theory exhibits pathological short-distance behavior for $j=|m|-1/2$ channel, due to the vanishing of centrifugal barrier~\cite{Kazama1977,shnir2006magnetic}. This is circumvented by defining self-adjoint extension of Hamiltonian that modifies boundary conditions with \emph{an external parameter} $\vartheta$~\cite{Goldhaber1977,Callias1977,yamagishi1983fermion,Grossman1983}. Only $\vartheta=0, \pi$ can preserve $\mathcal{CP}$ symmetry. The results for Witten effect is obtained by identifying $\vartheta$ with axion angle $\theta$~\cite{yamagishi1983fermion,Grossman1983}. The parameter $\vartheta$ can only be determined from a detailed analysis of UV-complete, microscopic models. 
Alternatively, Kazama \emph{et al.}~\cite{Kazama1977} have used an infinitesimal, anomalous magnetic moment or Zeeman coupling 
\begin{equation}\label{AZ}
H_{AZ}= \frac{\kappa \;\hbar e  }{2 m_e} \; g \; \sum_{j=1}^{3} \frac{\hat{\bs{r}}_j}{r^2} \psi^\dagger \Gamma_{j4} \psi  = \frac{\kappa \; \hbar e  }{2 m_e} \; g \; \sum_{j=1}^{3} \frac{\hat{\bs{r}}_j}{r^2} \chi^\dagger \Gamma_{j5} \chi, 
\end{equation}
which maintains a centrifugal barrier during intermediate steps of calculations, and $\kappa \to 0^+$ limit is taken at the end of calculations. The anomalous Zeeman coupling term obeys $\mathbb{Z}_2$ particle-hole symmetry, as $\{ H_{AZ}, \Gamma_5 \}=0$. While these UV regulators allow normalizable FZMs, the number of FZMs cannot be clearly related to any unambiguous, bulk topological invariant.

When $B \neq 0$, we have a natural short-distance regulator. As $r \to 0$, $H[\bs{a}] \to -B \bs{D}^2 \Gamma_4$ describes decoupled, non-relativistic Schr\"{o}dinger Hamiltonians, supporting non-vanishing centrifugal barriers. Consequently, the radial wave functions for all angular momentum channels vanish as $r \to 0$. Only in the large distance limit $r \to \infty$, the asymptotic behavior can be approximated by linearized Dirac theory.  
For $j=(|m|-1)/2$, $l=j+1/2=|m|/2$ channel, we substitute the ansatz 
\begin{eqnarray}
  && \chi^T_{m,j,j_z}(\mathbf{r})=[
   F_+(r)\eta^{T}_{m,j,j_z}, 
   F_-(r)\eta^{T}_{m,j,j_z}], \\
 && \eta^T_{m,j,j_z}(\theta,\phi)=\bigg[
-\bigg(\frac{j-j_z+1}{2j+2}\bigg)^{\frac{1}{2}}Y_{m/2,j+1/2,j_z-1/2}(\theta,\phi), \nn \\
&& \bigg(\frac{j+j_z+1}{2j+2}\bigg)^{\frac{1}{2}}Y_{m/2,j+1/2,j_z+1/2}(\theta,\phi)
 \bigg],
\end{eqnarray}
and obtain coupled radial equations 
\begin{eqnarray}\label{eq:conteigenstates}
&& \Big[M^\prime \tau_{3}-i \; \hbar v \; \mathrm{sgn}(m) \; \left(\partial_{r} +\frac{1}{r} \right)\tau_{1} + \Big \{ M-B \Big (\partial_{r}^2  \nn \\ && +\frac{2}{r}  \partial_{r}- \frac{|m|}{2r^2} \Big )  \Big \}\tau_2 \Big]\begin{pmatrix}
F_+(r)\\
F_-(r)
\end{pmatrix}=E \begin{pmatrix}
F_+(r)\\
F_-(r)
\end{pmatrix}.
\end{eqnarray}
Here, $F_\pm(r)$ are two independent radial functions, $\eta_{m,j,j_z}$ is the spinor monopole harmonics of the third type, and $Y_{m/2,j+1/2, j_z \pm 1/2}$ are spherical monopole harmonics~\cite{Kazama1977}. 

When $M^\prime=0$, only TIs can support normalizable FZMs, with 
\begin{eqnarray}
F_s (r)&=&  [1 -s \; \mathrm{sgn}(m \mathcal{N}_3) ] \; \frac{e^{-\frac{r\Lambda }{2} }}{2\sqrt{r}} \; [c_1 \; J_{\nu}(\tilde{r}) \; \Theta(4-\xi \Lambda) \nn \\
&&+ \; c_2 \; I_{\nu}(\tilde{r}) \; \Theta(\xi \Lambda-4)+c_3 \; (r\Lambda)^{\nu} \; \delta_{4,\xi \Lambda}]
\end{eqnarray} 
where $c_i$'s are normalization constants, $J_{\nu}(\tilde{r})$ is the Bessel function of first kind, $\nu=\frac{1}{2} \sqrt{1+2 |m|}$, and $\Theta(x)$ is the Heaviside step function. We have defined dimensionless radial variable $\tilde{r}=- \frac{i}{2} \; \Lambda r f(\xi\Lambda)$, and
$f(\xi \Lambda)= \sqrt{|1- 4 (\xi \Lambda)^{-1}|}$, 
where $\xi=\hbar |v|/|M|$ is the correlation length, and $\Lambda^{-1}= |B|/(\hbar |v|)$ is a short-distance scale.

When $r \to 0$, wave functions vanish as $r^{\nu -1/2}$. For large distance ($r \to \infty$), the decay of wave functions follows $F_s \sim e^{-r/\xi}/r$, when $\xi \Lambda > 4$, which also describes the long-wavelength behavior in the vicinity of TQCP ($\xi \Lambda \to \infty$). Far from the TQCP ($\xi \Lambda \leq 4$), the exponential decay is controlled by $\Lambda$, and $F_s \sim e^{-r \Lambda/2}$. The scaling behavior of normalization constants is given by
\begin{eqnarray}
&&|c_j| = \Big[\frac{\Lambda^2 \; 2^{2\nu-1} \; \Gamma(1+\nu) \; \Gamma(\frac{1}{2}) \; f^{-2\nu} \; |m|^{-1}}{ \Gamma (\nu +\frac{3}{2}) \; _2F_1(\nu+\frac{1}{2}, \nu+\frac{3}{2}, 2\nu+1, (-1)^jf^{2})} 
\Big]^{1/2}, \nn \\ \end{eqnarray}
with j=1,2, and
\begin{eqnarray}
|c_3|=\frac{\Lambda^{3/2}}{\sqrt{4 |m| \Gamma(2+2\nu)}}.
\end{eqnarray}

With explicit solutions of normalizable FZMs, we can formulate the following index theorem
\begin{eqnarray}
\langle \psi^\dagger \Gamma_4 \psi \rangle = \langle \chi^\dagger \Gamma_5 \chi \rangle =  - m \mathcal{N}_3. \label{index1}
\end{eqnarray}
When $M^\prime \neq 0$, the positive (negative) chirality zero-modes move to $E= M^\prime$ ($-M^\prime$). We also note that the infinitesimal, anomalous Zeeman coupling term of Eq.~\ref{AZ} leaves the structure of FZMs unaffected. Only the index of Bessel function is modified: $\nu \to \frac{1}{2} \sqrt{1+2 |\tilde{m}|}$, where $\tilde{m}=m (1+ \frac{\kappa \hbar^2}{2 m_e B} ) $. 

\begin{figure*}
\centering
\subfigure[]{
\includegraphics[scale=0.44]{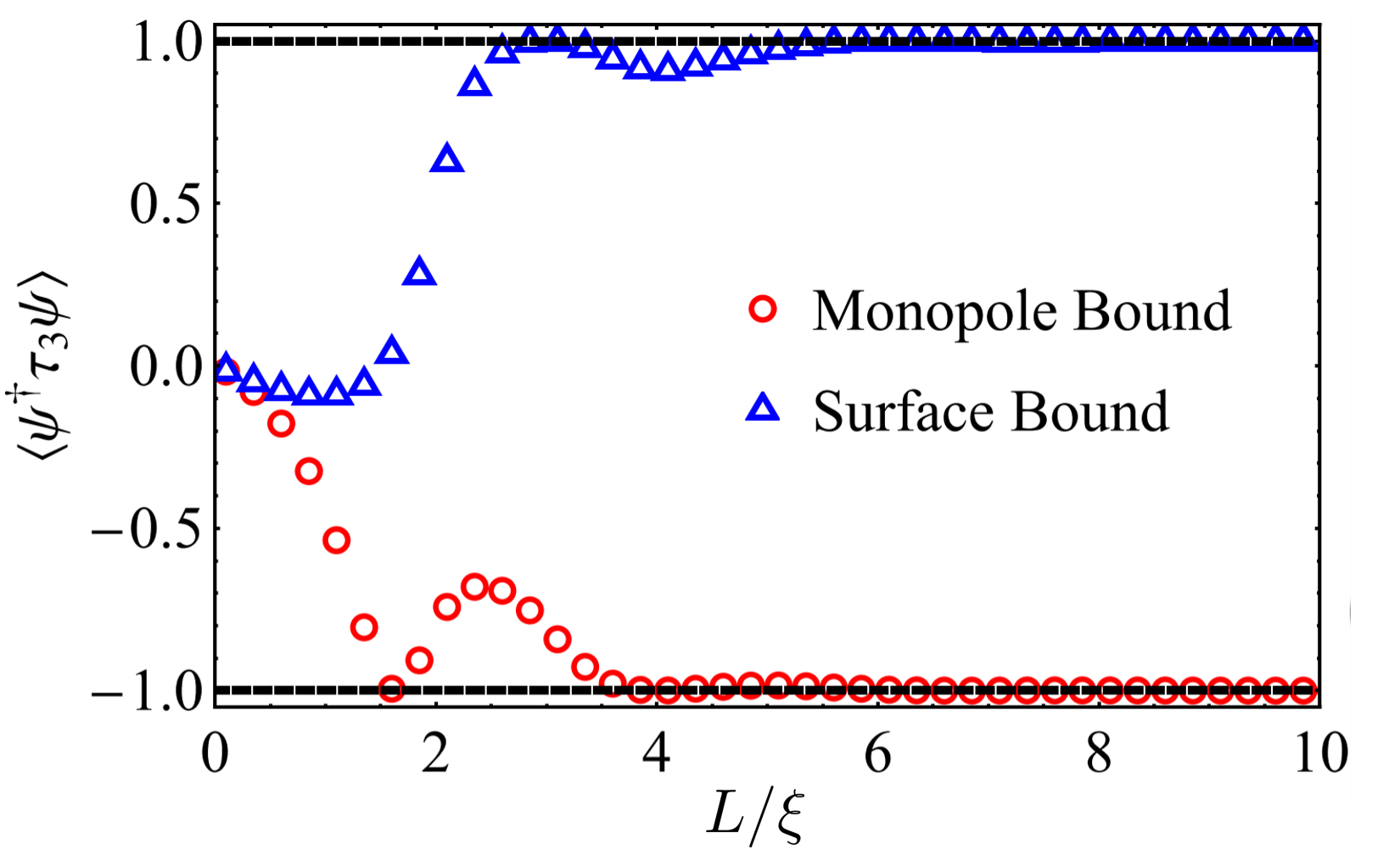}
\label{fig:conta}}
\subfigure[]{
\includegraphics[scale=0.43]{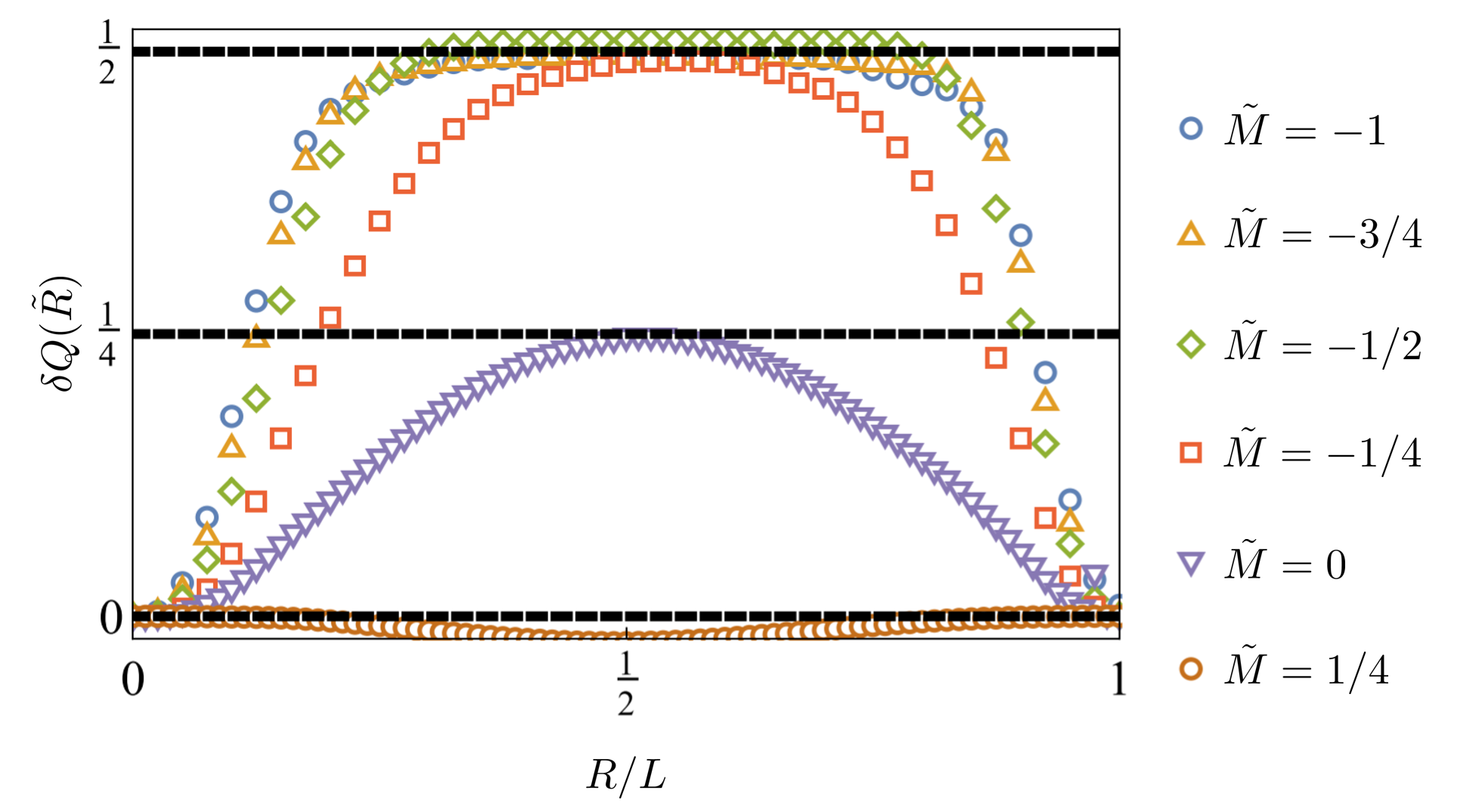}
\label{fig:contb}}
\subfigure[]{
\includegraphics[scale=0.75]{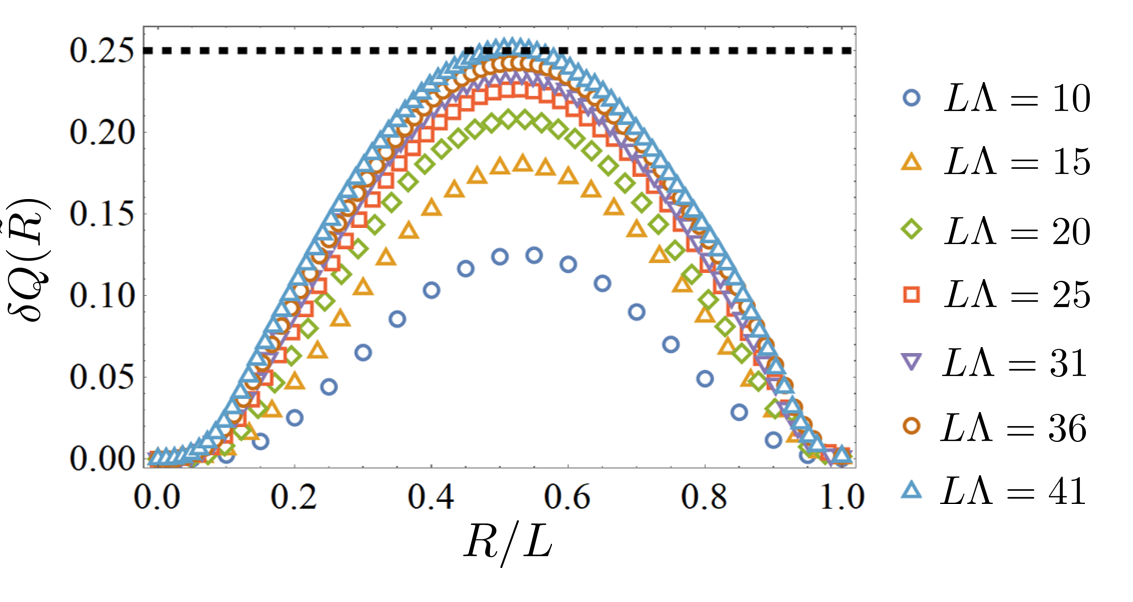}
\label{fig:contc}}
\caption{Numerical results for spin-charge separation for a spherically-symmetric, first-order topological insulator [Eq.~\ref{cont1} ], with winding number $\mathcal{N}_3=-1$, in the presence of unit monopole $m=+1$, under open boundary conditions. All lengths are measured in units of short-distance scale $\Lambda^{-1}=|B|/(\hbar |v|)$. (a) The system-size dependence of chirality of near-zero-modes, obtained from Eq.~\ref{eq:conteigenstates}, in the presence of a small pseudo-scalar mass $M^\prime>0$. Here, $L$ is the radius of the system and $\xi=\hbar v/|M|$ is the correlation length. When $L/\xi \gg 1$, the chirality (acting as the third component of spin) is clearly resolved. (b) The induced electric charge on monopole $\delta Q(\tilde{R})$ (in units of $-e$), enclosed by a Gaussian sphere of radius $R$ [see Eq.~\ref{dopedcont} ], when $M^\prime=0$, and the system is doped with one electron [$N_e=+1$ case of Fig.~\ref{fig:presentation1} ]. The dimensionless ratios $\tilde{R}=R \Lambda$ and $\tilde{M}= \text{sgn}(MB) /(\xi \Lambda)$. For a topological insulator ($\tilde{M}<0$), the maximum induced charge saturates to $-e/2$. (c) At the topological quantum critical point ($\tilde{M}=0$), the maximum induced charge saturates to $-e/4$, when $L \geq 40 \Lambda^{-1}$, confirming the half-integer winding number of quantum critical, Dirac semimetal. The quantized values of maximum induced charge have been calculated up to a numerical accuracy $10^{-4}$. Notice that the maximum values of induced charge occur for $R^\ast \approx L/2$.
}
\label{fig:ContMono}
\end{figure*}

Intriguingly, Eq.~\ref{index1} describes a topological mechanism for breaking $\mathcal{P}$, $\mathcal{T}$, $\mathcal{CP}$, and $\mathcal{CT}$ symmetries. Using the exact eigenstates of gauged Hamiltonian, fermion field operators can be expanded as 
\begin{eqnarray}
\Psi(\bs{r},t)= \sum_n \; c_n \psi_n(\bs{r}) e^{- i E_n t/\hbar}, \\
\Psi^\dagger(\bs{r},t)= \sum_n \; c^\ast_n \psi^\dagger_n(\bs{r}) e^{ i E_n t/\hbar}, 
\end{eqnarray}
where $c_n$ ($c^\ast_n$) is the fermion annihilation (creation) operator, and $\{ c_n, c^\ast_l \} =\delta_{n,l}$. Therefore, the vacuum expectation value of pseudo-scalar mass operator is given by
\begin{eqnarray}
 \langle \Psi^\dagger \Gamma_4 \Psi \rangle = \int d^3r \; \Psi^\dagger(\bs{r},t) \Gamma_4 \Psi(\bs{r},t)=-\frac{m \mathcal{N}_3}{2} ,
 \end{eqnarray}
and the factor of $1/2$ arises from the half-filling of FZMs. The calculation of induced charge is more involved, as all occupied states must be taken into account. Following Refs.~\cite{yamagishi1983fermion,Grossman1983}, we can show that $\delta Q=-\frac{m e \mathcal{N}_3}{2} $, when the radius of Gaussian surface is sent to $\infty$.

\subsection{Flavor-symmetry} If we consider $N_f$ copies of spherical topological insulators, the model will support $SU(N_f)$ flavor symmetry. In the presence of monopoles, each flavor will give rise to FZMs, and the flavor-singlet, topological response will be determined by 
$\sum_{i=1}^{N} \langle \Psi_i^\dagger \Gamma_4 \Psi_i \rangle = -\frac{ N_f m \mathcal{N}_3}{2}$,
and the total induced electric charge $\delta Q=-\frac{N_f m e \mathcal{N}_3}{2}$. By introducing hybridization between different species, flavor-symmetry will be reduced to a subgroup of $SU(N_f)$. The generic form of particle-hole-symmetric, flavor-symmetry-breaking is described by 
\begin{eqnarray}
&&H=\mathbb{1}_{N_f \times N_f} \otimes H_{sph}(\bs{k}) + H_{FSB}, \nn \\
&&H_{FSB}= \sum_{i=1}^{N^2-1} \; M_i (\bs{k}) \; \hat{\lambda}_i \otimes \Gamma_4,
\end{eqnarray} where $\hat{\lambda}_i$'s are the Hermitian generators of $\mathfrak{su}(N_f)$ algebra, and $M_i(\bs{k})$ are momentum-dependent hybridization parameters. When $M_i$'s are momentum independent constants, $H_{FSB}$ can be diagonalized in flavor-space with $SU(N_f)$ rotations, to obtain
\begin{equation}
H_{FSB} \to \text{diag} [m_{11},..,m_{NN}] \otimes \Gamma_4,
\end{equation} where $m_{jj}$'s are $N_f$ eigenvalues of the matrix $\sum_{i} M_i (\bs{k}) \; \hat{\lambda}_i$. If all eigenvalues are distinct, $SU(N_f)$ symmetry is reduced to $U(1) \times U(1) \times...\times U(1) = [U(1)]^{N-1}$. Therefore, the resulting theory is made out of $N_f$ species of decoupled ME insulators, with different strengths of pseudo-scalar mass $M^\prime_j=m_{jj}$, and non-quantized ME coefficients $\theta_j$. When probed with monopoles, FZMs will be split by $H_{FSB}$ and appear at reference energies $m_{jj} \text{sgn} (m \mathcal{N}_3)$. Naturally, such theories give rise to non-quantized ME response. 

Important question is whether \emph{$k$-dependent hybridization terms can qualitatively alter such conclusions and display quantized, flavor-singlet, ME response}. There is no generic answer to this question and one must perform explicit calculations. By inserting monopoles and measuring WE one can avoid all technical difficulties associated with the direct computation of Chern-Simons coefficient. In Sec.~\ref{TOTI1}, we will demonstrate that the octupolar HOTI breaks $SU(2)$ flavor-symmetry with momentum-dependent hybridization terms. But the hybridization terms obey underlying cubic symmetry and monopoles bind $N_f |m \mathcal{N}_3|$ FZMs,  and $\delta Q=-\frac{N_f m e \mathcal{N}_3}{2}$, with $N_f=2$. Next, we address finite-size effects for spherical FOTIs, with $N_f=1$.

 \begin{figure*}[t]
\centering
\subfigure[]{
\includegraphics[scale=0.18]{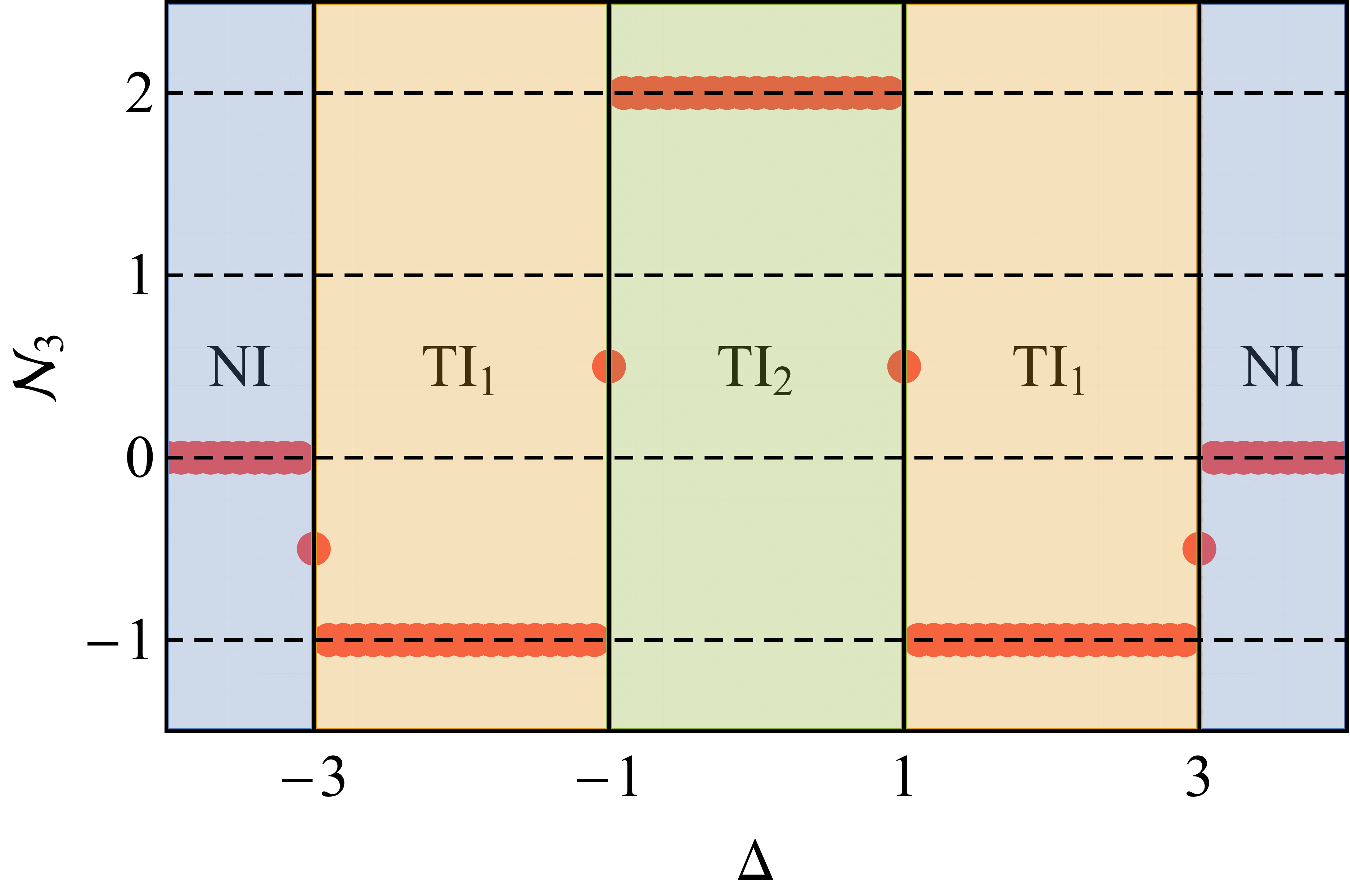}
\label{fig:1a}}
\subfigure[]{
\includegraphics[scale=0.47]{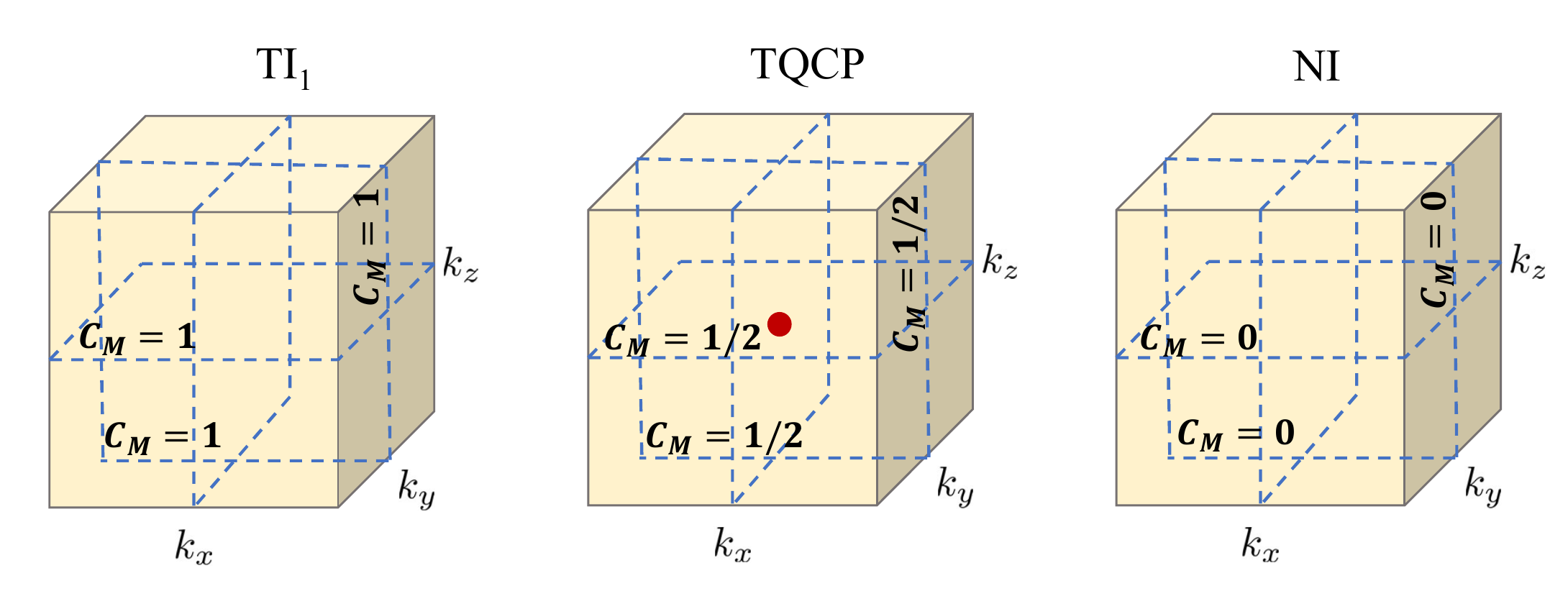}
\label{fig:1b}}
\caption{(a) Phase diagram for the lattice model of first-order topological insulators [see Eq.~\ref{eq:tbmodel} ], where $\Delta$ is a dimensionless, tuning parameter, and $\mathcal{N}_3$ is the bulk topological invariant [see Eq.~\ref{winding} ]. Topological quantum critical points support half-integer winding numbers $\mathcal{N}_3=\pm 1/2$. (b) Illustration of tunneling configurations of $SU(2)$ Berry connection of valence bands in terms of mirror Chern numbers of four-fold symmetric mirror planes~\cite{tyner2021symmetry}, as $\mathcal{N}_3=C_M(k_j=\pi) -C_M(k_j=0)$, with $j=1,2,3$. At $\Delta=-3 $, all three mirror planes, passing through the zone center exhibit quantum Hall plateau transitions, with $C_M(k_j=0)=1/2$. TI$_2$ phase supports $C_M(k_j=\pi)=+1$ and $C_M(k_j=0)=-1$, leading to $\mathcal{N}_3=+2$, which cannot be deduced from weak $\mathbb{Z}_2$ indices~\cite{tyner2021symmetry}.
 }
\label{fig:Berry}
\end{figure*}

\subsection{Finite-size effects, spin-charge separation, and quantum criticality} In the absence of monopole, the surface-states of a large finite-size system of radius $L$ ($L \Lambda >>1$) under spherical-symmetry-preserving OBC are described by two-component, Dirac fermions living on $S^2$. Employing the eigenstates of $\hat{\bs{r}} \cdot \boldsymbol \sigma$, the surface Hamiltonian can be written as
\begin{eqnarray}
H_{surface}(\theta, \phi)= -i \frac{\hbar v}{L} [\sigma_2 (\partial_\theta + \frac{1}{2} \cot \theta) - \sigma_1 \frac{1}{\sin \theta} \partial_\phi  ], \nn \\
\end{eqnarray}
leading to the minimum, spectral gap $2 \hbar v /L$~\cite{SphericalTI1}. In the presence of monopole, $H_{surface}[\bs{a}]$ supports FZMs~\cite{,PhysRevB.78.195426,JELLAL2008361}, with degeneracy $|m \mathcal{N}_3|$, and the chirality $\text{sgn}(m \mathcal{N}_3)$.

With numerical solutions of Eq.~\ref{eq:conteigenstates}, we can study hybridization between FZMs of opposite chirality and spin-charge separation. The energy splitting due to hybridization is given by
\begin{equation} \label{xi0}
E_h \approx \frac{\hbar v}{L} \exp \left[-\frac{L}{\xi_0} \right], 
\end{equation}
where $\xi_0 \sim \max \{ \xi, \Lambda^{-1} \}$ is the localization length of FZMs. In Fig.~\ref{fig:conta}, we show the chirality of NZMs for a minimal monopole $m=+1$, as a function of $L/\xi=L|M|/(\hbar v)$. We have introduced a small pseudo-scalar mass $M^\prime \neq 0$ to control hybridization effects. In the thermodynamic limit, when $M^\prime > E_h$, the chirality of NZMs can be clearly resolved. For small system-size, $M^\prime \leq E_h$, and the chirality is suppressed. 

When $M^\prime=0$, \emph{the half-filled state is two-fold degenerate} [see Fig.~\ref{fig:presentation1} ]. Due to the obscured chirality of NZMs, monopoles cannot exhibit WE, i.e., $\delta Q(L, R, \xi, M^\prime=0)=0$. This happens due to a precise cancelation between the contributions of occupied NZM and Dirac sea, which will be precisely described for lattice models. 
To address the induced charge for doped $SU(2)$ singlets, we calculate the contributions of NZMs to the regulated charge density operator: 
\begin{equation}
\delta \rho(\bs{r},L,\xi,M^\prime=0)=\rho_m(\bs{r}) -\rho_0(\bs{r}), 
\end{equation}
where $\rho_m(\bs{r}) $ [$\rho_0(\bs{r})$] denotes the charge density operator in the presence [absence] of monopole. By integrating $\delta \rho(\bs{r}^\prime)$ over a Gaussian surface of radius $R$, we obtain the induced charge 
\begin{equation}\label{dopedcont}
\delta Q(L,R,\xi,M^\prime=0)= \int_{|\bs{r}^\prime | < R} \; d^3r^\prime  \delta \rho(\bs{r}^\prime,L,\xi,M^\prime=0).
\end{equation}
When the ratio $L/\xi \gg 1$, the maximum induced charge for TI saturates to $\delta Q_{max}=  \frac{-e}{2} \; \text{sgn}(N_e)$ [see Fig.~\ref{fig:contb}]. Since by doping, we have introduced electric charge $-e \; \text{sgn}(N_e)$, the surface-states will support $\delta Q_s = -\frac{e}{2} \; \text{sgn}(N_e)$. 

If the quantum critical, Dirac semimetal is doped by a single electron or hole, $\delta Q_{max}$ saturates to $-\frac{e}{4} \; \text{sgn}(N_e)$. As the critical system does not support any surface-states, rest of the doped charge $-\frac{3e}{4} \text{sgn}(N_e)$ will be distributed over the entire sample. In Fig. ~\ref{fig:contc}, we show the system-size dependence of $\delta Q(L,R,\xi=\infty,M^\prime=0)$ at TQCP. The saturation of maximum induced charge to $-\frac{e}{4} \; \text{sgn}(N_e)$ corroborates half-integer winding number for TQCP.  With analytical and numerical insights gained from continuum theory, we now consider lattice models of FOTIs that can exhibit odd and even integer winding numbers. 

\section{Lattice model of first order topological insulators}\label{FOTI} We will work with the following model on a simple cubic lattice, 
\begin{eqnarray}\label{eq:tbmodel}
H_1(\bs{k})=\bs{N}(\bs{k}) \cdot \boldsymbol \Gamma=  t \sum_{j=1}^{3} \; \sin k_j \Gamma_j+ t^\prime[\Delta +  \sum_{j=1}^{3} \cos k_j ] \Gamma_5. \nn \\
\end{eqnarray}
The Bloch Hamiltonian $H_1$ still operates on a four-component spinor $\Psi(\bs{k})$, and the definitions of gamma matrices, and discrete symmetry operations $\mathcal{P}$, $\mathcal{T}$, and $\mathcal{C}$ are unchanged. Here, $t$ and $t^\prime$ are two independent hopping parameters, and the dimensionless tuning parameter $\Delta$ controls TQPTs. For convenience, we have set the lattice constant $a$ to be one. 
The $O(4)$ unit vector $\hat{N}(\bs{k})$ now describes maps from Brillouin zone three-torus $T^3$ to $S^3$, and instanton configurations are again classified by the third spherical homotopy group $\pi_3(S^3)=\mathbb{Z}$. The winding number $\mathcal{N}_3$ now 
counts the number of times $T^3$ can wrap around $S^3$. The phase diagram is shown in Fig.~\ref{fig:1a}. The TI$_2$ phase is a crystalline-symmetry-protected insulator, possessing higher winding number. Thus, many physical properties of TI$_2$ cannot be properly described by continuum theories.

At TQCPs $\Delta= \pm 3 \; (\pm 1)$, the bulk band gap vanishes at one (three) time-reversal invariant momentum points $\bs{k}=\pi(n_1,n_2,n_3)$, where $n_j=0,1$. This gives rise to three-dimensional, massless Dirac fermions as quantum critical excitations. For $\Delta=-3 \; (+3)$, the Dirac point is located at the center (corner) of Brillouin zone. In contrast to this, the inequivalent Dirac points for $\Delta= -1 \; (+1)$ are located at three $X$ [$M$] points, with $\bs{k}=(\pi,0,0), \; (0,\pi, 0), \; (0, 0, \pi)$ [$\bs{k}=(\pi,\pi,0), \; (\pi,0,\pi), \; (0, \pi, \pi)$]. Notice that the number of inequivalent massless Dirac fermions and the winding number at a TQCP are given by the difference and the average of winding numbers of adjacent insulating states, respectively. Hence, TQCPs of lattice models also correspond to \emph{sphaleron configurations of $\hat{N}(\bs{k})$, possessing half-integer winding numbers}. 

As illustrated in Fig.~\ref{fig:1b}, $\mathcal{N}_3$ is directly related to the tunneling of mirror Chern numbers (or second homotopy class) of four-fold symmetric planes. We note that the mirror symmetry operations $H_1(k_a, k_b, k_c) = \Gamma_{c4} H_1(k_a, k_b, -k_c) \Gamma_{c4}$ are implemented by non-commuting matrices $\Gamma_{c4}=\frac{1}{2} \epsilon_{cab} \Gamma_5 \Gamma_{ab}$, reflecting the non-Abelian nature of the physical problem. At $\Delta = 3$ ($-3$), all mirror planes, passing through the zone corner (center) simultaneously display quantum Hall plateau transitions.

In literature, the TI$_2$ phase is often referred to as a weak TI~\cite{FuKane,FuKaneMele2007}, with strong $\mathbb{Z}_2$ index $\nu_0=0$, and weak $\mathbb{Z}_2$ indices $[1,1,1]$. The weak indices only describe the parity (even vs. odd integer) of mirror Chern numbers for $k_j=\pi$ planes. Being insensitive to the sign of mirror Chern numbers, the weak indices cannot address 3D tunneling configuration of TI$_2$ phase. This is generally true for any state whose third homotopy class is described by an even integer winding number~\cite{tyner2021symmetry}.

Since $\{H(\bs{k}), \Gamma_5 \} =0$, the model of Eq.~\ref{eq:tbmodel} exhibits $\mathbb{Z}_2$ particle-hole symmetry, which can introduce large gauge ambiguities for Chern-Simons invariant. We avoid such ambiguities by regulating our model Hamiltonian with a pseudo-scalar mass: $H(\bs{k}) \to H(\bs{k}) + M^\prime \Gamma_4$. After obtaining $ \mathcal{CS} (M^\prime)$, using Eq.~\ref{A11}, we can show that 
\begin{eqnarray}
\mathcal{CS}_\pm=\mathcal{CS}(M^\prime \to 0^\pm)-\mathcal{CS}(M \to \pm \infty)= \pm \frac{\mathcal{N}_3}{2}. \label{ChernSimonscont}
\end{eqnarray}
\emph{In our numerical calculations, we will consistently implement $M^\prime >0$, and show the measured induced charge for half-filled system tracks} $\theta_+ = 2 \pi \mathcal{CS}_+ = \pi \mathcal{N}_3$.

\subsection{Fermion spectrum with monopoles} To diagnose bulk topology of lattice model with magnetic monopoles, we first obtain a tight-binding Hamiltonian in real space, by performing Fourier transformation of $H(\bs{k})$. Subsequently, all hopping parameters ($t_{ij}$'s ), connecting two different sites $\bs{r}_i$ and $\bs{r}_j$ are modified by Peierls phase factors $e^{i\nu_{ij}}$, where $\nu_{ij}=(e/\hbar)\int_{\bs{r}_i}^{\bs{r}_j}\mathbf{a}\cdot d\mathbf{l}$, and $\mathbf{a}$ is the vector potential. The monopole will be placed at the center of the system $\bs{r}=(0,0,0)$, such that the sites of cubic lattice are labeled by $\bs{r}_i=\frac{a}{2}(n^x_i,n^y_i,n^z_i)$, and $n^a_i \in \mathbb{Z}$. All calculations will be performed under OBC, by employing the singular, north-pole gauge
 \begin{eqnarray}
 \bs{a}^N(\bs{r}_i)= \frac{g}{r_i} \cot \frac{\theta_i}{2} \; \hat{\phi}_i=g \; \frac{-y_i \hat{x} + x_i \hat{y}}{r_i(r_i+z_i)}, 
 \end{eqnarray}
with the Dirac string oriented along negative $z$-axis. Since the Dirac quantization condition $2eg/\hbar=m \in Z$ makes the string unobservable, the fermion spectrum remains unchanged for 
the south-pole gauge: 
\begin{equation}
\bs{a}^S(\bs{r}_i)= -\frac{g}{r_i} \tan \frac{\theta_i}{2} \; \hat{\phi}_i=g \; \frac{y_i \hat{x} - x_i \hat{y}}{r_i(r_i-z_i)},
\end{equation}
with the Dirac string, along positive $z$-axis, and (ii) the patch-wise, smooth gauge of Eq.~\ref{eq:7} that avoids Dirac strings. To reduce finite-size effects, we have considered system-size up to $(L/a)^3=30^3$, yielding $(4\times 30^3)=108000$ eigenstates. 

\begin{figure*}[t]
\centering
\subfigure[]{
\includegraphics[scale=0.55]{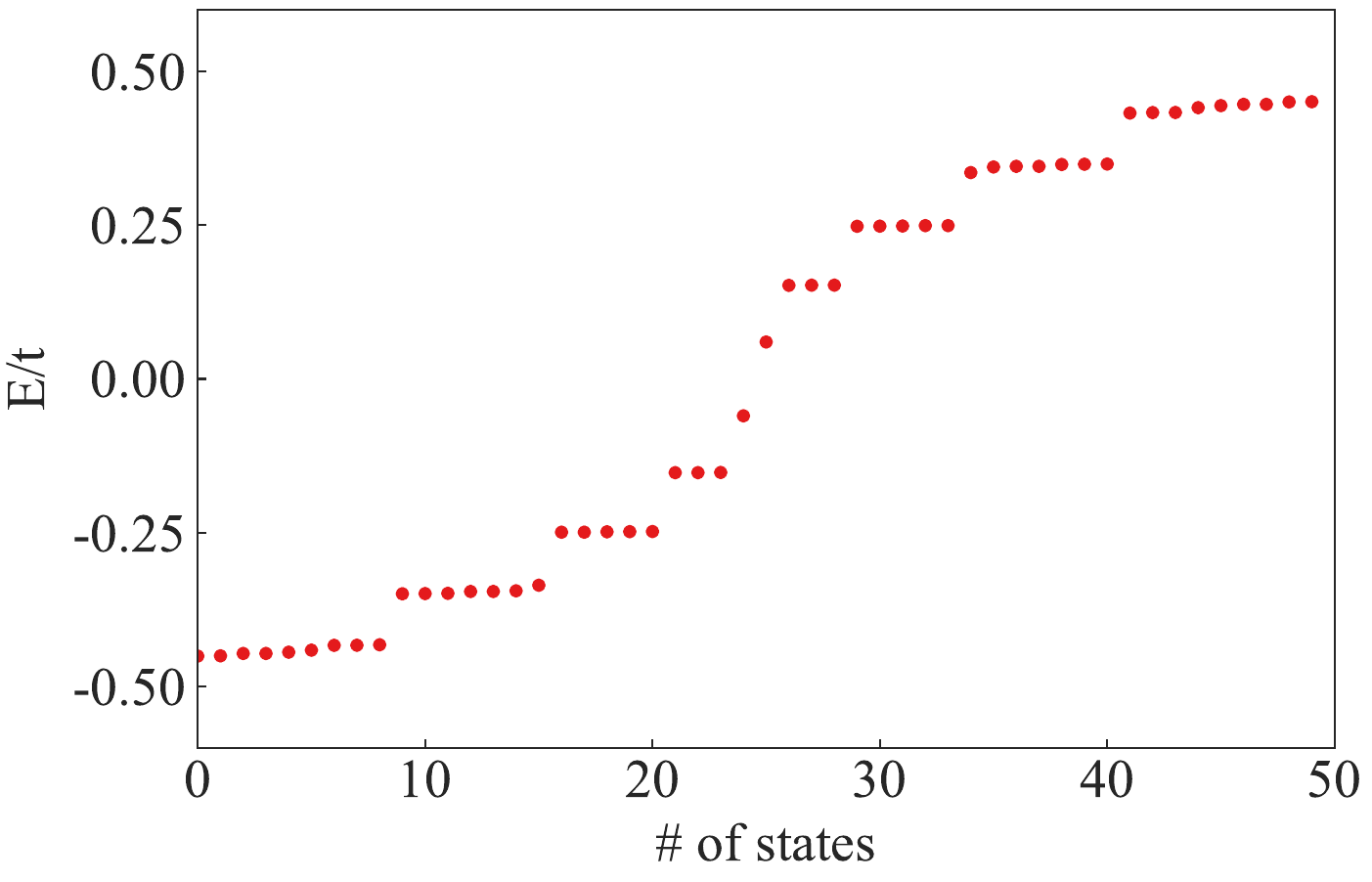}
\label{fig:2a}}
\subfigure[]{
\includegraphics[scale=0.55]{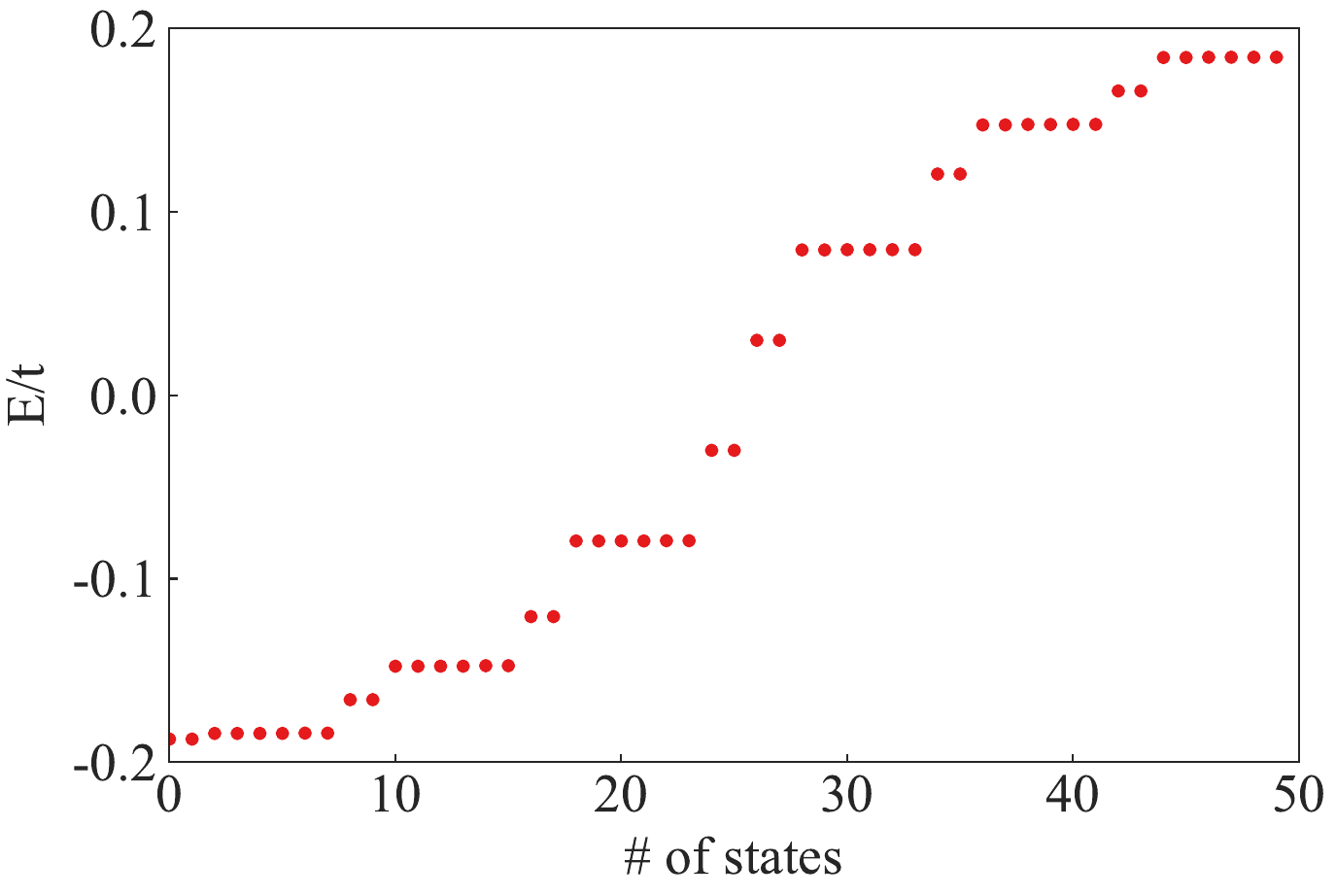}
\label{fig:2b}}
\subfigure[]{
\includegraphics[scale=0.45]{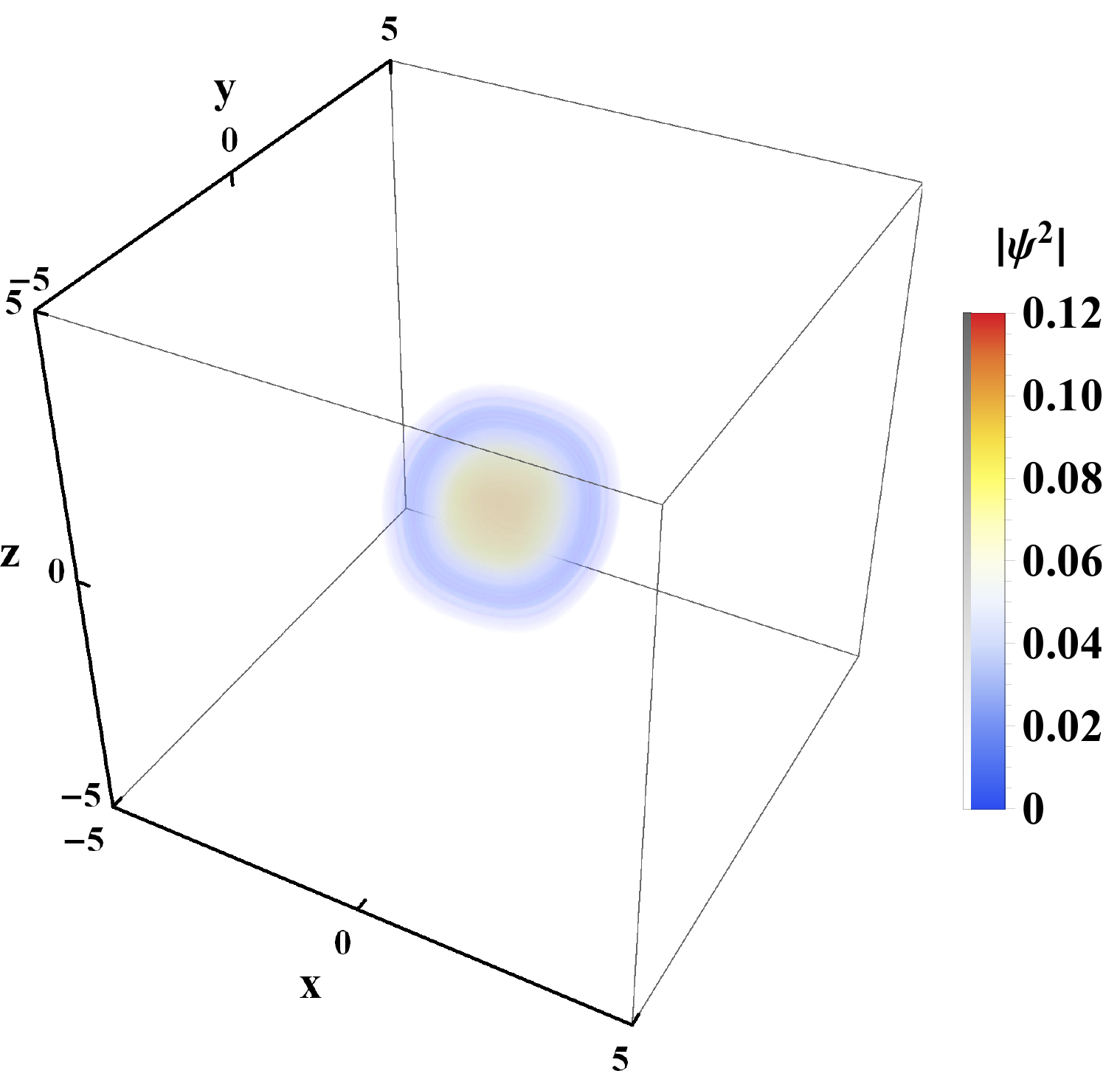}
\label{fig:2c}}
\subfigure[]{
\includegraphics[scale=0.45]{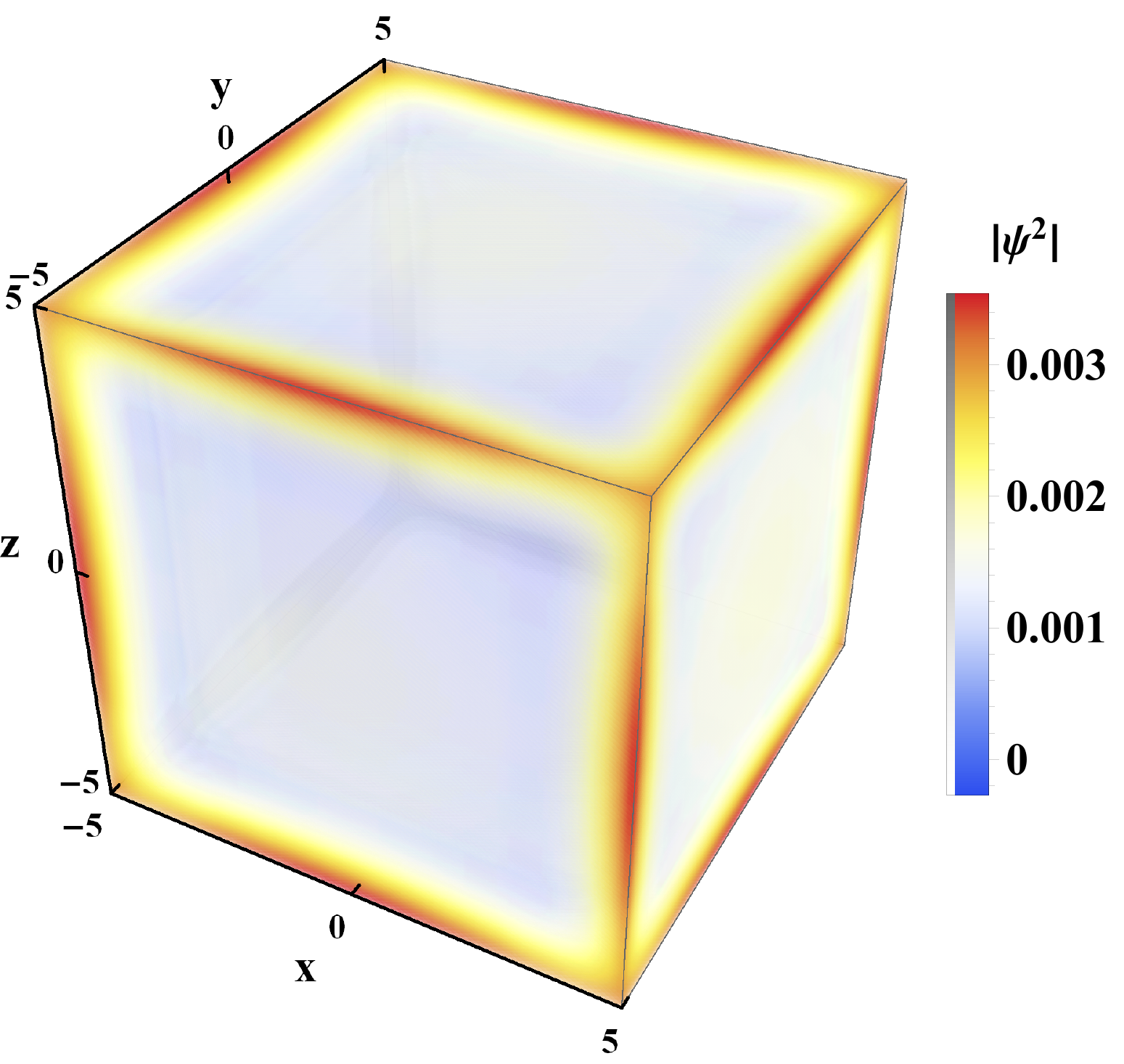}
\label{fig:2d}}
\subfigure[]{
\includegraphics[scale=0.55]{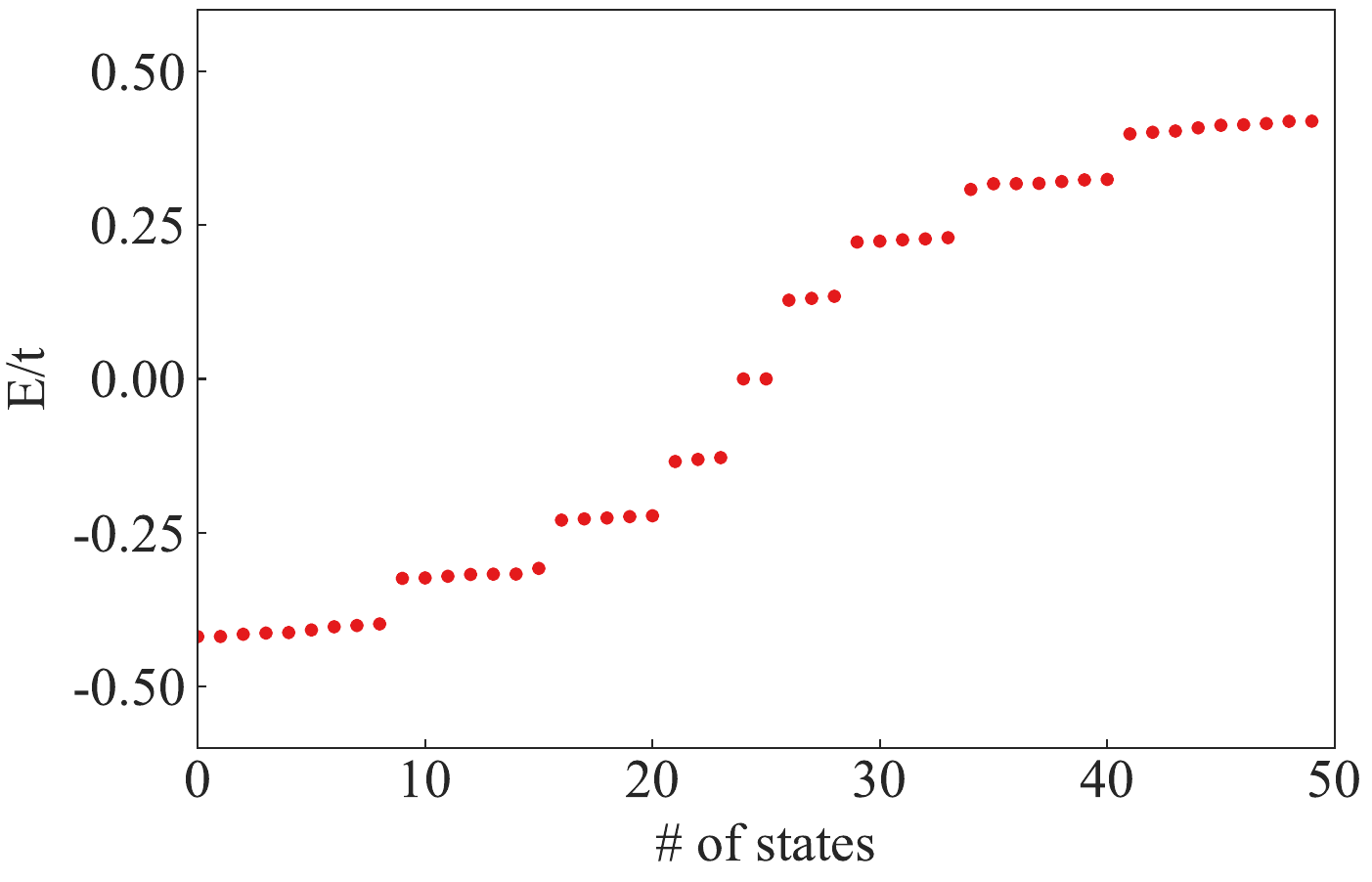}
\label{fig:2e}}
\subfigure[]{
\includegraphics[scale=0.55]{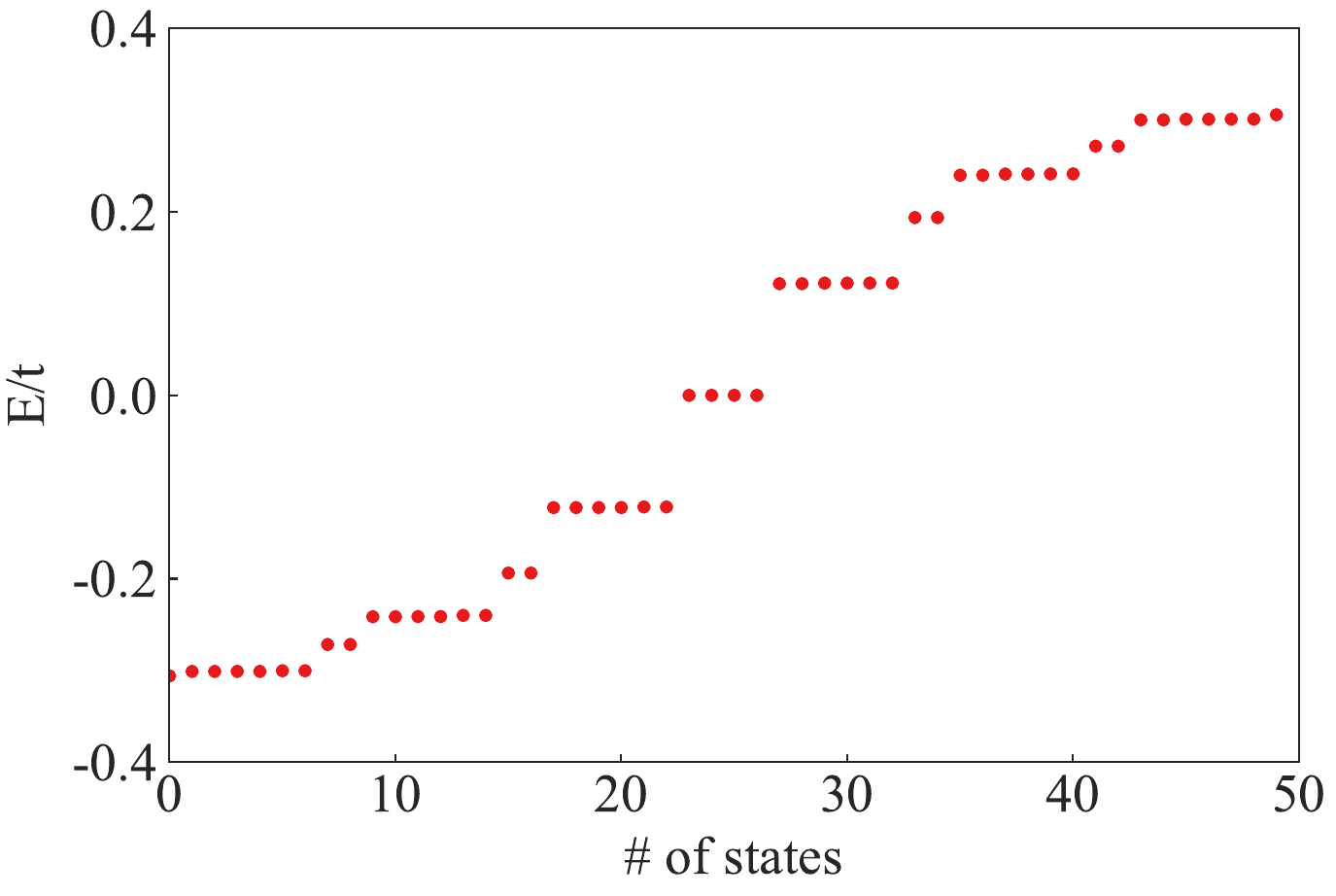}
\label{fig:2f}}
\caption{Fermion spectrum for the lattice model of first-order topological insulators [see Eq.~\ref{eq:tbmodel} ] in the presence of unit monopole $m=+1$. We have used band parameters $t=t^\prime$, and $t>0$. Only fifty low lying states near zero-energy are shown. (a) Energy vs. number of states plots for TI$_1$ phase, with tuning parameter $\Delta=2$, perturbed by a small, pseudo-scalar mass, $M^\prime=+0.05 t$. (b) Energy vs. number of states plots for TI$_2$ phase, with $\Delta=0$, $M^\prime=+0.05 t$. The degeneracy of the states closest to $E=0$, with energy $E \approx \pm M^\prime$ is $|\mathcal{N}_3|$. (c)-(d) For the TI$_1$ phase, these states are localized on the monopole and the boundary of the sample, respectively. The maximum probability density of boundary-localized state around twelve hinges of cubic sample is not related to higher-order topology. (e)-(f) When $M^\prime=0$, these bound states transform to near zero-energy modes, with an exponentially small hybridization gap due to finite-size effects.  }
\label{fig:monopolespectra}
\end{figure*}

Our results for fermion spectrum are demonstrated in Fig.~\ref{fig:monopolespectra} for $m=+1$. With $M^\prime >0$, the degeneracy of bound states with $E \approx \pm |M^\prime|$ is determined by $|\mathcal{N}_3|$, as shown in Fig.~\ref{fig:2a} and \ref{fig:2b}. In Fig.~\ref{fig:2c} and \ref{fig:2d}., we plot the localization patterns of these states for the TI$_1$ phase, with $\mathcal{N}_3=-1$. Since $-\text{sgn}(m \mathcal{N}_3) =+1$, the positive (negative) energy state remains bound to the monopole (surface), in agreement with Eq.~\ref{index1}. As we set $M^\prime=0$, these finite-energy, bound states transform into topologically protected, NZMs. As long as $|M^\prime| > E_h \approx 2 t a/L \exp(-L/\xi_0)$, the identity or chirality of the bound states can be clearly revealed. The topologically trivial insulator (NI) does not support NZMs. By studying non-minimal monopoles with $|m|>1$, we have found that the total number of NZMs is given by $2|m \mathcal{N}_3|$.

\begin{figure*}[t]
\centering
\subfigure[]{
\includegraphics[scale=0.45]{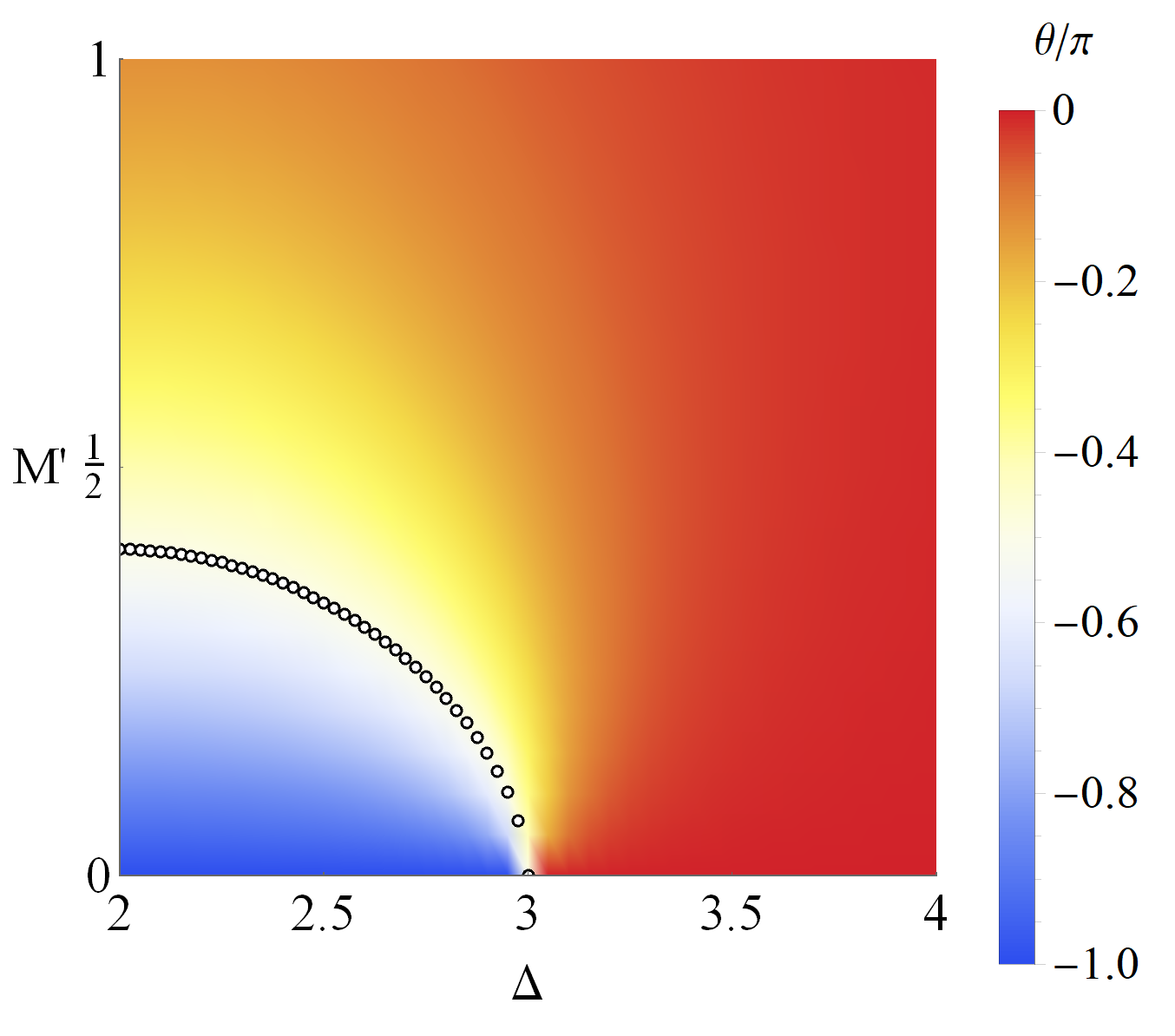}
\label{fig:3a}}
\subfigure[]{
\includegraphics[scale=0.55]{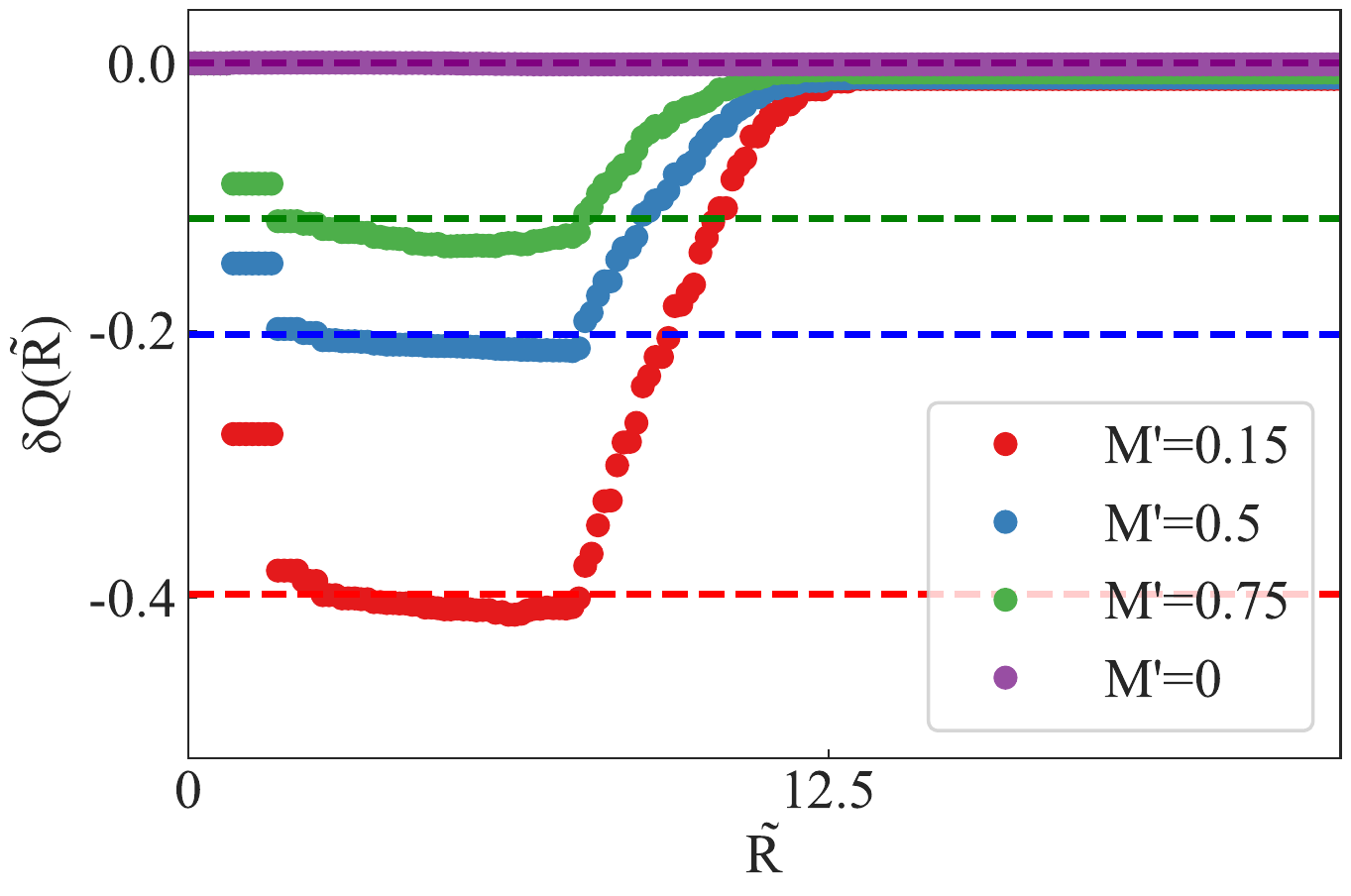}
\label{fig:3b}}
\subfigure[]{
\includegraphics[scale=0.55]{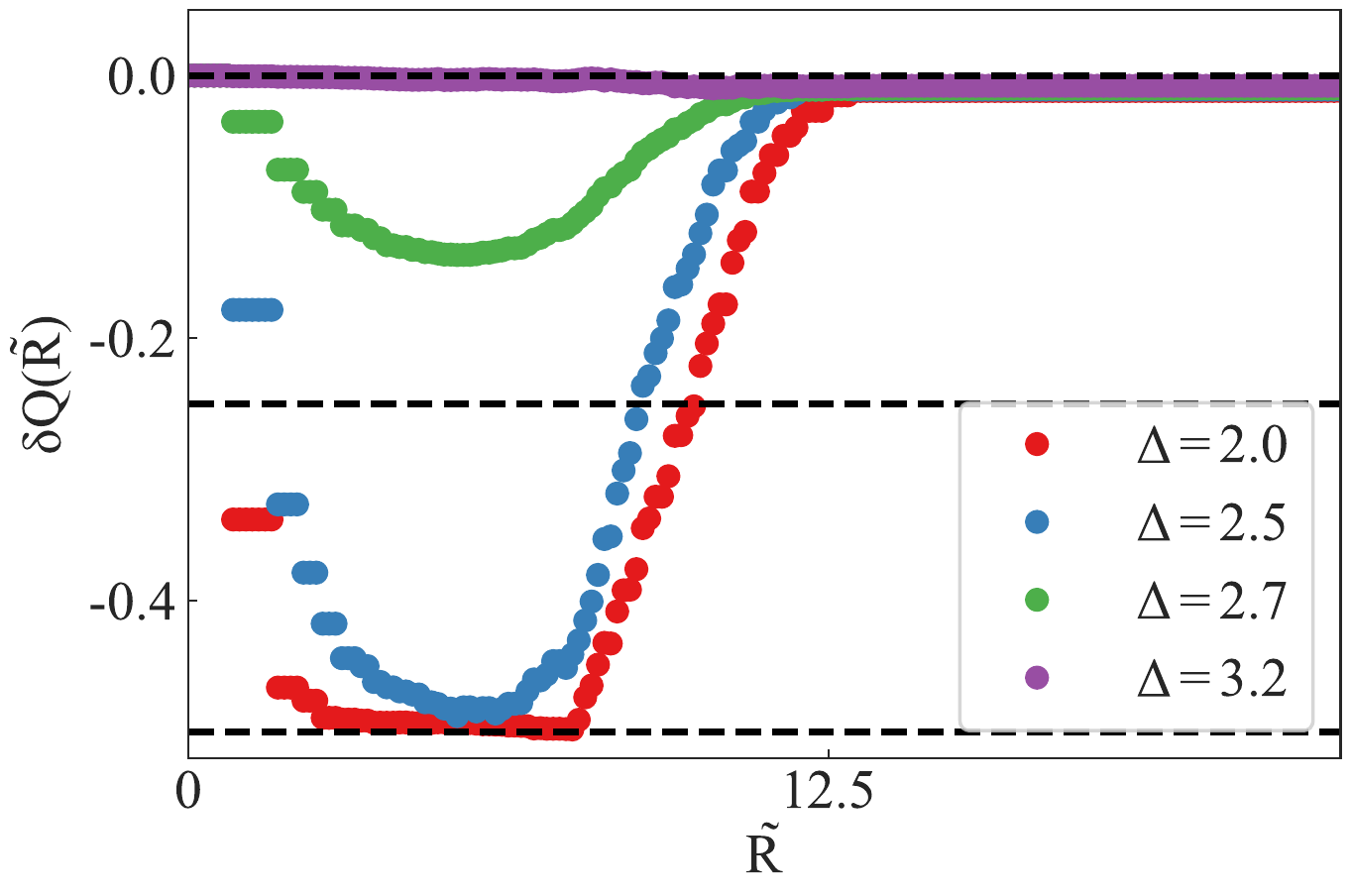}
\label{fig:3c}}
\subfigure[]{
\includegraphics[scale=0.55]{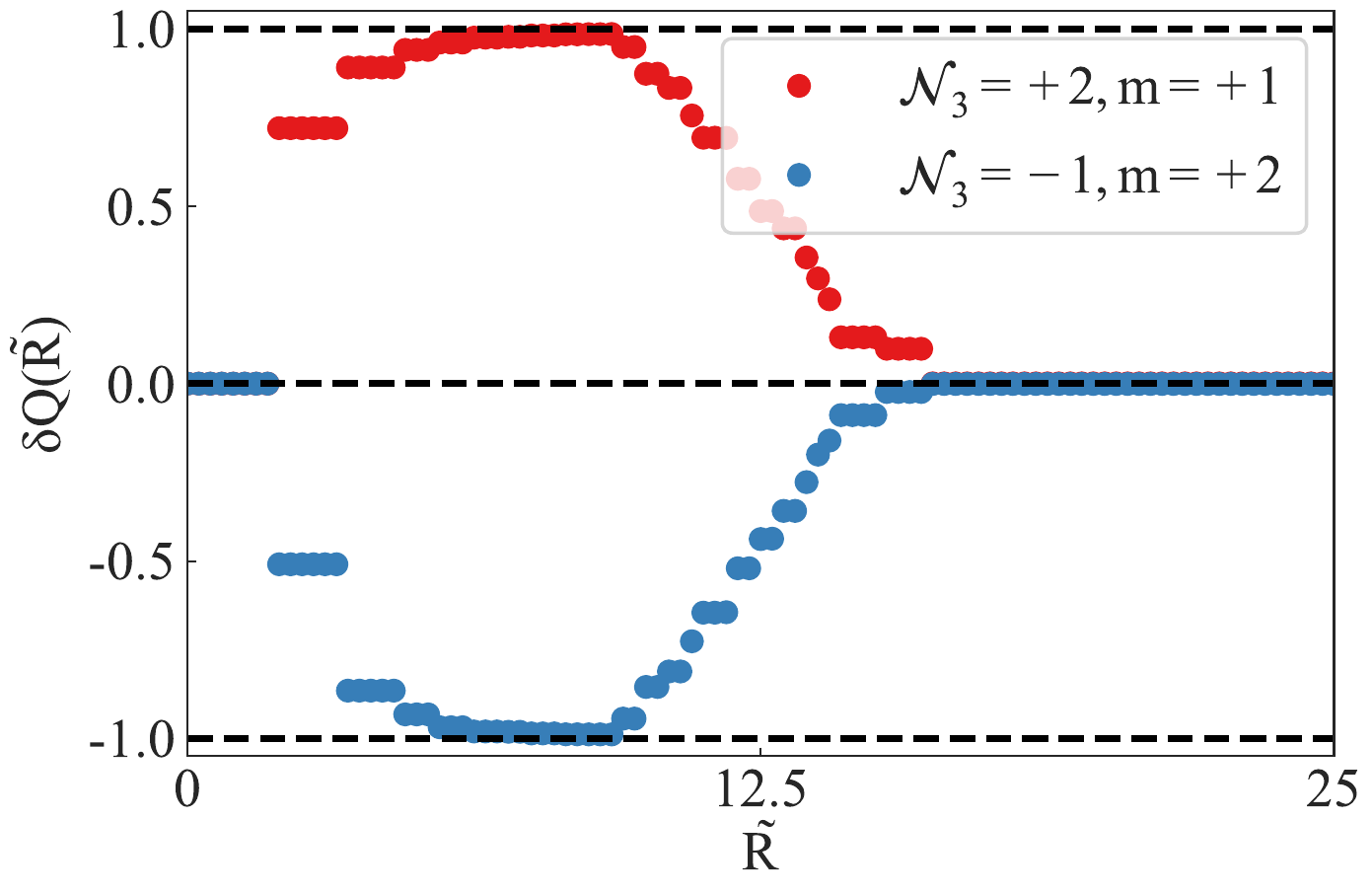}
\label{fig:3d}}
\caption{Witten effect for half-filled systems, in the presence of unit monopole $m=+1$, and a pseudo-scalar mass $M^\prime>0$ (in units of hopping parameter $t$). Numerical calculations of induced charge have been performed for system-size $(L/a)^3=30^3$. (a) The band theory results for $\theta$ of magneto-electric insulators, in the thermodynamic limit ($L \to \infty$) [see Eq.~\ref{MEhalffilled} ]. The dotted line tracks $\theta=-\pi/2$, and it terminates at the quantum critical point ($\Delta=3$, $M^\prime=0$) . (b) The induced electric charge $\delta Q(\tilde{R})$ (in units of $-e$), enclosed by a Gaussian sphere of radius $R$, which is centered at the monopole [see Eq.~\ref{eqdeltaq} ], and $\tilde{R}=R/a$. We have used $t>0$, $t^\prime=t$ and $\Delta=2$. When $M^\prime$ is bigger than the hybridization scale $E_h$ of bound states, the maximum values of $\delta Q$ agree with the predictions of band theory (dashed lines). For $M^\prime=0$, the ground state is two-fold degenerate, and $\delta Q=0$ implies the absence of Witten effect. (c) The induced electric charge (in units of $-e$) for $M^\prime =+0.05 t$, and different values of $2<\Delta<4$. Far from the quantum critical point, $\xi \approx a |3-\Delta|^{-1}$ is of the order of lattice spacing, and $M=t|(3-\Delta)|>M^\prime>E_h$. In this adiabatic regime, the maximum value of $\delta Q/(-e)$ approaches $\mathcal{N}_3/2 =-1/2$. For larger system-size, $M^\prime$ can be gradually reduced to zero. (d) Witten effect for TI$_2$ phase with unit monopole $m=+1$, and TI$_1$ phase with double monopole $m=+2$. The maximum induced charge follows Eq.~\ref{gen1}. Notice that the maximum values of charge is found for $4 a \leq R^\ast \leq 6 a$.}
\label{fig:Witten_Num1}
\end{figure*}

\begin{figure*}[t]
\centering
\subfigure[]{
\includegraphics[scale=0.4]{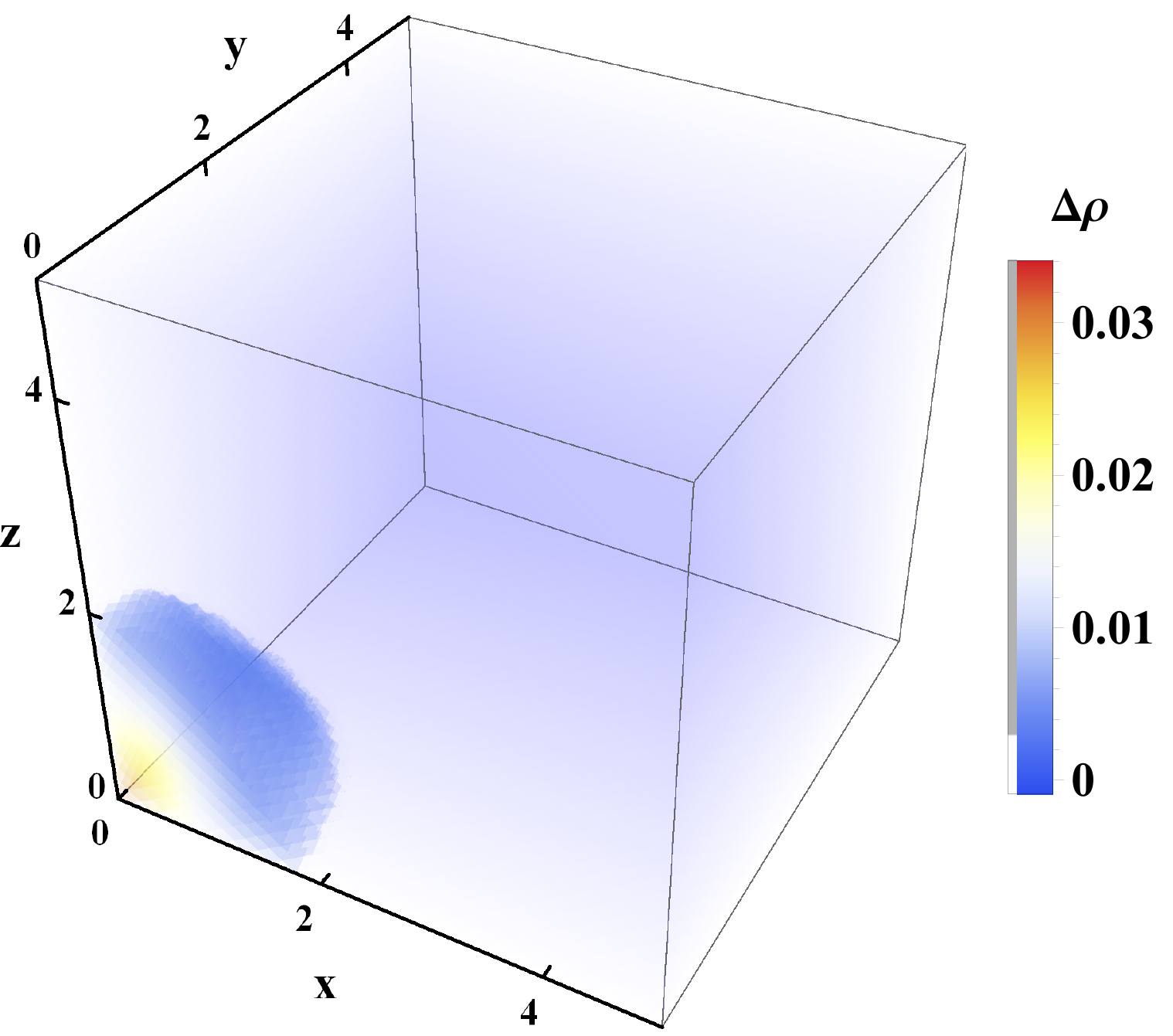}
\label{fig:5a}}
\subfigure[]{
\includegraphics[scale=0.55]{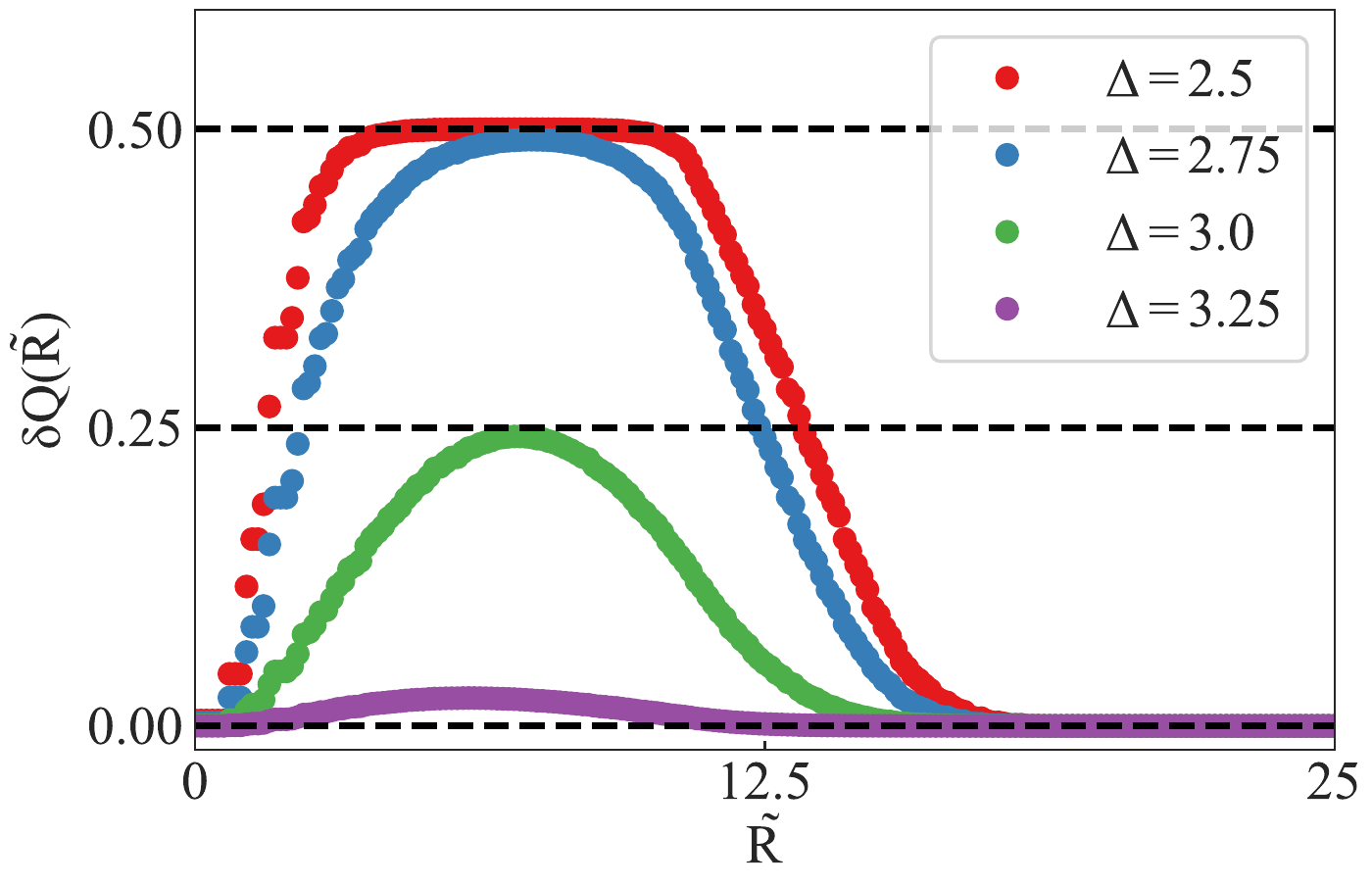}
\label{fig:5b}}
\subfigure[]{
\includegraphics[scale=0.2]{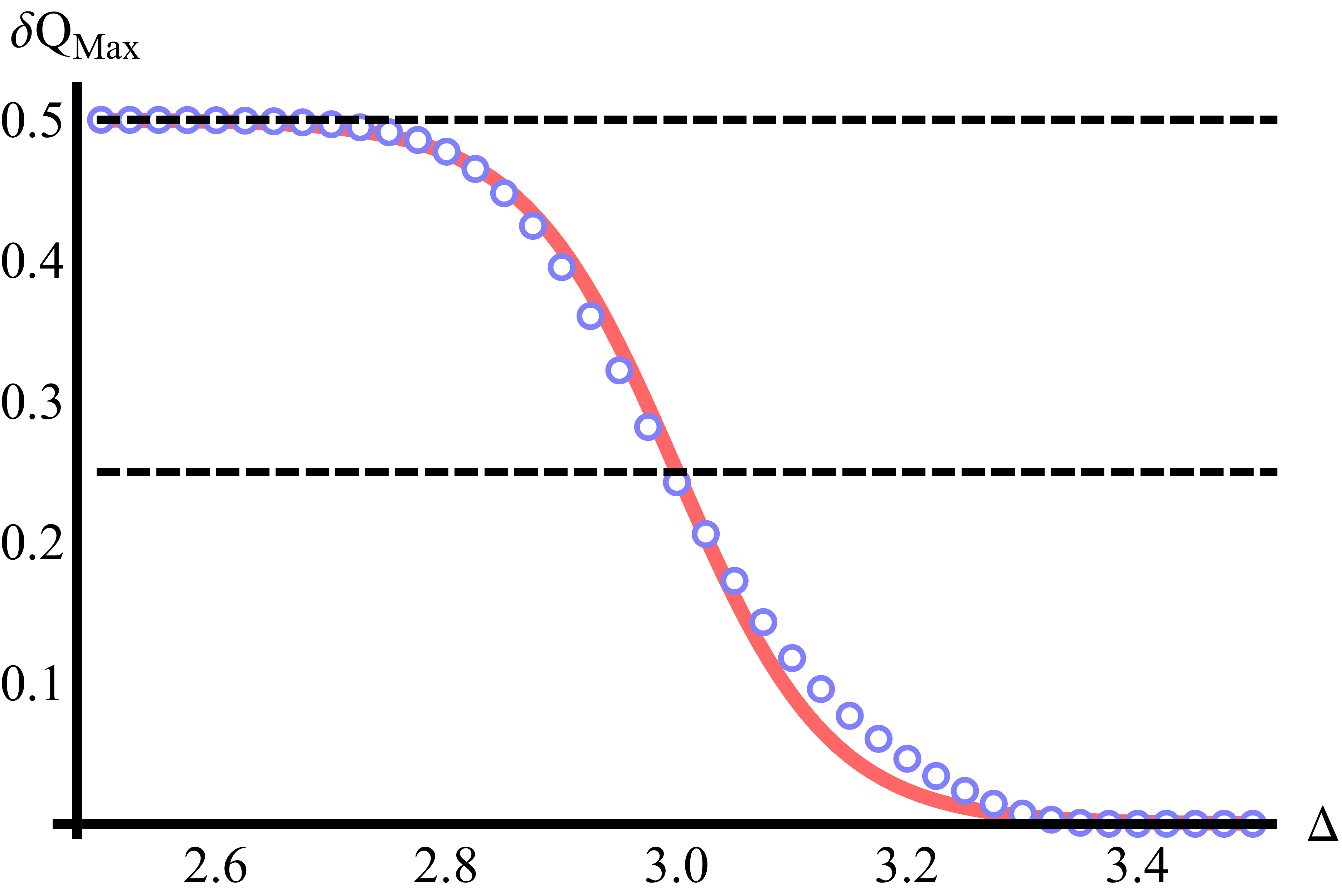}
\label{fig:5c}}
\caption{Spin-charge separation for $\mathcal{P}$ and $\mathcal{T}$ symmetric first-order topological insulator TI$_1$, without pseudo-scalar training field, in the presence of unit monopole $m=+ 1$, under fully open boundary conditions. The system is doped with one electron [$N_e=+1$ case in Fig.~\ref{fig:presentation1} ]. All numerical calculations have been performed for a system-size $(L/a)^3=30^3$. (a) Intensity plot of induced charge density around the monopole for TI$_1$ phase, with $\Delta=2.5$. For clarity, only the first octant is shown. (b) The total induced charge (in units of $-e$), localized within a Gaussian sphere of radius $R$, which is centered the monopole [described by Eq.~\ref{eqdeltaq} with $M^\prime=0$], for different values of $\Delta$, and $\tilde{R}=R/a$. The data clearly identifies the maximum induced charge $-e/2$ ($-e/4$) for TI$_1$ phase (quantum critical point). (e) The finite size scaling of maximum induced charge (in units of $-e$). In the vicinity of critical point and inside TI$_1$ phase, the data (blue circles) can be well approximated by $f(L,\Delta)= \frac{1}{2} \left \{1+ \exp \left[(\Delta -3)\frac{L}{2a}\right]\right \}^{-1}$ (solid, red line). Notice that the maximum values of induced charge for topological insulator and quantum critical point are found for $R^\ast \approx  6 a$.}
\label{fig:Witten_Num2}
\end{figure*}

\subsection{Witten effect at half-filling} \label{WEHF}To determine the induced charge on monopole, we closely follow the procedure of Ref.~\onlinecite{rosenberg2010witten}. The exact diagonalization of Hamiltonian is performed with and without the monopole. Using normalized eigenstates $\psi_{n, m}(\bs{r}_i)$ and energy eigenvalues $\epsilon_{n,m}$, we compute the charge density 
\begin{equation}
\rho_{m}(\bs{r}_i,L,\xi,M^\prime)=-e \sum_n |\psi_{n, m}(\bs{r}_i)|^2 \Theta(E_{F,m}-\epsilon_{n,m}),
\end{equation}
in the presence of monopole, where $E_{F,m}$ is the Fermi energy. Similarly, we define the charge density 
\begin{equation}
\rho_{0}(\bs{r}_i,L,\xi,M^\prime)= -e \sum_n |\psi_{n,  0}(\bs{r}_i)|^2 \Theta(E_{F,0}-\epsilon_{n,0}),
\end{equation} in the absence of monopole. The Fermi energies are obtained by keeping the total number of particles fixed and in general $E_{F,m} \neq E_{F,0}$. As we are implementing the particle-hole symmetric regulator $M^\prime \to 0^+$, the half-filled condition is enforced by $E_{F,m}=E_{F,0}=0$. The induced charge density on monopole is found from 
\begin{equation}
\delta \rho(\bs{r}_i,L,\xi,M^\prime)= \rho_m(\bs{r}_i,L,\xi,M^\prime) - \rho_0(\bs{r}_i,L,\xi,M^\prime),
\end{equation}
and the total induced charge inside a spherical Gaussian surface of radius $R$, centered around the monopole is obtained from 
\begin{equation}\label{eqdeltaq}
\delta Q(R,L,\xi,M)=\sum_{|\bs{r}_{i}|<R} \delta \rho(\bs{r}_{i},L,\xi,M^\prime).
\end{equation}  
All lengths will be measured in units of lattice spacing.

To benchmark our results for non-degenerate half-filled state, we first consider $\mathcal{P}$ and $\mathcal{T}$ breaking ME insulators, with $M^\prime > 0 $, and $2<\Delta<4$. The regularized ME coefficient 
\begin{equation} \label{MEhalffilled}
\theta_+(M^\prime)=2\pi [\mathcal{CS}_+(M^\prime)-\mathcal{CS}(M^\prime=+\infty)] 
\end{equation}
from band theory calculations is shown in Fig.~\ref{fig:3a}. Notice that the TQCP at $\Delta=3$, $M^\prime=0$ is a multi-critical point, where the ME insulator with $\theta_+(M^\prime \neq 0) =-\pi/2$, the TI$_1$ phase with $\theta_+(M^\prime \to 0^+)=-\pi$, and the NI with $\theta_+(M^\prime \to 0^+)=0$ meet. 

The results for $\delta Q(\tilde{R})$ for $\Delta=2$ and various $M^\prime$ are shown in Fig.~\ref{fig:3b}.  Since $\Delta=2$ is far from TQCP, $\xi_0 \sim a$, and finite-size effects are very weak, when $L/a=30$. Consequently, the maximum values of induced electric charge $\delta Q_{max} /(-e)$ agree with $\theta_+(M^\prime)$, obtained in thermodynamic limit $L \to \infty$. \emph{When $M^\prime=0$, the induced charge for 2-fold-degenerate, half-filled ground state vanishes, for any $\Delta$}, implying a precise cancelation between the contributions from occupied NZMs and Dirac sea.

In Fig.~\ref{fig:3c}, we show the dependence of $\delta Q(\tilde{R})$ on $\Delta$ for a fixed $M^\prime=+0.05 t$. Deep inside the TI$_1$ phase, we find $\delta Q_{max}/(-e) \to \mathcal{N}_3/2=-1/2$. When TQCP is approached, $\delta Q_{max}$ can significantly deviate from $-e \mathcal{N}_3/2$ due to strong finite-size corrections. By varying the winding number $\mathcal{N}_3$ and the monopole strength $m$, we have confirmed that the maximum induced charge for thermodynamically large systems is given by [see Fig.~\ref{fig:3d} ]
\begin{equation}\label{gen1}
\delta Q_{max, TI}=  - e \; m \; \mathcal{CS}_+ =  -\frac{e}{2}  \; m \; \mathcal{N}_3. 
\end{equation}
As the overall half-filled system is charge-neutral, compensating charge 
\begin{equation}
\delta Q_s = -\delta Q_{max, TI}=+\frac{e}{2}  \; m \; \mathcal{N}_3
\end{equation} appears around  the sample boundary. The localization pattern of surface-localized NZMs in Figs.~\ref{fig:2d} implies that the highest intensity of surface-bound charge density under OBC occurs around $12$ hinges. By changing boundary conditions, the pattern of surface-bound charge density can be altered. But the monopole-bound charge density and $\delta Q_{max, TI}$ will remain unaffected.

We have also found that the TQCP between two distinct insulators leads to 
\begin{equation}\label{gen2}
\delta Q_{max, TQCP}=- \frac{e}{4} m \; (\mathcal{N}_{3,1}+ \mathcal{N}_{3,2}).
\end{equation}
This is a three-dimensional analog of the law of corresponding states for quantum Hall plateau transition~\cite{,Kivelson1}. Usually, the quantum critical phase does not support normalizable surface-states, and the compensating charge $-\delta Q_{max, TQCP}$ stays distributed over the entire sample. 

\begin{figure*}
\centering
\subfigure[]{
\includegraphics[scale=0.55]{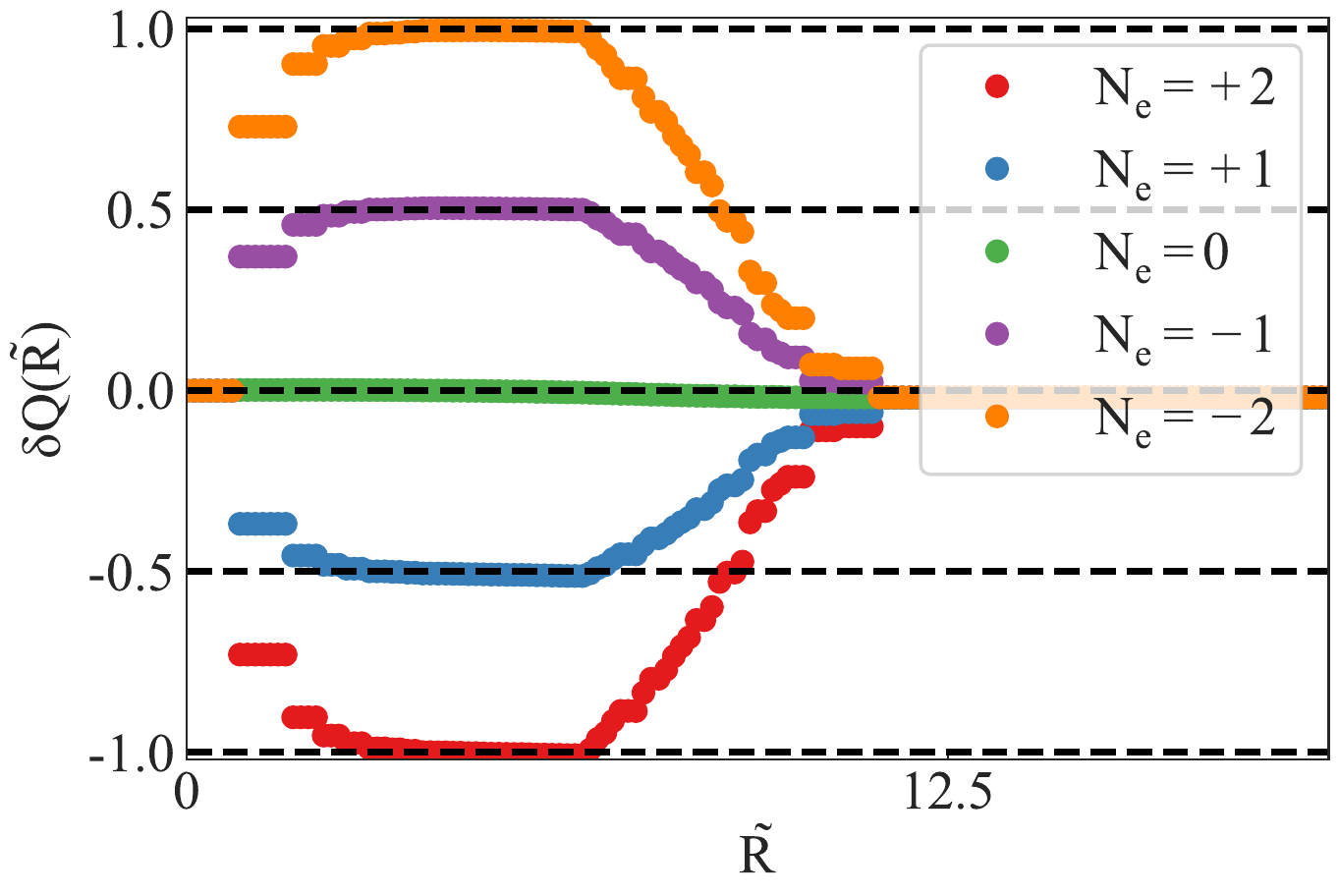}
\label{fig:6a}}
\subfigure[]{
\includegraphics[scale=0.55]{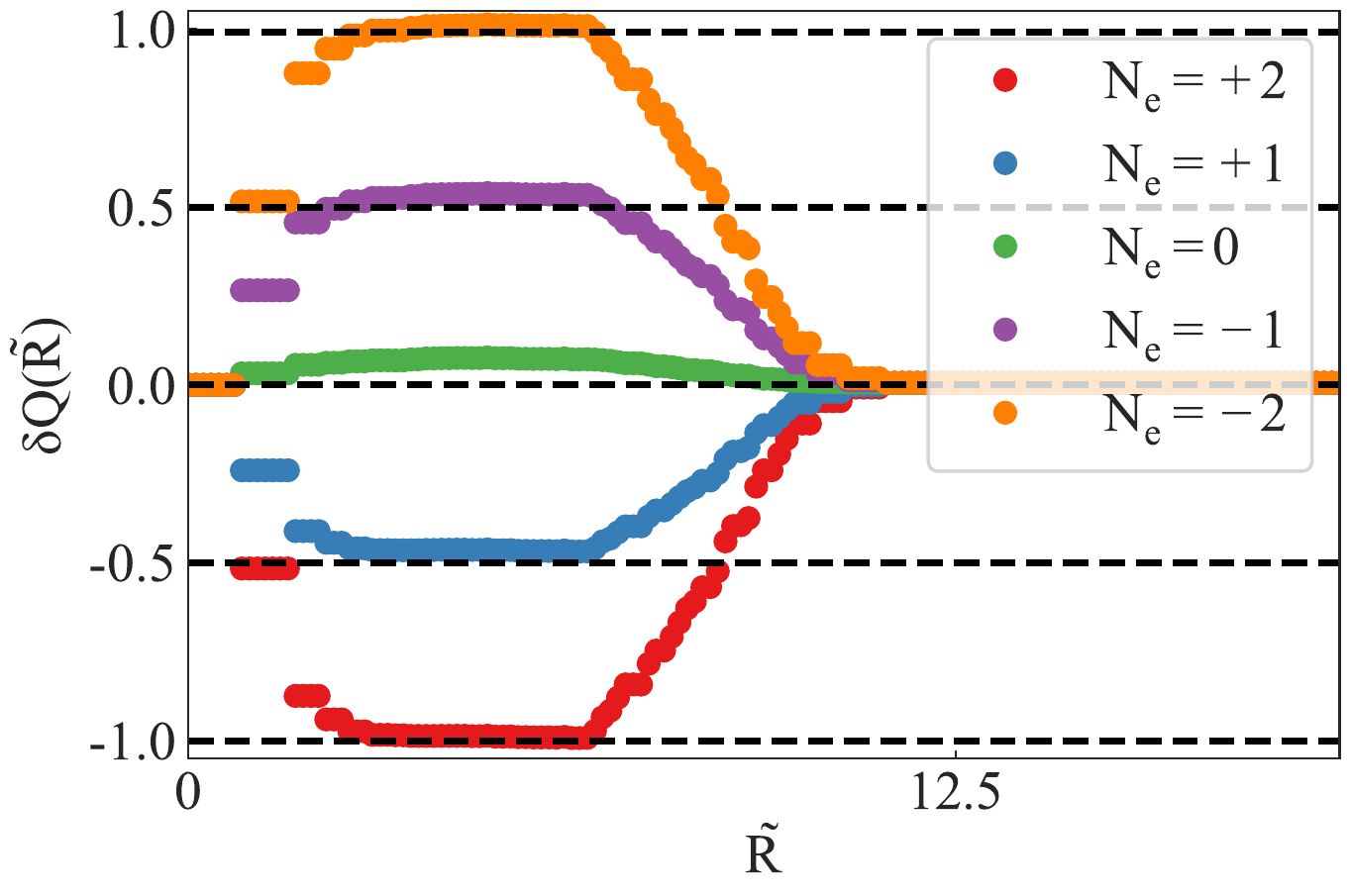}
\label{fig:6b}}
\subfigure[]{
\includegraphics[scale=0.55]{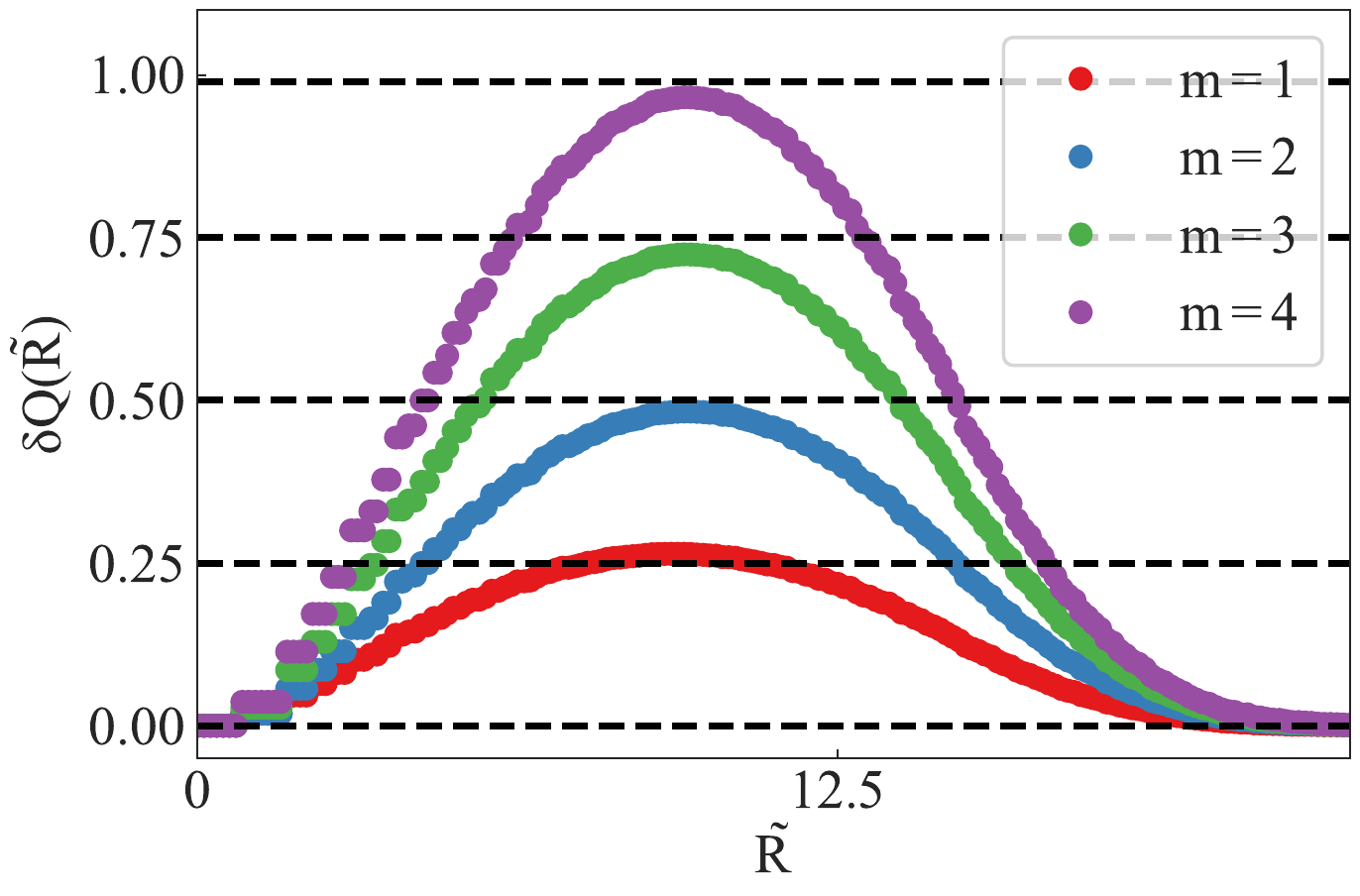}
\label{fig:6c}}
\caption{Spin-charge separation of first-order topological insulators for different winding numbers and monopole strengths, in the absence of a pseudo-scalar mass. All numerical calculations have been performed for a system-size $(L/a)^3=30^3$. As outlined in Fig.~\ref{fig:presentation2}, (a) TI$_2$ phase with winding number $\mathcal{N}_3=+2$, (b) TI$_1$ phase with winding number $\mathcal{N}_3=-1$ show identical structure of spin-charge separation, in the presence of unit monopole $m=+1$, and double monopole $m=+2$, respectively. We are accessing all charged $SU(4)$ multiplets by doping $-2 \leq N_e \leq +2$ electrons. The maximum induced charge oscillates between half-integer and integer values, following the general rule of Eq.~\ref{gen3}. (c) For topological quantum critical point at $\Delta=+3$, varying the monopole strength between $m=+1,+2,+3,+4$, and adding $N_{e}=m$ electrons, we observe four distinct values of maximum induced charge that follow Eq.~\ref{scTQCP}. }
\label{fig:Witten_Num_22}
\end{figure*}

\subsection{Doped system and spin-charge separation}\label{WED}
In the absence of pseudo-scalar training field, we can study spin-charge separation for TI$_1$ and TI$_2$ phases. We first consider the simplest case of TI$_1$ phase in the presence of unit monopoles $m= \pm1$. The results for one doped electron are shown in Fig.~\ref{fig:Witten_Num2}, which follows the schematic of Fig.~\ref{fig:presentation1} and the analysis of continuum theory. The results for one doped electron are shown in Fig.~\ref{fig:Witten_Num2}.  At TQCP,  $\delta Q_{max}=-e/4$ ($+e/4$) for single electron (hole) doping again corroborates our prediction $|\theta_c|=\pi/2$. In Fig.~\ref{fig:5c}, we elucidate the quantum critical scaling behavior of induced charge by plotting $\delta Q_{max}/(-e)$ as a function of $\Delta$. 

We have also performed such calculations for $|\mathcal{N}_3|>1$ and $|m|>1$. The spin-charge separation is controlled by $SU(2|n m|)$-multiplets. As we dope $N_e$ electrons, the maximum induced charge on monopoles will oscillate between half-integer and integer multiples of $e/2$, according to the formula
\begin{eqnarray}\label{gen3}
\delta Q_{max}=-\frac{e}{2} N_e, \; \text{with} \; -|m \mathcal{N}_3| \leq N_e \leq +|m \mathcal{N}_3|. \nn \\
\end{eqnarray}
Rest of the doped charge $\delta Q_s=-\frac{e}{2} N_e$ will be localized around the boundary. 
In Figs.~\ref{fig:6a} and \ref{fig:6b}, we show that (i) $\mathcal{N}_3=-1$, $m=+2$ and (ii) $\mathcal{N}_3=+2$, $m=+1$ support identical structure of spin-charge separation, as outlined in Fig.~\ref{fig:presentation2}.

Furthermore, by doping with $N_e=\pm |m|$ electrons at TQCP, we have observed that the monopole supports maximum induced charge 
\begin{equation}\label{scTQCP}
\delta Q_{max,TQCP}=- \frac{e}{4} N_e
\end{equation}
for $m= 1,  2, 3, 4$, respectively [see Fig.~\ref{fig:6c} ]. For the quantum critical system, rest of the doped charge $-(3/4) e N_e$ stays distributed over the entire sample. 
Next, we address the stability of third homotopy class when the number of discrete symmetries is reduced.

\section{Chiral higher-order topological insulators}\label{SOTI}
In this regard, we consider 3D, second-order TIs~\cite{Schindlereaat0346}. A four-band model for such systems is given by 
\begin{equation}\label{eq:CH}
H_2(\mathbf{k})=H_1(\bs{k})+t^{\prime \prime}(\cos k_{1}-\cos k_{2})\Gamma_{4}=\sum_{j=1}^{5} N_j(\bs{k}) \Gamma_j,
\end{equation}
which describes $\mathcal{P}$-, $\mathcal{T}$-, and cubic-symmetry- breaking, but $\mathcal{C}$-preserving, ME insulators. While the $C^z_4$ symmetry is broken for generic values of $k_z$, it remains unbroken for the high-symmetry planes, with $k_z=0, \pi$. The generator of $C^z_4$-symmetry for these planes is $\Gamma_{12}$. At these planes $H_2$ reduces to two-dimensional, Benalcazar-Bernevig-Hughes model of second-order insulators~\cite{Benalcazar61,BBHPrb}. In Ref.~\onlinecite{Schindlereaat0346}, the perturbed TI$_1$ phase ($1<|\Delta|<3$) was shown to possess quantized ME coefficient $\theta=\pi$. 

\begin{table}[t]
	\def\arraystretch{1.5}
	\begin{tabular}{|c|c|c|}
		\hline
		\begin{tabular}{c}Tuning \\
		parameter\end{tabular} & \begin{tabular}{c} FOTI ($t^{\prime \prime}=0$)\\ $(\mathfrak{C}^{\pi}_{12},  \mathfrak{C}^0_{12},n_3)$ \end{tabular} & \begin{tabular}{c} ch-HOTI ($t^{\prime \prime} \neq 0$)\\ $(\mathfrak{C}^{\pi}_{12},  \mathfrak{C}^0_{12},n_3)$ \end{tabular} \\
		\hline
		$1<\Delta<3$ & $(-1,0,-1)$ & $(-1,0,-1)$ \\
		\hline
		$-1<\Delta<1$ & $(+1,-1,+2)$ & $(-1,-1,0)$ \\
		\hline
		$-3<\Delta<-1$ & $(0,+1,-1)$ & $(0,+1,-1)$\\
		\hline
	\end{tabular}
	\caption{Classification of tunneling configurations of $SU(2)$ Berry connection for first order topological insulator (FOTI) and second order, chiral topological insulator (ch-HOTI). The $\mathfrak{C}^\pi_{12}$, and $\mathfrak{C}^0_{12}$ are the relative Chern numbers for $k_z=\pi$, $k_z=0$ planes, respectively [see Eq.~\ref{relChern} ]. The third homotopy class is identified by $n_3=\mathfrak{C}^\pi_{R}-\mathfrak{C}^0_{R}$, which leads to the regularized Chern-Simons coefficient $\mathcal{CS}_+=n_3/2$.} \label{fig:CHTable}
\end{table}

\begin{figure*}
\centering
\subfigure[]{
\includegraphics[scale=0.2]{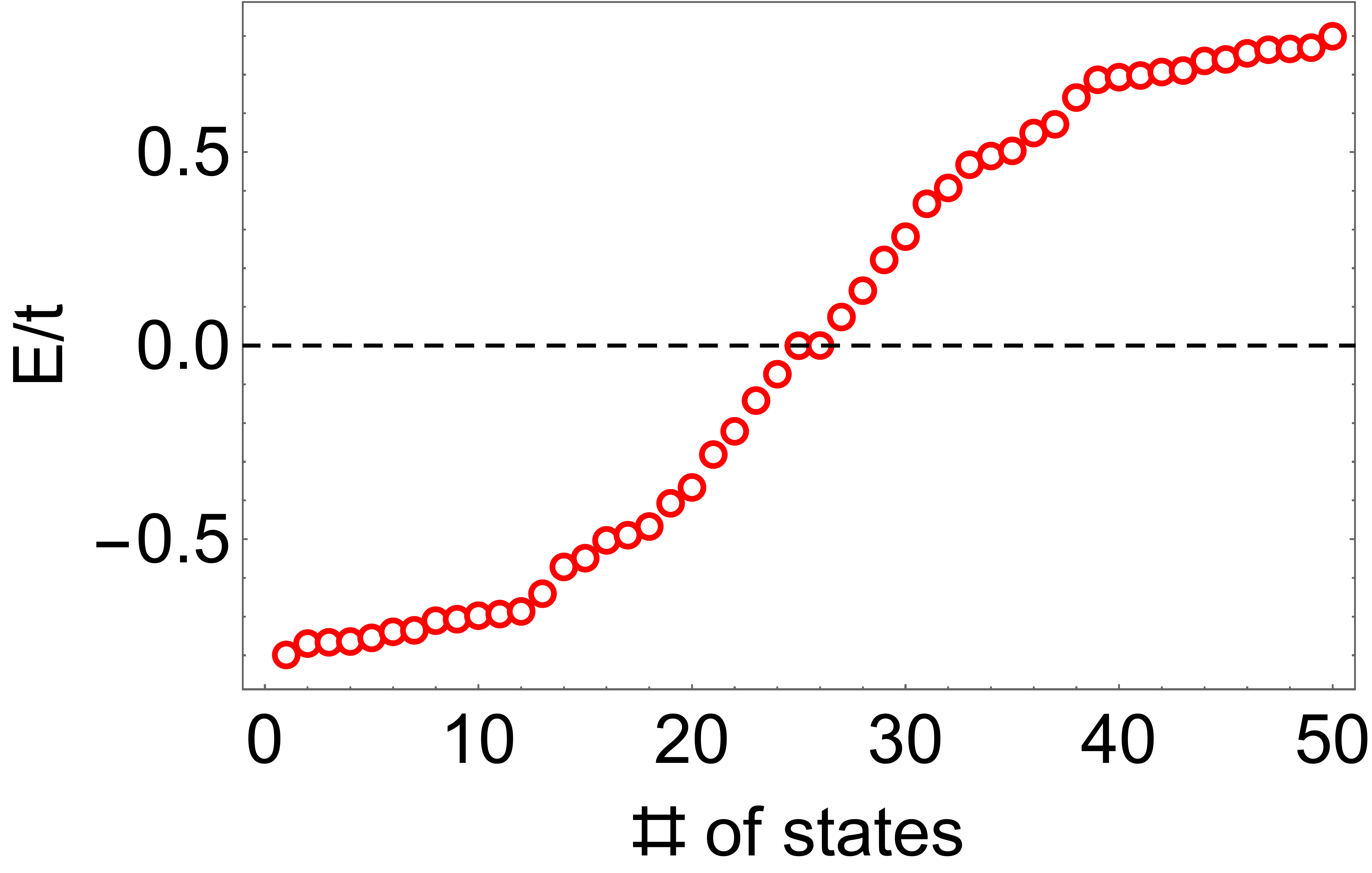}
\label{fig:CHState1}}
\subfigure[]{
\includegraphics[scale=0.2]{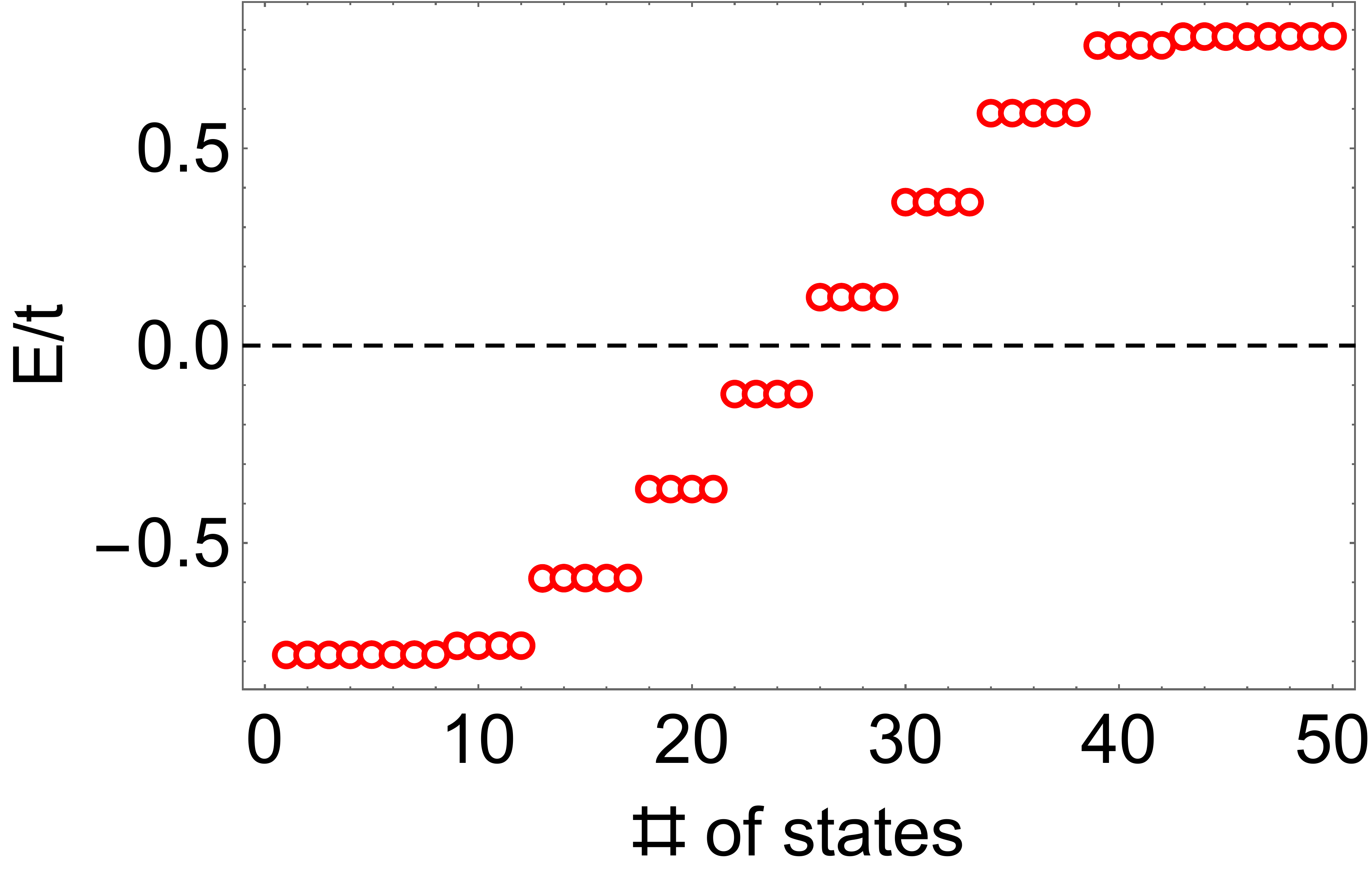}
\label{fig:CHState2}}
\subfigure[]{
\includegraphics[scale=0.2]{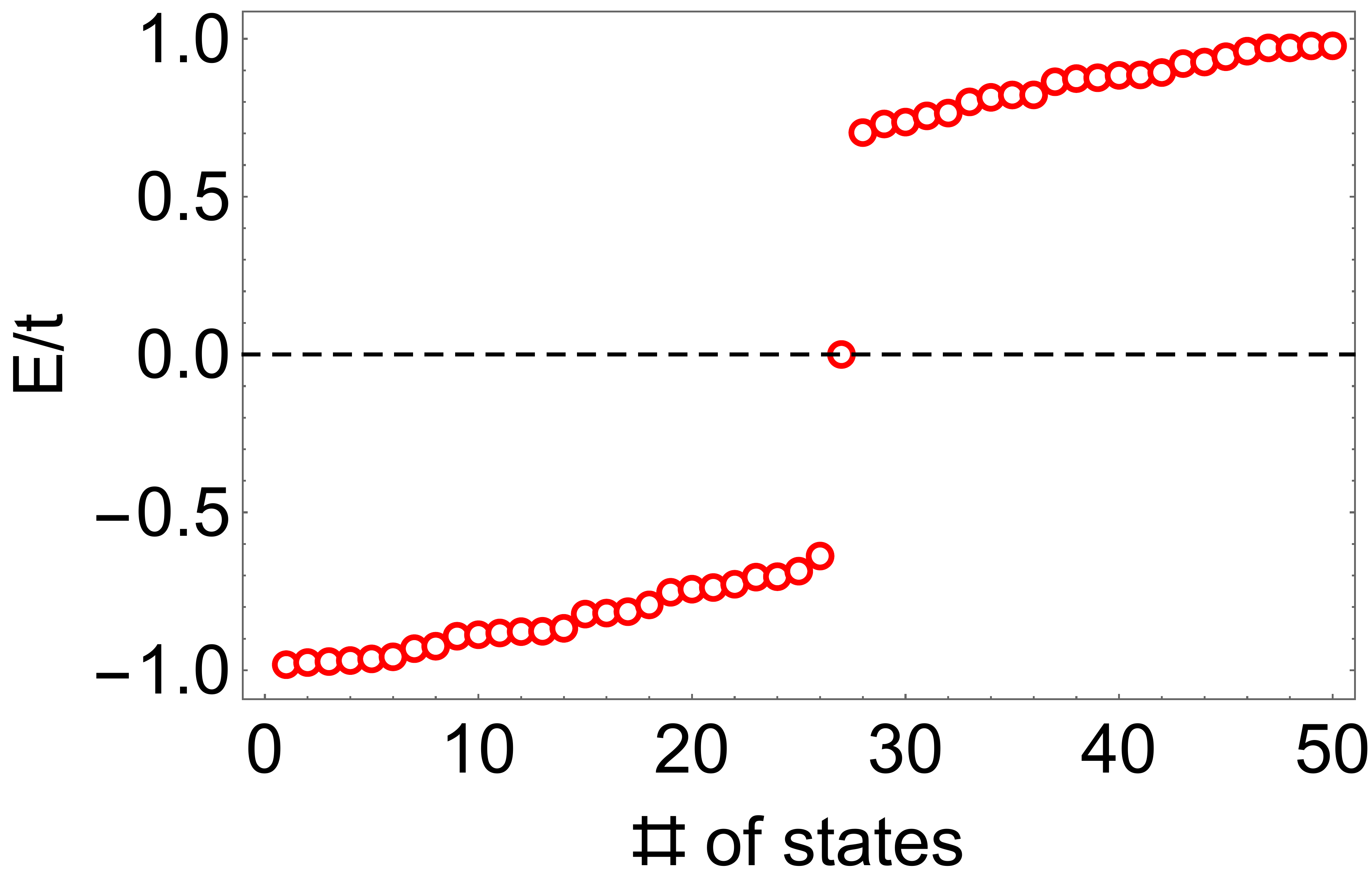}
\label{fig:CHLoc}}
\subfigure[]{
\includegraphics[scale=0.53]{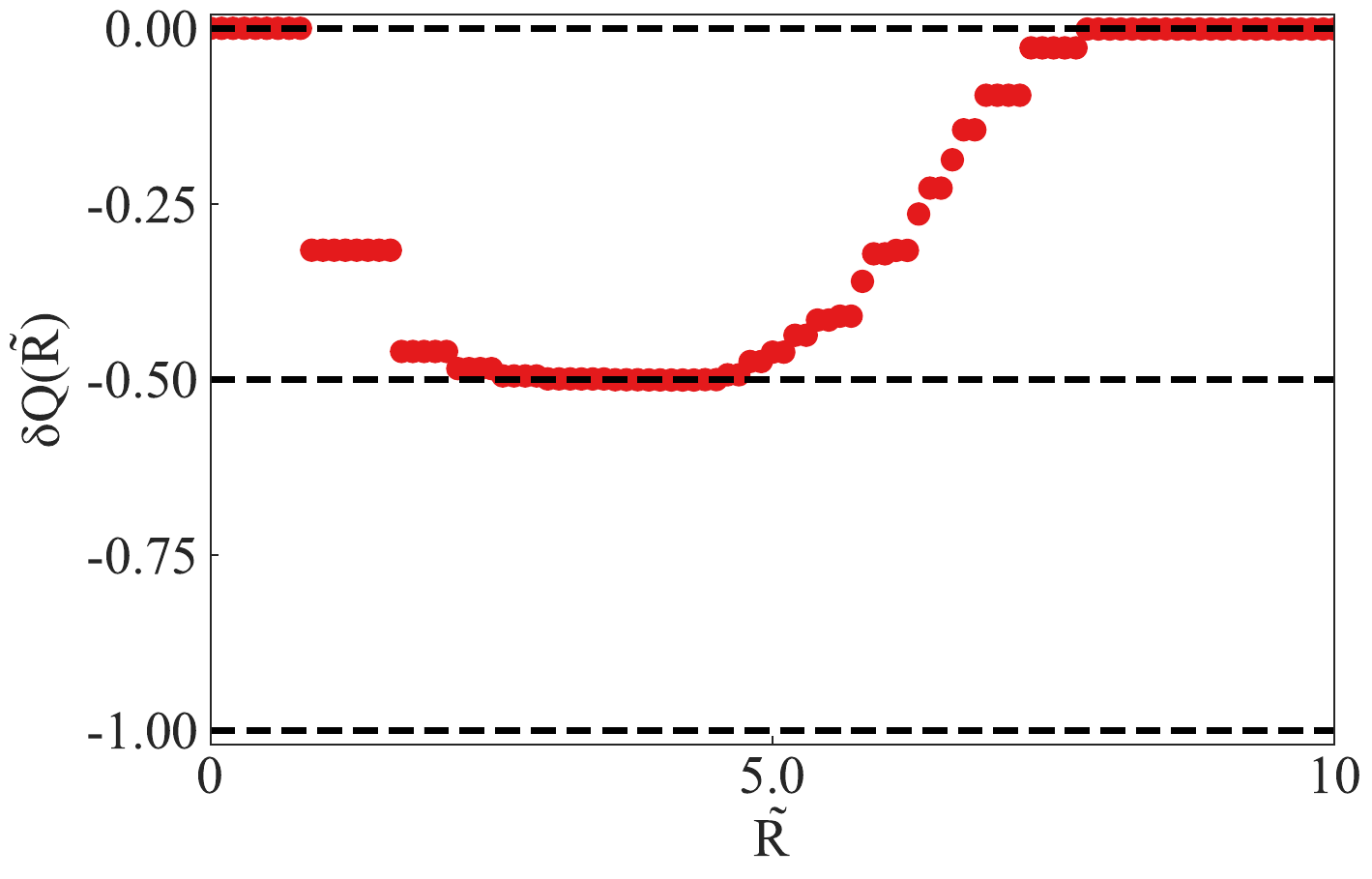}
\label{fig:CHWitten}}
\caption{Thought experiments on chiral higher-order topological insulators [see Eq.~\ref{eq:CH} ] with unit monopole $m=+1$. Fifty low-lying states are shown for a system-size $(L/a)^3=10^3$. (a) Under fully open boundary conditions, the perturbed TI$_1$ phase (with tuning parameter $\Delta=2$) supports two near-zero-modes, as three-dimensional, winding number $n_3=-1$ (see Table~\ref{fig:CHTable} ). While one mode is localized on the monopole, the other one is localized on the boundary. (b) Perturbed TI$_2$ state (with tuning parameter $\Delta=0$), does not support any near-zero-modes, as this phase has $n_3=0$ (see Table~\ref{fig:CHTable} ). (c) By implementing mixed boundary conditions, surface-localized fermion zero-mode can be eliminated. As the total number of states is an even integer, for a half-filled system, the monopole-localized zero-mode remains unoccupied. (d) Under open boundary conditions, Witten effect can only be found with a pseudo-scalar training field ($M^\prime \to 0^+$). In contrast to this, Witten effect can be observed under mixed boundary conditions, without any pseudo-scalar training field ($M^\prime=0$). Both systems show identical maximum induced charge on monopole $-\frac{e}{2} n_3$. By changing $t \to -t$ in the tight-binding Hamiltonian, while holding all other parameters fixed, we can change the sign of winding number $n_3$ and the induced charge. }
\label{fig:Witten_Num_CHHOTI}
\end{figure*}
\par 

Witten's original result $\delta Q= -\frac{m n e }{2}$ was obtained for isotropic $\mathcal{P}$ and $\mathcal{T}$ symmetric vacuum~\cite{witten1979dyons} by combining the constraints of $\mathcal{CP}$ symmetry and the Schwinger-Zwanziger quantization condition for electric and magnetic charges of dyons~\cite{Schwinger1,Schwinger2,Zwanziger1,Zwanziger2}. When the $\mathcal{CP}$-symmetry is broken, the induced electric charge can be arbitrary. The topological quantization of $\theta$ for chiral HOTI was shown to be a consequence of the combined $C^z_4 \mathcal{T}$ symmetry~\onlinecite{Schindlereaat0346}. Furthermore, the perturbed TI$_2$ ($|\Delta|<1$) and NI ($|\Delta|>3$) states were identified as trivial insulators.

A comprehensive topological analysis of the entire phase diagram can be performed by following Refs.~\onlinecite{tyner2021symmetry,sur2022mixed}. The Bloch Hamiltonian describes map from $T^3$ to the coset space $SO(5)/SO(4)=USp(4)/[SU(2) \times SU(2)]$. The main idea is to consider (i) the first homotopy class of high-symmetry lines, and (ii) the second homotopy class or quantized Berry flux of high-symmetry planes. Such axes and planes correspond to topological defects of the map. (iii) Finally, the third homotopy class of Bloch Hamiltonian (or texture) can be identified from the tunneling of quantized Berry flux for high-symmetry planes. A convenient form of intra-band $SU(2) \times SU(2)$ Berry connection is given by 
\begin{eqnarray}\label{Berryconnection}
&&A_{i}(\bs{k})=\sum_{a<b} \; A_{i,ab} \Gamma_{ab} \nn \\
&=&\sum_{a<b} \; \frac{[N_a (\bs{k})\partial_i N_b (\bs{k})- N_b(\bs{k})\partial_i N_a(\bs{k})]\Gamma_{ab}}{2[|\bs{N}(\bs{k})|+ N_5(\bs{k})]}.
\end{eqnarray}
where $\partial_i = \frac{\partial}{\partial k_i}$, $a$ and $b$ can take values from $1,2,3,4$. By projecting with $P_\pm=\frac{1}{2}(\mathbb{1} \pm \Gamma_5)$ we obtain $SU(2)$ connections of conduction ($+$) and valence ($-$) bands, respectively. Following Ref.~\cite{tyner2020topology}, we can calculate the relative Chern numbers, 
\begin{equation}\label{relChern}
\mathfrak{C}_{ab}(k_z) = \frac{1}{2\pi} \; \int dk_x dk_y \; (\partial_x A_{y,ab} -\partial_y A_{j,ab}),
\end{equation}
measuring the flux of various non-Abelian color components through $k_x-k_y$ planes, as a function of $k_z$. Only the $\Gamma_{12}$ component supports quantized flux $\pm 2\pi$ for $C^z_4$-symmetric planes, with $k_z=0, \pi$. The tunneling configuration is classified by
\begin{equation}\label{tunnelingrelChern}
n_3=\mathfrak{C}_{12}(k_z=\pi) - \mathfrak{C}_{12}(k_z=0),
\end{equation}
leading to $\mathcal{CS}_+=n_3/2$. The results are summarized in Table~\ref{fig:CHTable}. Notice that the third homotopy class of perturbed TI$_1$ (TI$_2$) phase remains unaffected (becomes trivial). 

\subsection{Thought experiments} The results of Table~\ref{fig:CHTable} can be verified by thought experiments with unit magnetic monopole $m=+1$ [see Fig.~\ref{fig:Witten_Num_CHHOTI} ]. The results under OBC for perturbed TI$_1$ phase are similar to FOTIs, as $[001]$-surfaces can support gapless states. Only in the presence of a pseudo-scalar training field, we can observe WE.

 As chiral HOTI exhibits gapped spectrum for $[100]$ and $[010]$ surfaces, we can eliminate surface-bound zero-mode with MBC [see Fig.~\ref{fig:CHLoc} ]. We implemented the following MBC: (i) OBC along $[100]$-direction, with Dirac string oriented along $x$-axis, and (ii) PBC along $y$ and $z$-axes. As the monopole-bound FZM becomes non-degenerate, WE can be observed without any pseudo-scalar training field. As monopole unambiguously detects bulk winding number, $\delta Q_{max}=-\frac{e n_3}{2}$ remains unchanged under different boundary conditions. Only the details of compensating charge density and bulk-boundary correspondence are affected by different types of boundary conditions.

\begin{figure}
    \subfigure[]{
    \includegraphics[scale=0.65]{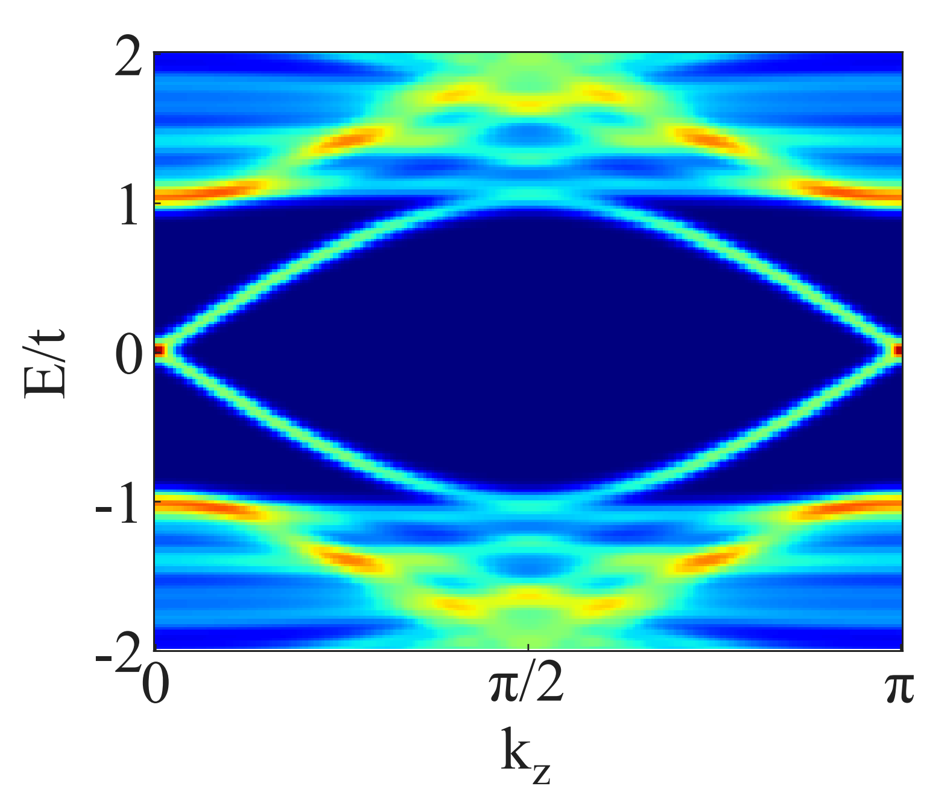}
    \label{fig:FDOS}}
    \subfigure[]{
    \includegraphics[scale=0.65]{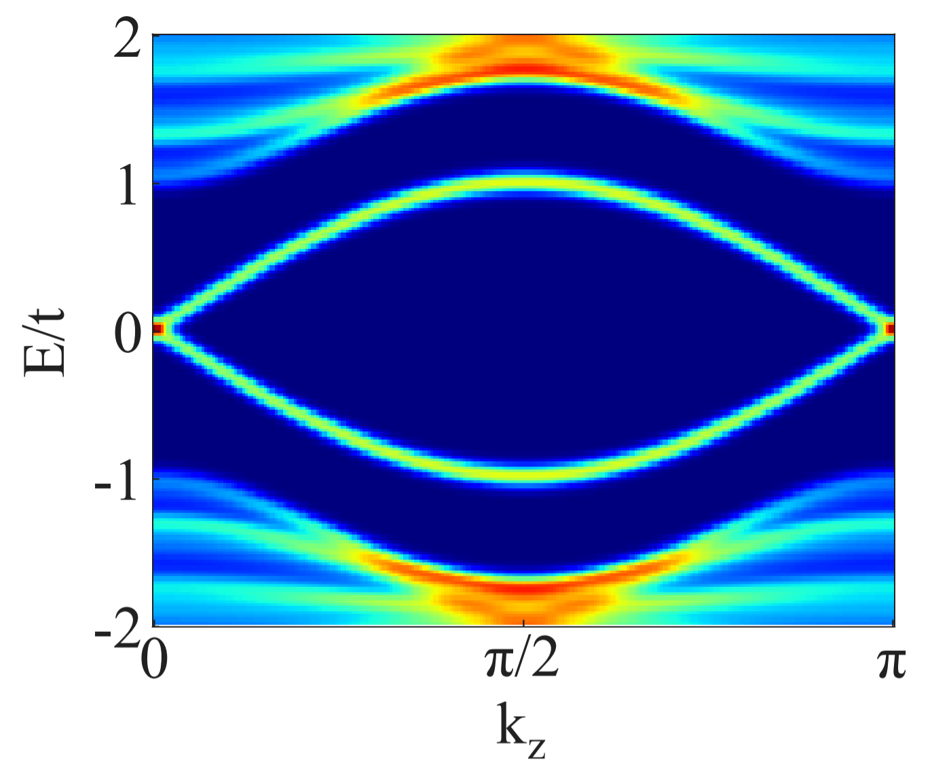}
    \label{fig:FStates}}
    \caption{Diagnosis of the second homotopy class of $xy$ planes with magnetic $\pi$-flux tube. (a) Local density of states on flux tube for first-order, TI$_2$ phase, with $\Delta=t^{\prime \prime}=0$ as a function of $k_{z}$. Both $k_z=0, \pi$ planes support two-fold degenerate zero-energy bound states. As a function of $k_{z}$, they evolve as finite-energy mid-gap states, and merge with the bulk states for $k_{z} \approx \pm \pi/2$. This is consistent with a non-trivial stacking of two-dimensional, first-order insulators (see Table~\ref{fig:CHTable}). (b) For the perturbed TI$_2$ phase, with $\Delta=0$, $t^{\prime \prime}=t^\prime=t$, the mid-gap states remain isolated, which is consistent with the trivial stacking of two-dimensional, second-order topological insulators, and $n_3=0$ (see Table~\ref{fig:CHTable}). However, the merger of mid-gap states with continuum is not an essential criterion for the existence of non-trivial third homotopy class, which will be demonstrated with the example of octupolar topological insulators. Only monopoles can unambiguously detect the third homotopy class.}
    \label{fig:FxInsert}
\end{figure}

\begin{figure*}[t]
\centering
\subfigure[]{
\includegraphics[scale=0.55]{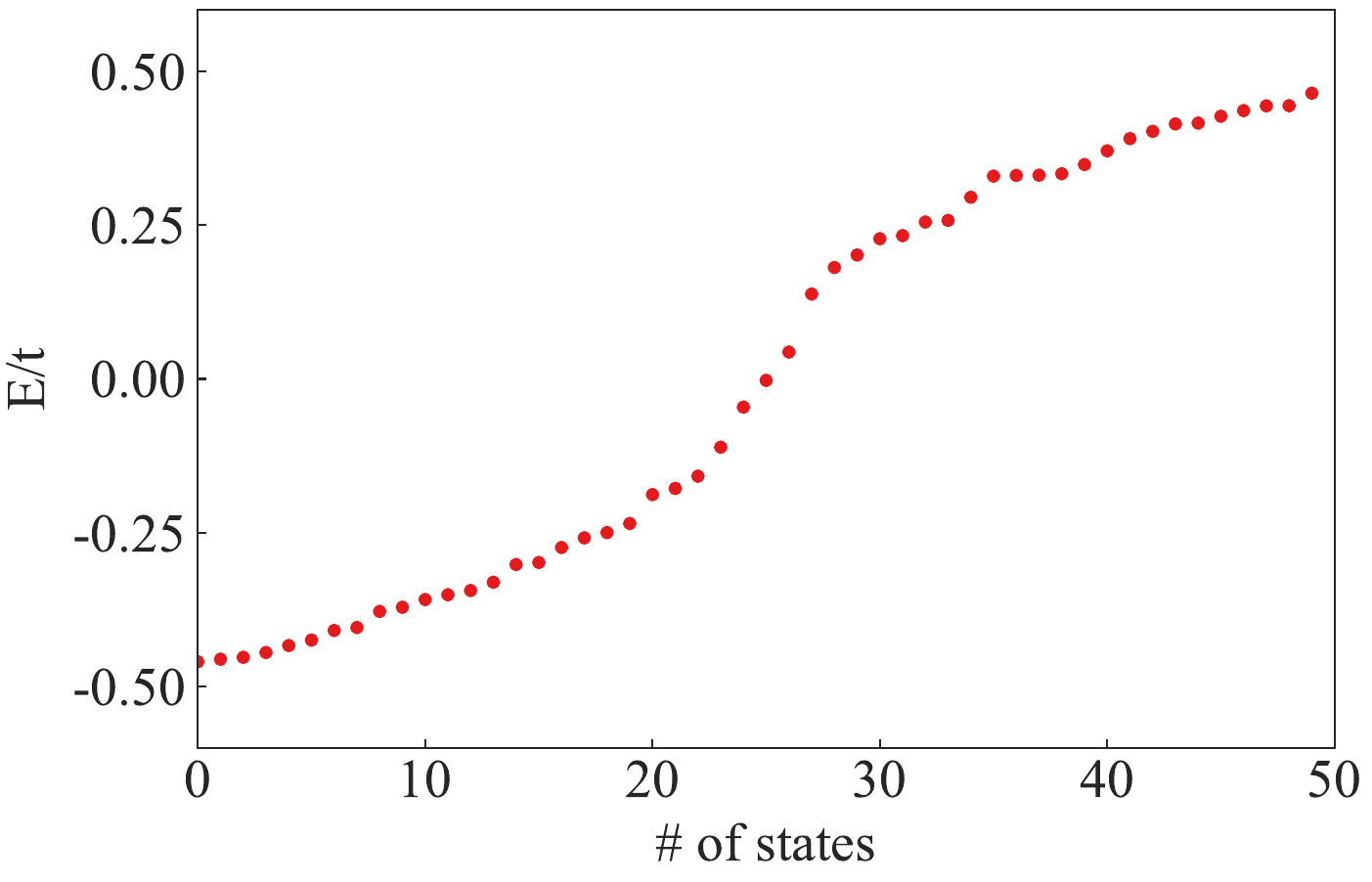}
\label{fig:8a}}
\subfigure[]{
\includegraphics[scale=0.55]{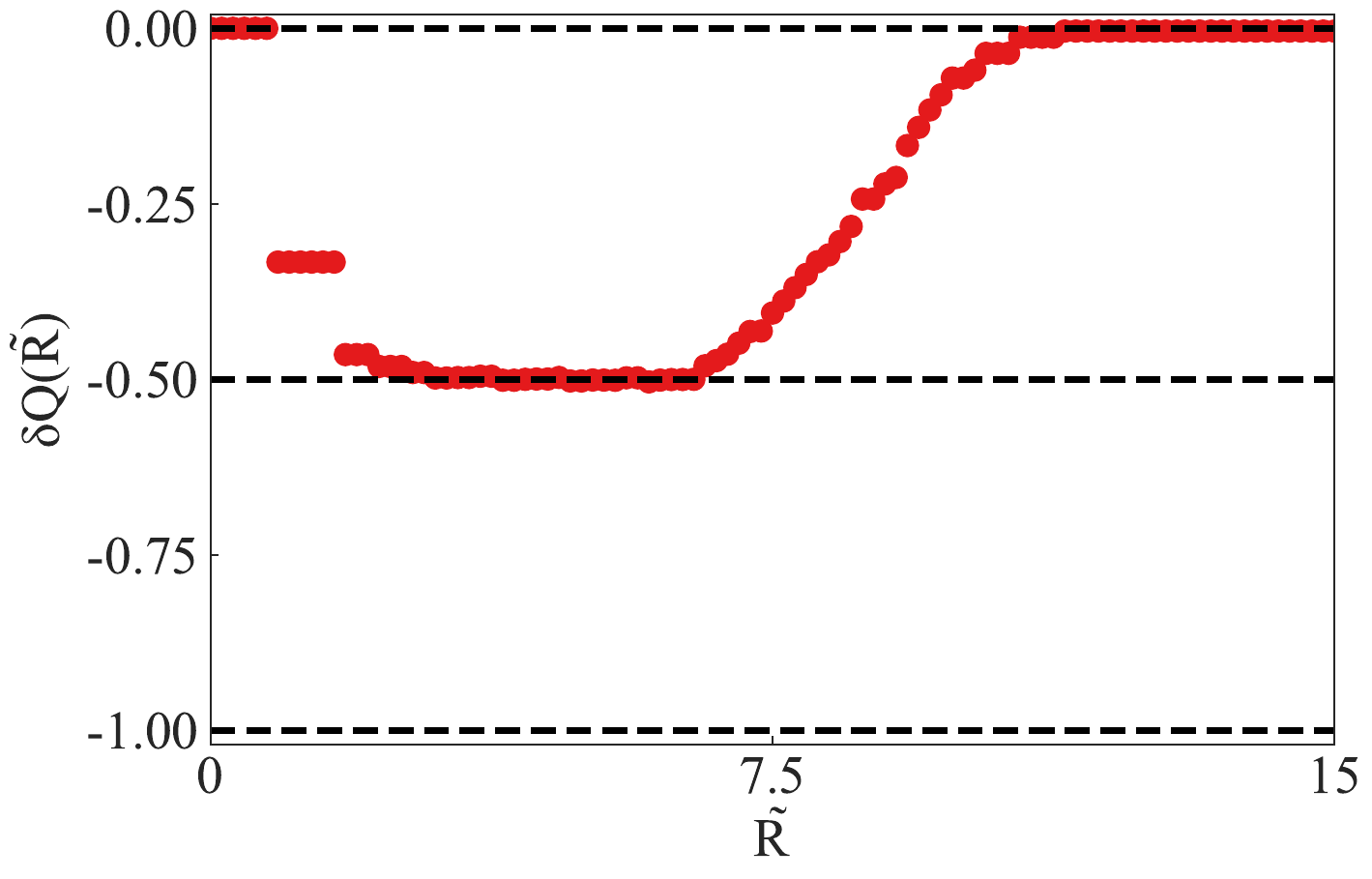}
\label{fig:8b}}
\subfigure[]{
\includegraphics[scale=0.55]{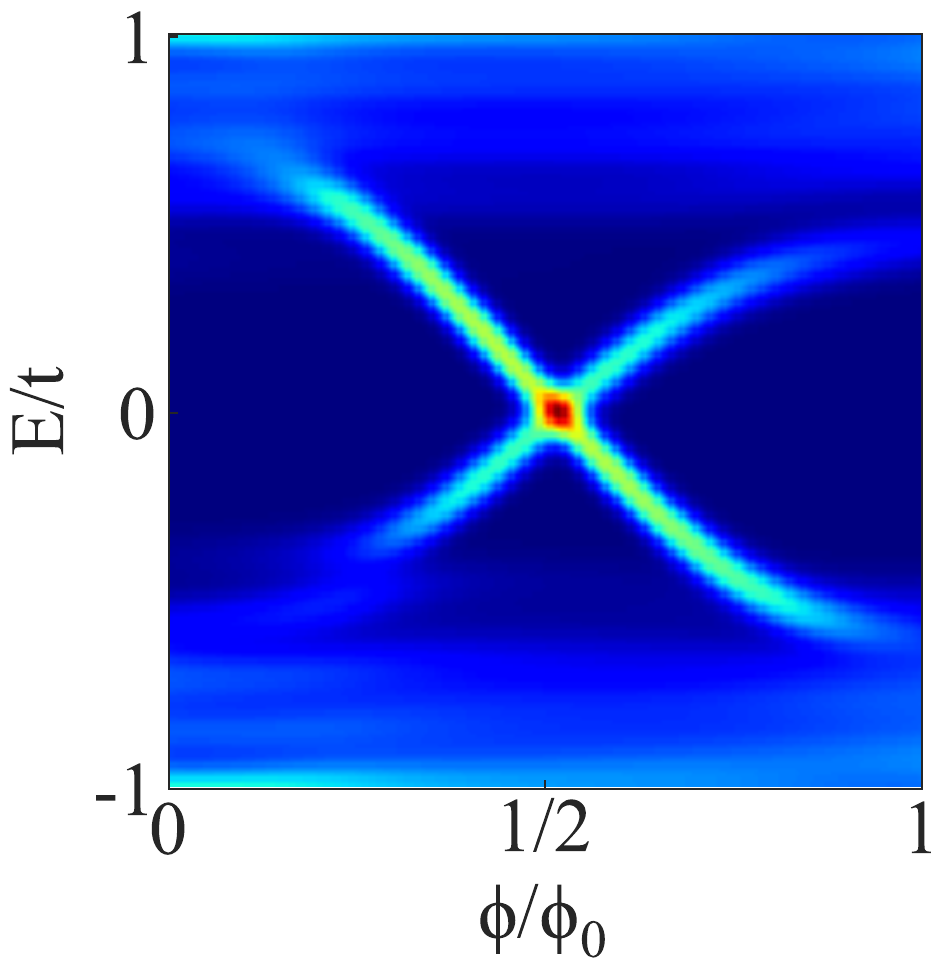}
\label{fig:8c}}
\caption{Fermion spectrum and Witten effect for $C^z_4$-symmetric, magnetic topological insulators [see Eq.~\ref{eq:MagneticTI} ] that break all fundamental discrete symmetries. The calculations are performed for system-size $(L/a)^3=(15)^3$, as it produces numerical accuracy $10^{-4}$ for the quantized value of maximum induced electric charge. We are showing the data for $m_{12}=m_{34}=m_{45}=0.1 t$, $t=t^\prime$, and $\Delta=2$. (a) Fifty lowest energy states, in the presence of unit monopole $m=+1$. Due to the lack of particle-hole symmetry, the bound states move away from $E=0$. For magnetic topological insulators, the monopole- and surface- localized modes are not energetically degenerate. Thus, half-filled system can exhibit Witten effect without any pseudo-scalar training field. (b) The maximum induced electric charge (in units of $-e$) for half-filled system is given by $-1/2$, which identifies $Tr[\theta_{ij}]=-3\pi$ [see Eq.~\ref{tracetheta} ]. (c) In the presence of a magnetic flux tube, along the $z$-axis, only the $k_z=\pi$ plane shows two-fold degenerate fermion zero-modes, when $\phi=\phi_0/2$. This identifies non-trivial (trivial) second homotopy class of $k_z=\pi$ ($0$) plane, arising from $C^z_4$-symmetry protected Berry flux. The degeneracy of zero-modes and subsequent splitting for $\phi \neq \phi_0/2$ show that the net Berry flux of two occupied valence bands for the $k_z=\pi$ plane is zero (an example of 2D, non-quantum Hall insulator). Therefore, the Berry flux is carried by a traceless generator other than $\Gamma_5$. The non-trivial third homotopy class arises from the tunneling configuration of $C^z_4$-symmetry protected Berry flux. }
\label{fig:MagneticTI}
\end{figure*}

Table~\ref{fig:CHTable} also implies that the perturbed TI$_2$ is not a topologically trivial phase. It is a HOTI, which is distinguished from the trivial phase by the non-trivial second homotopy class of high-symmetry planes. Vanishing of $n_3$ indicates that the perturbed TI$_2$ can be considered as 2D, second-order TIs, trivially stacked along the $z$-direction. To confirm this assertion, we have probed the second homotopy class of $C^z_4$-symmetric planes with a magnetic $\pi$-flux tube~\cite{SpinChargeSCZ,VishwanathPiFlux,Juricic2012,mesaros2013zero}. After performing Fourier transformation of $H_2$ along $x$ and $y$ directions, we insert a magnetic $\pi$-flux tube at $(x,y)=(0,0)$, oriented along the $z$ direction. Thus, the hopping matrix elements are modified according to $t_{ij} \to t_{ij} e^{i\nu_{ij}}$,
where $\nu_{ij}=(e/\hbar)\int_{\bs{r}_i}^{\bs{r}_j}\mathbf{a}_t\cdot d\mathbf{l}$, and the vector potential
\begin{equation}
\bs{a}_t(\bs{r}_i)=\frac{\hbar}{2e} \frac{-y_i \hat{x} + x_i \hat{y}}{x^2_i + y^2_i}.
\end{equation}
As $k_z$ is a good quantum number, we have simulated a system of transverse-size $L_x/a=L_y/a=20$, under PBC along all three principal axes. The local density of states on the flux tube, for the first-order TI$_2$ phase and the perturbed TI$_2$ phase are shown shown in Fig.~\ref{fig:FxInsert}. The flux tube binds two-fold-degenerate zero-modes (Kramers pair) at $k_{z}=0, \; \pi$ planes, identifying these planes as generalized quantum spin Hall insulators. 
 
 The results on chiral HOTI do not imply that ME HOTIs with $C^z_4 \mathcal{T}$-symmetry can only possess $n_3 = 0, \pm 1$. Such results are specific to the model of Eq.~\ref{eq:CH}. Instead of $B_{1g}$ perturbation $t^{\prime \prime} [\cos (k_1) -\cos (k_2)]\Gamma_4$, we can consider (i) $t^{\prime \prime} [\cos (2k_1) -\cos (2k_2)]\Gamma_4$,  or (ii) $B_{2g}$-type perturbation $t^{\prime \prime} \sin k_{1} \sin k_{2} \Gamma_{4}$ to arrive at different models of ME HOTIs. By employing Eqs.~\ref{relChern}-\ref{tunnelingrelChern} and thought experiments for both models, we can show that the non-trivial third homotopy class of perturbed TI$_2$ phase is described by $n_3=\mathcal{N}_3=+2$. Therefore, $C^z_4 \mathcal{T}$-symmetric ME HOTIs can support $\theta = n \pi$, with $n \in \mathbb{Z}$.

\section{Magnetic topological insulators} \label{AppA}
Next we discuss the third homotopy class and quantized, ME response of magnetic TIs that require minimal protection of global discrete symmetries. In this regard, we consider the following four-band model
 \begin{eqnarray}
 H_{mag}=H_1(\bs{k})+m_{12}\Gamma_{12} + m_{34} \Gamma_{34} + m_{45} \Gamma_{45}, \label{eq:MagneticTI}
\end{eqnarray}
The fermion bilinear $\Psi^\dagger \Gamma_{12} \Psi$ is odd under $\mathcal{T}$ and even under $\mathcal{P}$ and $\mathcal{C}$. The bilinear $\Psi^\dagger \Gamma_{34} \Psi$ is odd under $\mathcal{T}$ and $\mathcal{C}$, but even under $\mathcal{P}$. Finally, the bilinear $\Psi^\dagger \Gamma_{45} \Psi$ is odd under $\mathcal{P}$, but even under $\mathcal{T}$, and $\mathcal{C}$. Therefore, Eq.~\ref{eq:MagneticTI} describes a generic member of unitary Altland-Zirnbauer class A. Such systems break all fundamental discrete symmetries $\mathcal{P}$, $\mathcal{T}$, $\mathcal{C}$, and the combined discrete symmetries $\mathcal{PT}$, $\mathcal{CP}$, $\mathcal{CT}$, and $\mathcal{CPT}$. For the present model only the $U(1)$ total number conservation law and the discrete four-fold rotational symmetry about the $z$ axis ($C^z_4$) remain intact. 

Within the Altland-Zirnbauer classification scheme, one flattens all occupied (empty) states to have identical dispersion. Thereby, it causes a fictitious degeneracy of occupied (empty) non-degenerate valence (conduction) bands. For the 4-band model of Eq.~\ref{eq:MagneticTI}, Altland-Zirnbauer scheme will identify the coset space to be $\frac{U(4)}{U(2) \times U(2)}$. Since the third homotopy group $\pi_3(\frac{U(4)}{U(2) \times U(2)})$ is trivial, one will conclude that the model cannot support topologically non-trivial, 3D insulators~\cite{RyuLudwigPRB,ryu2010topological}. An exception to this rule was considered in Ref.~\cite{moore2008topological}, while addressing two-band models of Hopf insulators, as $\pi_3(\frac{U(2)}{U(1) \times U(1)})=\mathbb{Z}$. Very recently, Ref.~\cite{Lapierre2021} has considered the correct coset space of $N$-band models of Hopf insulators $\frac{U(N)}{[U(1)]^N}$, which can support non-trivial third homotopy classification, as $\pi_3(\frac{U(N)}{[U(1)]^N})=\mathbb{Z}$. 

We emphasize that the correct coset space of non-degenerate bands is always described by $\frac{U(N)}{[U(1)]^N}$, and magnetic TIs (both Hopf and non-Hopf insulators) can possess non-trivial third homotopy class. Arbitrary deformation of Lie groups or coset space changes the homotopy groups, leading to incorrect results. This issue has been described for tight-binding models and \emph{ab initio} band structures in Ref.~\onlinecite{tyner2021symmetry}. In general, the $N$-band model of class A will possess a diagonal matrix of topological invariants,
$n_{ij} = n_i \delta_{ij}$,
where $n_i$ is the Abelian Chern Simons coefficient of $i$-th band.
 
Even though $H_{mag}$ is a sufficiently simple 4-band model, one cannot analytically obtain closed form expressions of band dispersions and eigenfunctions. Therefore, the bulk invariant must be computed numerically. Furthermore, numerical calculations show that the model exhibits direction-dependent gapped and gapless surface-states. The [001] ([100] or [001]) surface harbors gapped (gapless) surface-states. Hence, $H_{mag}$ describes magnetic HOTIs.  

Since $\mathcal{T}$-breaking operators reduce spatial rotational symmetries, the effective action for adiabatic electrodynamic response
 \begin{eqnarray}
S_{eff}&=&\int d^3x dt \bigg [\frac{1}{2} \epsilon_{ij} E_i E_j + \frac{1}{2} \mu^{-1}_{ij} B_i B_j +  \frac{\theta_{ij} e^2}{4 \pi^2 \hbar} E_i B_j \nn \\
&&+....\bigg],
\end{eqnarray}
will involve anisotropic permittivity ($\epsilon_{ij}$), permeability ($ \mu_{ij} $), and ME ($\theta_{ij}$) tensors, and higher-order, anisotropic couplings are indicated by ellipsis. A static magnetic monopole gives rise to the displacement field 
\begin{equation}
D_i (\bs{r})=  P_{ME,i}(\bs{r})=\frac{\theta_{ij} e m}{8 \pi^2 \hbar} \frac{\hat{r}_j }{r^2},
\end{equation}
and anisotropic volume charge density $\rho(\bs{r})= -\nabla \cdot \bs{P}_{ME}$. Since the total induced electric charge enclosed by a spherical Gaussian surface equals 
\begin{equation}\label{tracetheta}
Q=- \frac{e m}{2 \pi} \; \frac{ \text{Tr}[\theta_{ij} ]}{3}, 
\end{equation}
the Chern-Simons invariant for the ground state determines $\frac{ \text{Tr}[\theta_{ij} ]}{6 \pi}$. Can it be quantized for a system breaking all fundamental discrete symmetries and $C^z_4 \mathcal{T}$? 
 
To answer this question we performed thought experiments with magnetic monopoles. The results are shown in Fig.~\ref{fig:MagneticTI}. As the system lacks $\mathcal{C}$, $\mathcal{P}$ and $\mathcal{T}$ symmetries, the monopole and surface localized states are no longer degenerate. Furthermore, when $m_{12} \neq 0$, the Hamiltonian does not possess $\mathbb{Z}_2$ particle-hole symmetry, as $\{H_{mag},\Gamma_4\} \neq 0$. Therefore, monopole- and surface- localized, bound states move away from $E=0$. For sufficiently small $m_{12}$, $m_{34}$, and $m_{45}$, such that the band gap is not closed, the half-filled system exhibits quantized ME effect, without any pseudo-scalar training mass $M^\prime$. Thus, generic magnetic insulators with $n$-fold rotational symmetry can exhibit non-trivial third homotopy classification and topologically quantized, ME response. By tuning the band parameter $\Delta$, we also confirmed that magnetic TIs allow both odd and even integer invariants (i.e., $\mathbb{Z}$-classification).

\section{Octupolar topological insulators}\label{TOTI1}
Finally, we consider the third homotopy class of $\mathcal{C}$-, $\mathcal{P}$-, and $\mathcal{T}$- symmetric octupolar TIs, which exhibit gapped surface-states, and corner-localized, zero-energy, mid-gap states, under cubic-symmetry-preserving, OBC~\cite{Benalcazar61}. There are different but equivalent ways of writing $8 \times 8$ Bloch Hamiltonian of third order TIs. Ref.~\onlinecite{Benalcazar61} considered spin-less fermions to describe octupolar TIs. But the assumption of spin-less fermions is not essential for describing topological universality class of octupolar HOTIs. The general form of Bloch Hamiltonian
\begin{eqnarray}\label{TOTI}
&& H_3(\mathbf{k})=t\sum^{3}_{j=1}\sin k_{j}\gamma_{j}+t^\prime(\Delta +\sum^{3}_{j=1}\cos k_{j})\gamma_{5} + \nn \\ && t^{\prime \prime}(\cos k_{x}-\cos k_{y})\gamma_{4}+\frac{t^{\prime \prime}}{\sqrt{3}}(2\cos k_{z}-\cos k_{x}-\cos k_{y})\gamma_{6}, \nn \\
\end{eqnarray}
has been described in Ref.~\onlinecite{sur2022mixed}, and we have demonstrated that the octupolar HOTI can display gapless surface-states (four-component, 2D massless Dirac fermions) for $[111]$ surface. 

On a physical ground, we can think of hybridization between two four-band FOTIs, with $E_g$ type d-wave harmonics $(\cos k_x -\cos k_y)$ and $(2\cos k_z -\cos k_x -\cos k_y)$. The specific assignment of hopping constants $t^{\prime \prime}$ and $t^{\prime \prime}/\sqrt{3}$ guarantees that the cubic symmetry is preserved. Here $\gamma_j$'s are six mutually anti-commuting $8 \times 8$ matrices, and $\{H_3, \gamma_7 \}=0$ describes the $\mathbb{Z}_2$ particle-hole symmetry with respect to the absent matrix $\gamma_7$. The explicit form of $\gamma_j$'s depend on the choice of basis.  

When $t^{\prime \prime}=0$, the decoupled model displays $SU(2)$ flavor symmetry, generated by $\gamma_{46}$, $\gamma_{67}$, and $\gamma_{47}$, with $\gamma_{ij}=[\gamma_i, \gamma_j]/(2i)$. Such non-Abelian flavor symmetry is broken by the hybridization terms. Consequently, topological invariants and response of octupolar TI will require a clear understanding of important and inter-twined roles of FZMs, $\mathcal{CP}$-violation and flavor-symmetry-breaking. 

 \begin{figure*}
\centering
\subfigure[]{
\includegraphics[scale=0.195]{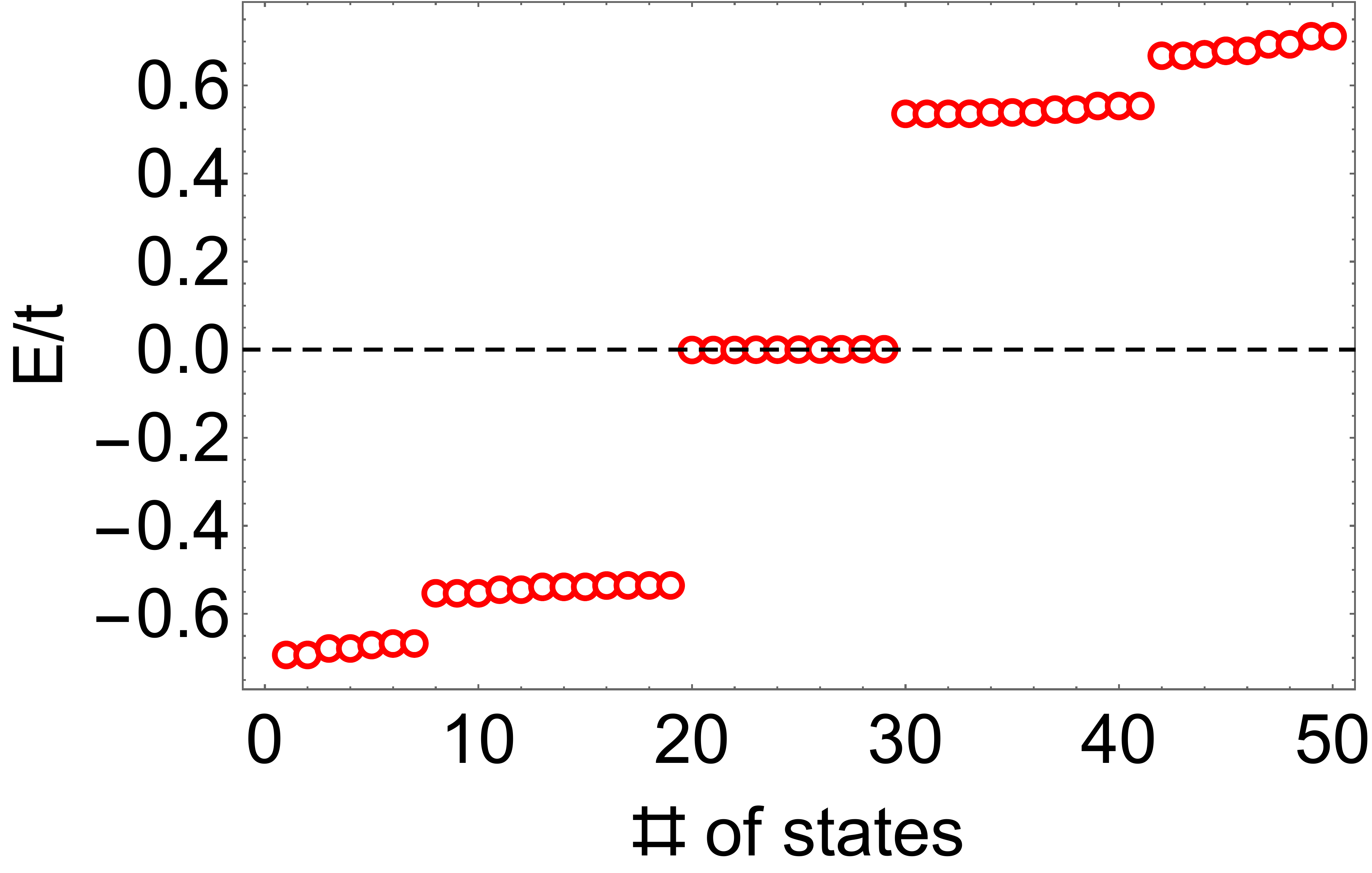}
\label{fig:OHState}}
\subfigure[]{
\includegraphics[scale=0.45]{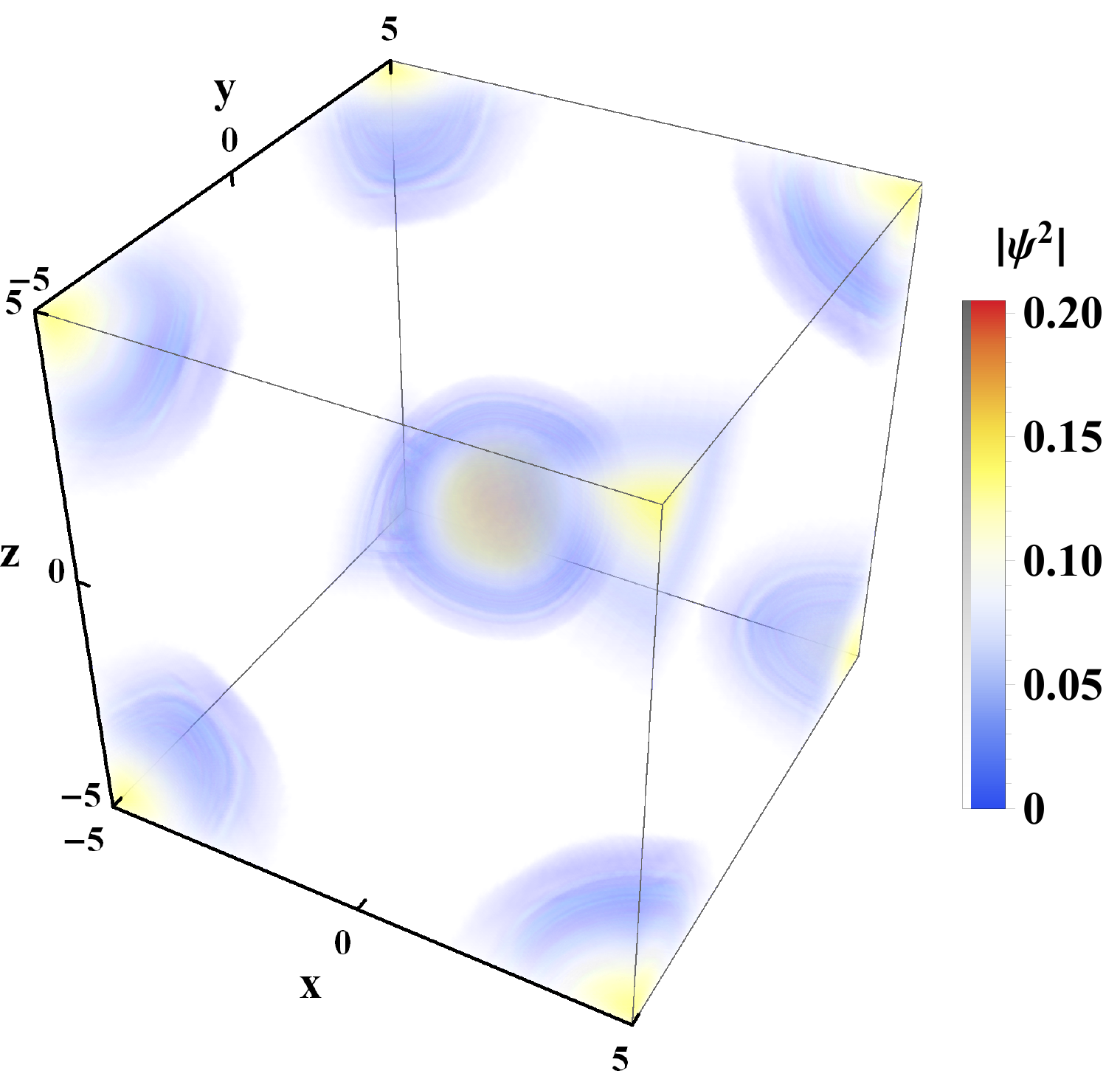}
\label{fig:OHLoc}}
\subfigure[]{
\includegraphics[scale=0.2]{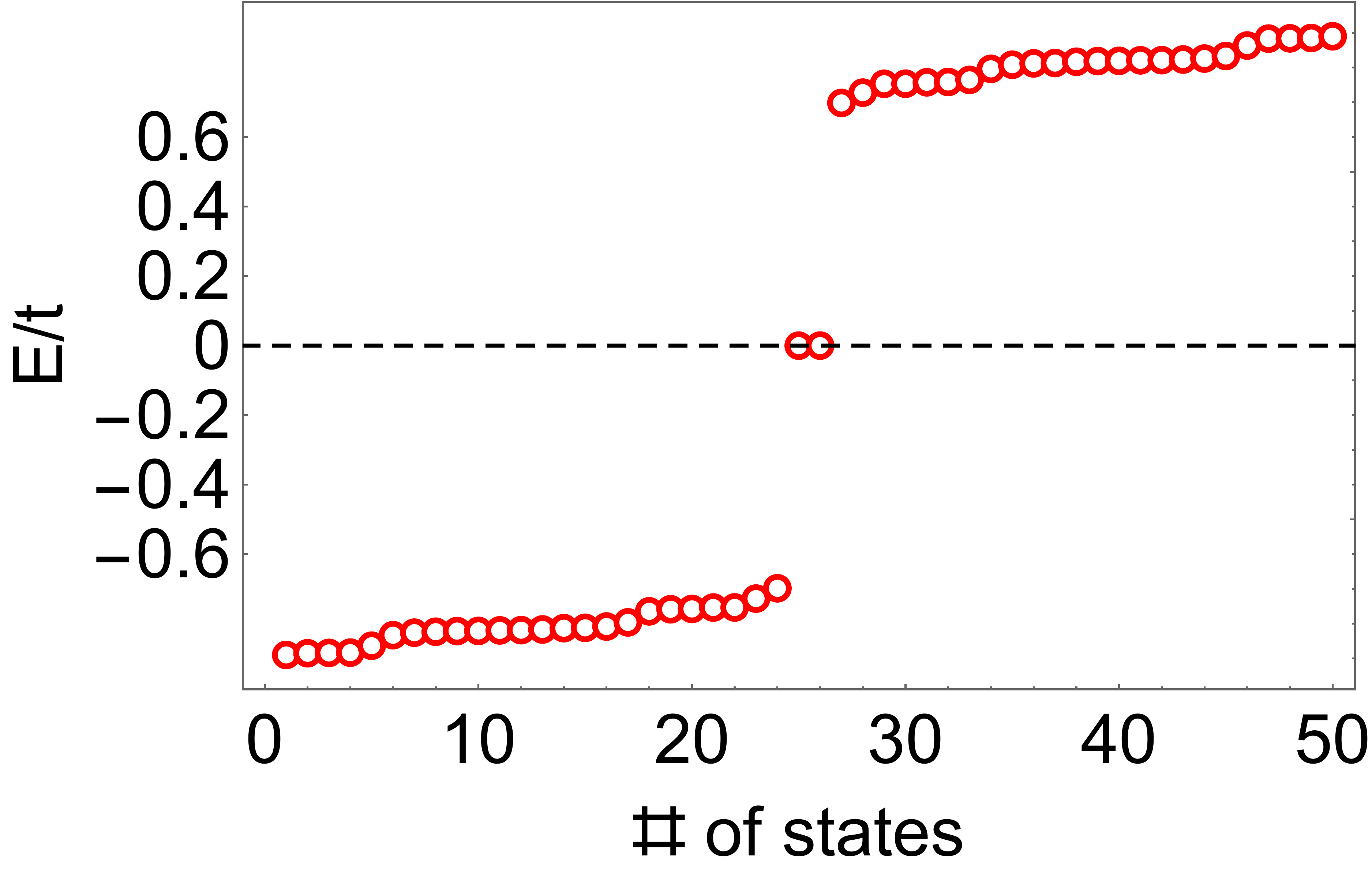}
\label{fig:OHPBC}}
\subfigure[]{
\includegraphics[scale=0.45]{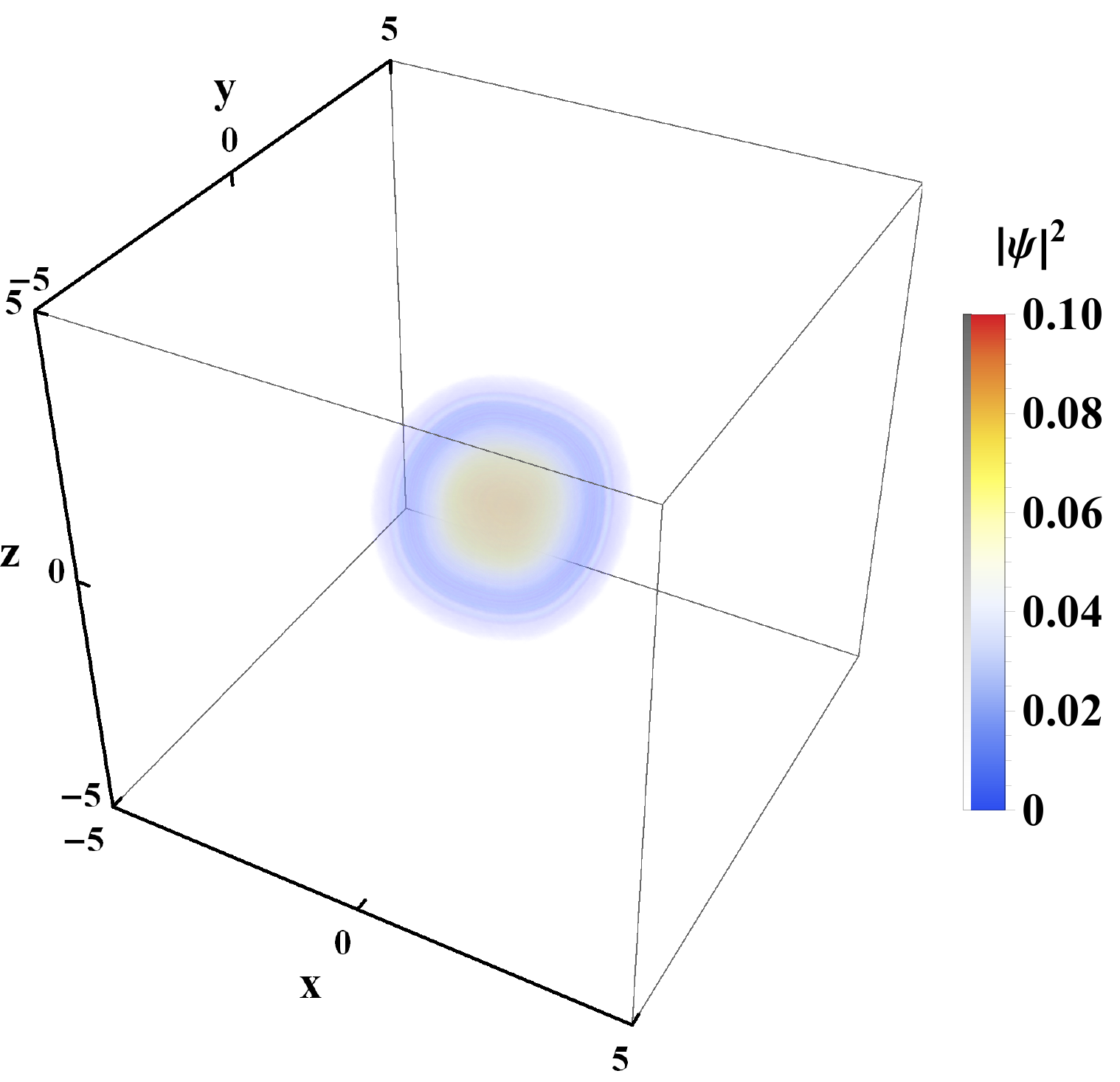}
\label{fig:OHLocPBC}}
\subfigure[]{
\includegraphics[scale=0.195]{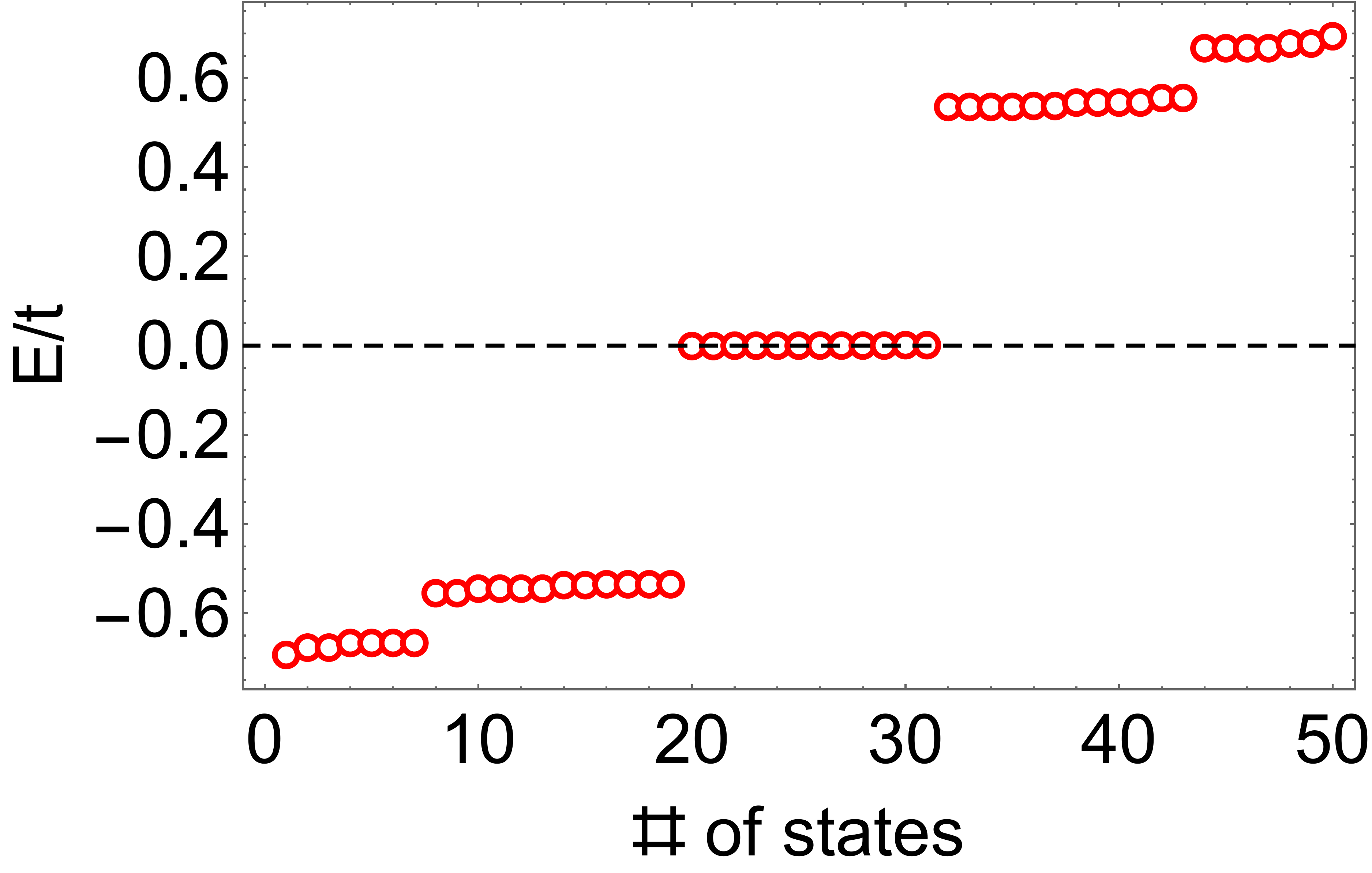}
\label{fig:OHStatedoubleopen}}
\subfigure[]{
\includegraphics[scale=0.195]{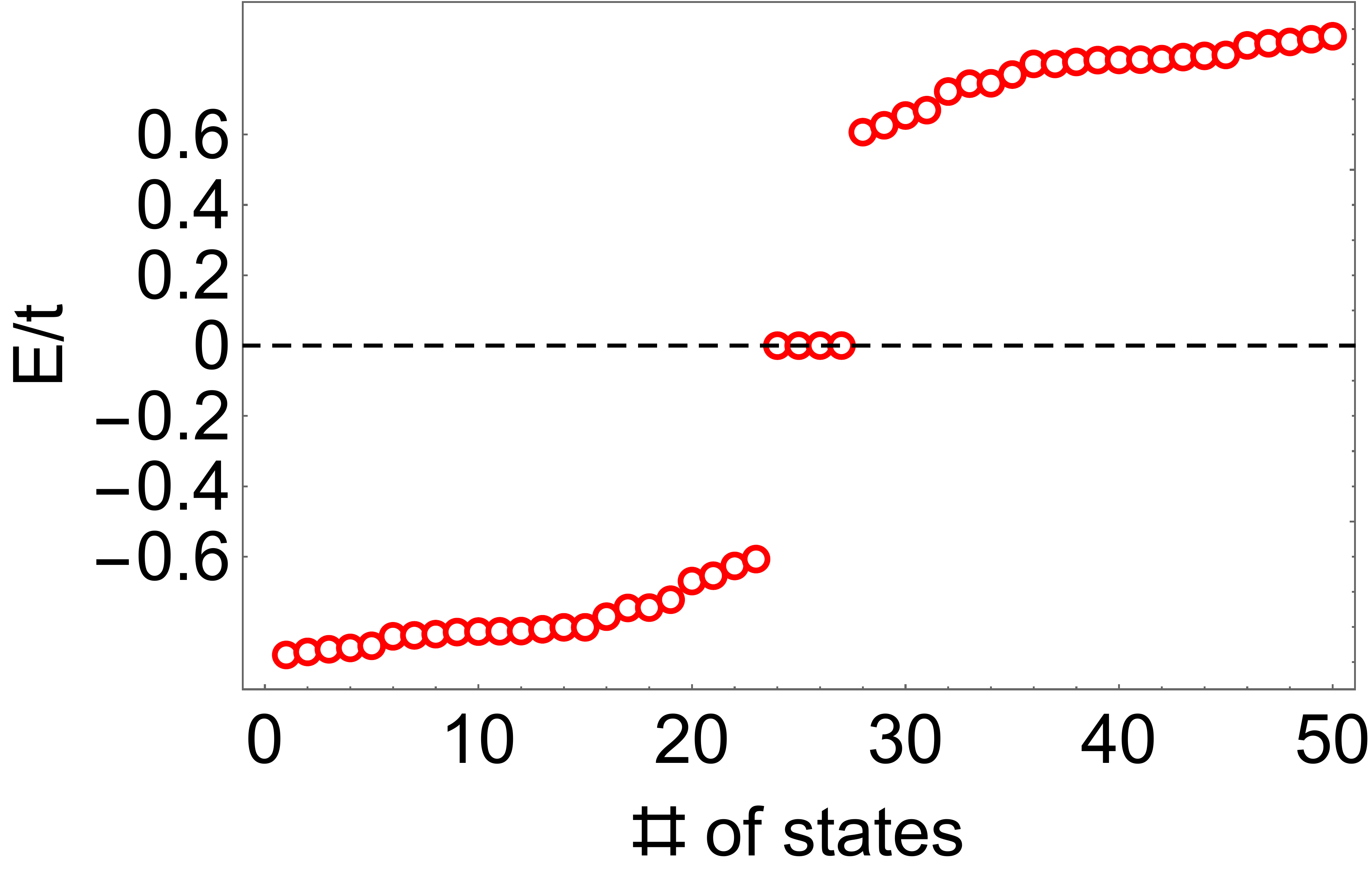}
\label{fig:OHStatedoublemixed}}
\caption{Fermion spectrum for octupolar topological insulators [see Eq.~\ref{TOTI} ] in the presence of monopoles. (a) Fifty low-lying states near $E=0$, under cubic-symmetry-preserving, open boundary conditions, in the presence of unit monopole $m=+1$. We used band parameters $t=t^\prime=t^{\prime \prime}$, and $\Delta=2$. (b) Localization pattern of ten-fold degenerate fermion-zero-modes, shown in (a). Eight (two) modes are bound to the corners (monopole). (c) Low lying states under mixed boundary conditions. Corner-states are removed by implementing periodic (open) boundary conditions along $x,y$ ($z$) directions. (d) Two-fold degenerate fermion-zero-modes, shown in (c), are bound to the monopole. For a general monopole $m$, the number of zero modes are given by $2|m| +8$ and $2|m|$ for open and mixed boundary conditions respectively. This is confirmed for a double monopole $m=+2$, under (e) open boundary conditions, and  (f) mixed boundary conditions. Clearly, the three-dimensional bulk invariant is tracked by the number of zero-modes, localized on monopole, irrespective of boundary conditions.}
\end{figure*}

As the model only allows anti-commuting $8 \times8$ gamma matrices, the conduction and valence bands are four-fold degenerate, and $H_3(\mathbf{k})$ describes map from $T^3$ to the coset space $SO(6)/SO(5)$. The gauge group of intra-band Berry connection is $SO(5)=USp(4)/\mathbb{Z}_2$ and the calculation of bulk topological invariants is complicated by the four-fold degeneracy. This model supports one topologically non-trivial phase, when $|\Delta| < 3$. 

Without any loss of generality, we will consider the following representation of gamma matrices: 
\begin{eqnarray}
&& \gamma_{j}=\eta_0 \otimes \Gamma_{j}, \: \text{with} \; j=1,2,3,5 \; \nn \\
&& \gamma_4=\eta_1 \otimes \Gamma_4, \; \Gamma_6=\eta_2 \otimes \Gamma_4, \; \Gamma_7=\eta_3 \otimes \Gamma_4,
\end{eqnarray}
where the $2 \times 2$ identity matrix $\eta_0$ and the Pauli matrices $\eta_j$'s operate on the flavor index. The $\mathcal{C}$, $\mathcal{P}$, $\mathcal{T}$, and the mirror symmetries are implemented as 
\begin{eqnarray}
&& \mathcal{C}^\dagger H^\ast_3(-\bs{k}) \mathcal{C} = - H_3(\bs{k}) , \; \mathcal{C}=i \gamma_3 \gamma_1 \gamma_4=\eta_1 \otimes \Gamma_{25}, \\
&& \mathcal{P}^\dagger H(-\bs{k}) \mathcal{P} = H(\bs{k}), \; \mathcal{P}=i\gamma_{4}\gamma_{5} \gamma_{6}=\eta_3 \otimes \Gamma_5, \\
&& \mathcal{T}^\dagger H^\ast_3(-\bs{k}) \mathcal{T} = H_3(\bs{k}), \; \mathcal{T}= i \gamma_2 \gamma_5 \gamma_6= \eta_2 \otimes \Gamma_{31}, \\
&&M^\dagger_j H_3(-k_j) M_j = H_3(k_j), \; M_j= \gamma_{j7}=\eta_3 \Gamma_{j4}, \nn \\ && \text{with} \; j=1,2,3.
\end{eqnarray} 
Consequently, the fermion bilinear $\Psi^\dagger \gamma_7 \Psi$ is odd under $\mathcal{C}$, $\mathcal{P}$, three mirror operations, and even under $\mathcal{T}$, like an $\mathcal{CP}$-symmetric, $xyz$-type, electric octupole moment. In contrast to this, $\Psi^\dagger (-i \gamma_4 \gamma_6 \gamma_7) \Psi $ is the $\mathcal{C}$ even, but $\mathcal{P}$-, $\mathcal{T}$-, $M_j$- odd, pseudo-scalar operator. 

\subsection{Monopoles and fermion spectrum} 
To unambiguously probe three-dimensional invariant and topological response, we have performed thought experiments with Dirac monopoles. Without monopole, the octpolar HOTI supports eight, corner-localized, NZMs, under \emph{cubic-symmetry-preserving}, OBC. In the presence of minimal monopole $m=+1$, two monopole-localized NZMs are introduced, raising the total number of NZMs to ten [see Fig.~\ref{fig:OHState} ]. The localization pattern of NZMs is shown in Fig.~\ref{fig:OHLoc}. 
Note that under OBC, the $SU(2)$-flavor-symmetric, decoupled model (with $t^{\prime \prime}=0$) would lead to four NZMs (two on monopole and two on surface). As the hybridization terms ($t^{\prime \prime} \neq 0$ ) gap out all surface- and hinge- localized states, only two monopole-localized NZMs can survive. Thus, the number of NZMs, bound to the monopole agrees with the assignment of a bulk winding number $|n_{3}|=2$. 

To separate the physics of bulk and corner-localized states, we can impose MBC: (i) PBC in $xy$ plane, and (ii) OBC along $z$ direction to maintain the invisibility of Dirac string. The resulting spectrum for a system-size $L/a=10$ is shown in Fig.~\ref{fig:OHPBC}. As MBC does not obey cubic symmetry, all corner-localized modes are eliminated. But the monopole-localized NZMs remain unaffected, as shown in Fig.~\ref{fig:OHLocPBC}. 

To further demonstrate the separation between bulk topological invariant and geometry-dependent corner-states, we have considered higher monopole strengths. The results for $m=+2$ are shown in Fig.~\ref{fig:OHStatedoubleopen} and \ref{fig:OHStatedoublemixed}. While the number of corner-states ($8$ vs. $0$) is fixed by the boundary conditions, the number of zero-modes localized on monopole is given by $2 |m|$, irrespective of boundary conditions. Thus, the number of monopole-localized zero-modes identifies $|n_{3}|=2$ and performs $\mathbb{N}$-classification of the third homotopy class of octupolar HOTI.

\subsection{Witten effect and spin-charge separation} 
\begin{figure}[t]
\centering
\includegraphics[scale=0.57]{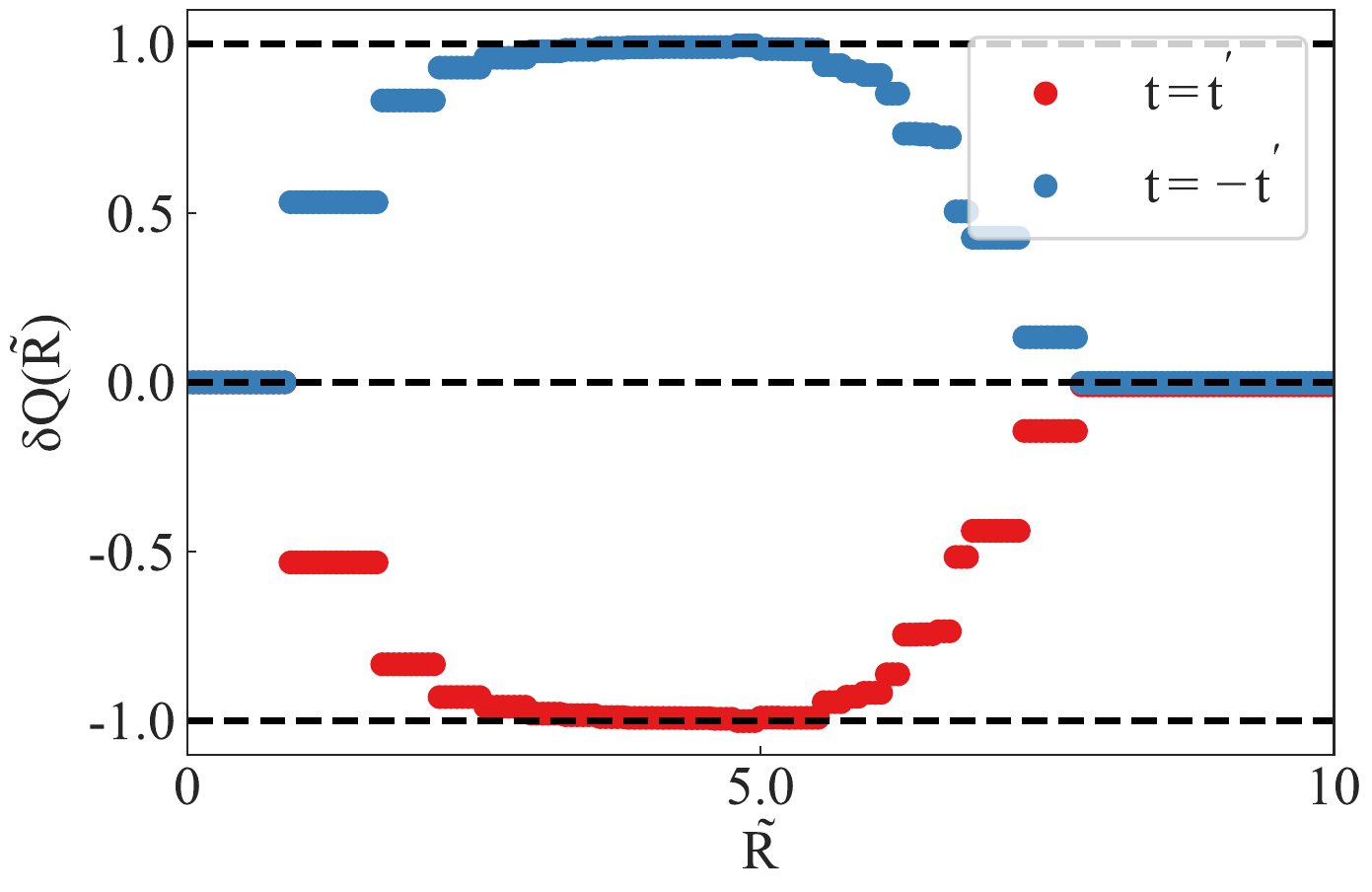}
\caption{Witten effect for half-filled, higher order octupolar topological insulator in the presence of a minimal monopole $m=+1$. The system has been trained with a small $\mathcal{CP}$-violating, pseudo-scalar field $M^\prime=+0.1 t$ (see Eq.~\ref{octpseudo1} ). Witten effect identifies $\mathbb{Z}$-classification of magneto-electric coefficient (see Eq.~\ref{octpseudo2} ). Identical quantization of induced electric charge is observed for open and mixed boundary conditions, which shows the bulk response is insensitive to the presence of corner-states. For numerical calculations we have used $t^{\prime}=t=t^{\prime \prime}$, $\Delta=2$, and system-size $(L/a)^3=10^3$. For larger system-size, we can gradually reduce the strength of $M^\prime$ toward zero.}
 \label{fig:OctIndtrainingCharge}
\end{figure}

To establish $\mathbb{Z}$-classification of quantized ME response, we have studied WE in the presence of a pseudo-scalar training field. Therefore, the Bloch Hamiltonian is modified as 
\begin{equation}\label{octpseudo1}
H_3 \to H^\prime_3 = H_3 + M^\prime (-i \gamma_4 \gamma_6 \gamma_7).
\end{equation}
and we calculate monopole-bound electric charge $\delta Q(L,R,\xi,M^\prime)$ by taking $M^\prime \to 0^+$ limit. The results for $m=+1$ monopole are shown in Fig.~\ref{fig:OctIndtrainingCharge}. The half-filled state exhibits WE and the maximum induced electric charge saturates to $\delta Q_{max} = -\frac{e m \theta}{2\pi}$, with
\begin{equation}\label{octpseudo2}
\theta= n_3 \pi, \; \text{and} \; n_3=-2 \; \text{sgn}(t) \; \Theta(3-|\Delta|) ,
\end{equation}
under OBC, and MBC. This implies the Chern-Simons coefficient of the ground state is given by $\mathcal{CS}_+ = \lim_{M^\prime \to 0+} \mathcal{CS}(M^\prime) = \frac{n_3}{2} \neq 0$. 
To maintain overall charge-neutrality of the half-filled state, compensating bound charge $-\delta Q_{max}$ will be localized in the vicinity of eight corners under OBC. Hence, each corner will support 
\begin{equation}
\delta Q_{corner}=+\frac{1}{8} \; e \; m \; \text{sgn}(n_3).
\end{equation} 
In contrast to this, under MBC, $-\delta Q_{max}$ will be shared by two gapped $[001]$-surfaces.

\begin{figure}[t]
\centering
\includegraphics[scale=0.55]{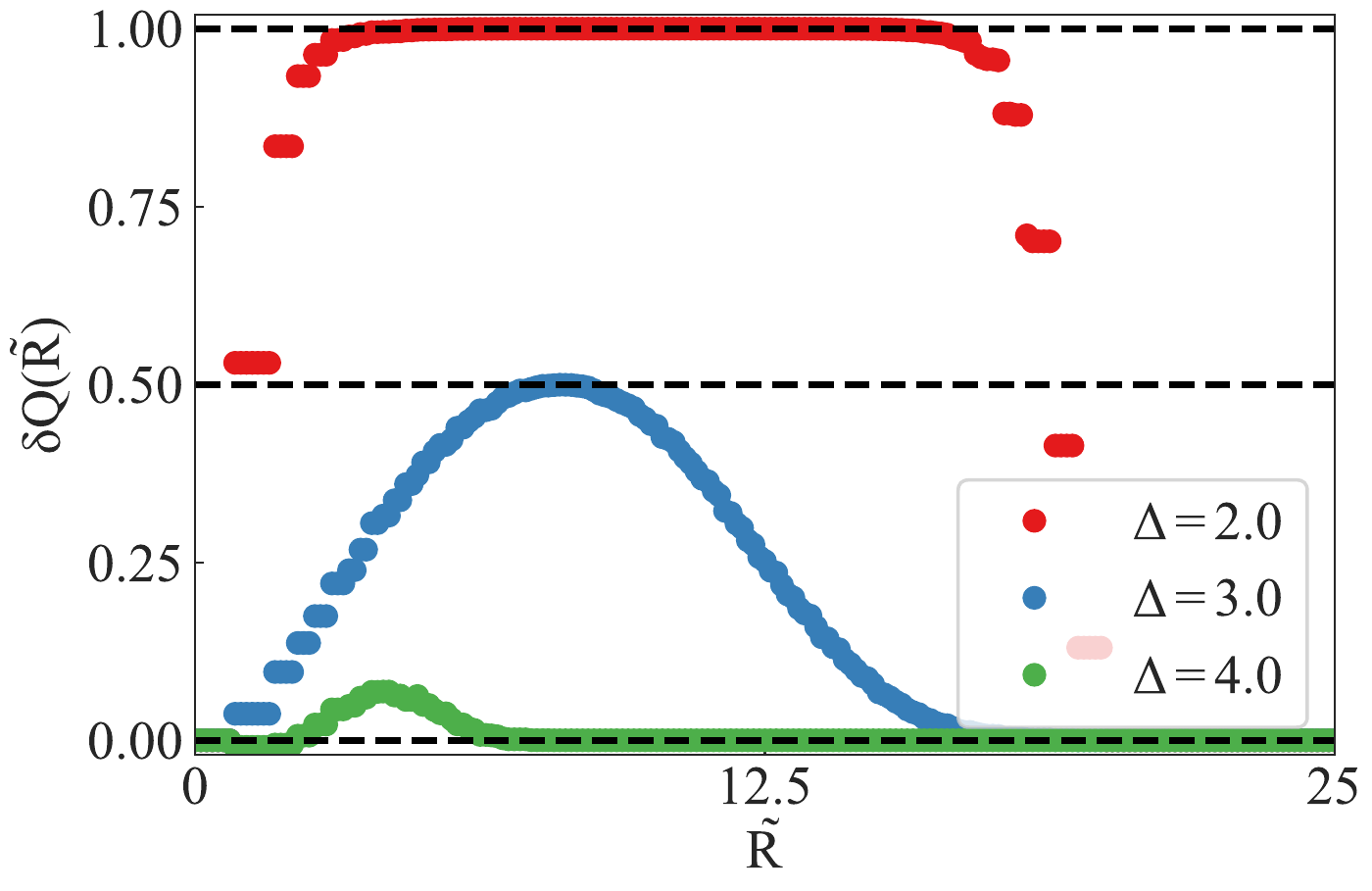}
\caption{Spin-charge separation of octupolar topological insulator in the presence of a minimal monopole $m=+1$, under mixed boundary conditions, without any pseudo-scalar training field. When the system is doped with one electron (hole), monopole ($m=+1$) binds maximum induced charge $-e$ ($+e$). The maximum induced charge at the topological quantum critical point equals $\mp e/2$ for electron and hole doping, respectively. All numerical calculations of induced charge have been performed for a system-size $(L/a)^3=30^3$ to achieve numerical accuracy $10^{-4}$ of quantized values.}
 \label{fig:OctIndCharge}
\end{figure}

Without any pseudo-scalar training field, $\delta Q(L,R,\xi,M^\prime=0)=0$. By maintaining the degeneracy of monopole bound zero-modes, we can study spin-charge separation. Under MBC with $m=+1$, we can dope one electron (hole) to completely occupy (empty) two NZMs. We have found  $\delta Q_{max}= -e $ ($+e$) [see Fig.~\ref{fig:OctIndCharge} ], which completely accounts for the charge of doped electron (hole).  At TQCP, we obtained $\delta Q_{max,TQCP}= -e/2 $ ($+e/2$). Similarly, under OBC, we can dope five electrons (or holes), to completely occupy (empty) ten NZMs. We have found $\delta Q_{max} = - e$ ($+e$) to be localized on monopole. Rest of the doped charge $-4e$ ($+4e$) will be distributed among eight corners, and each corner will exhibit $-e/2$ ($+e/2$). 

\begin{figure}[t]
\centering
\subfigure[]{
\includegraphics[scale=0.55]{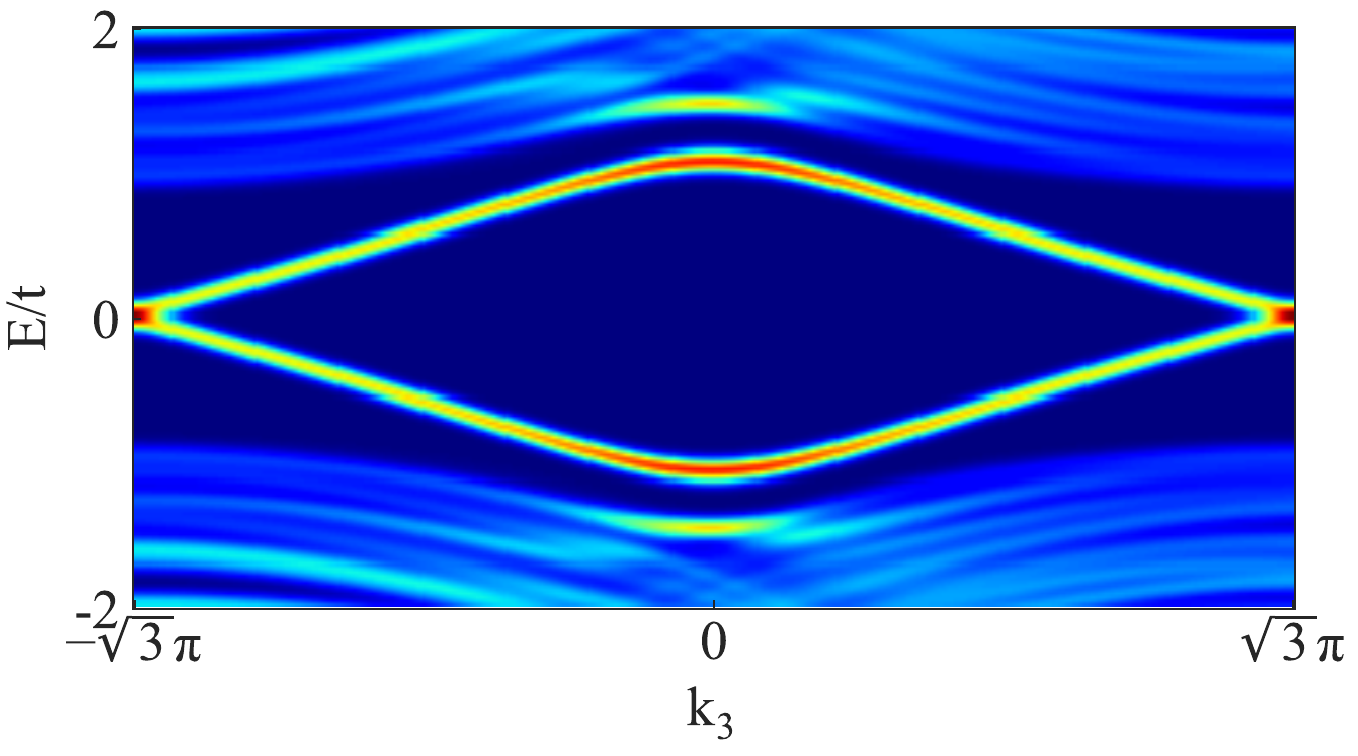}
\label{fig:OHStatevortex1}}
\subfigure[]{
\includegraphics[scale=0.55]{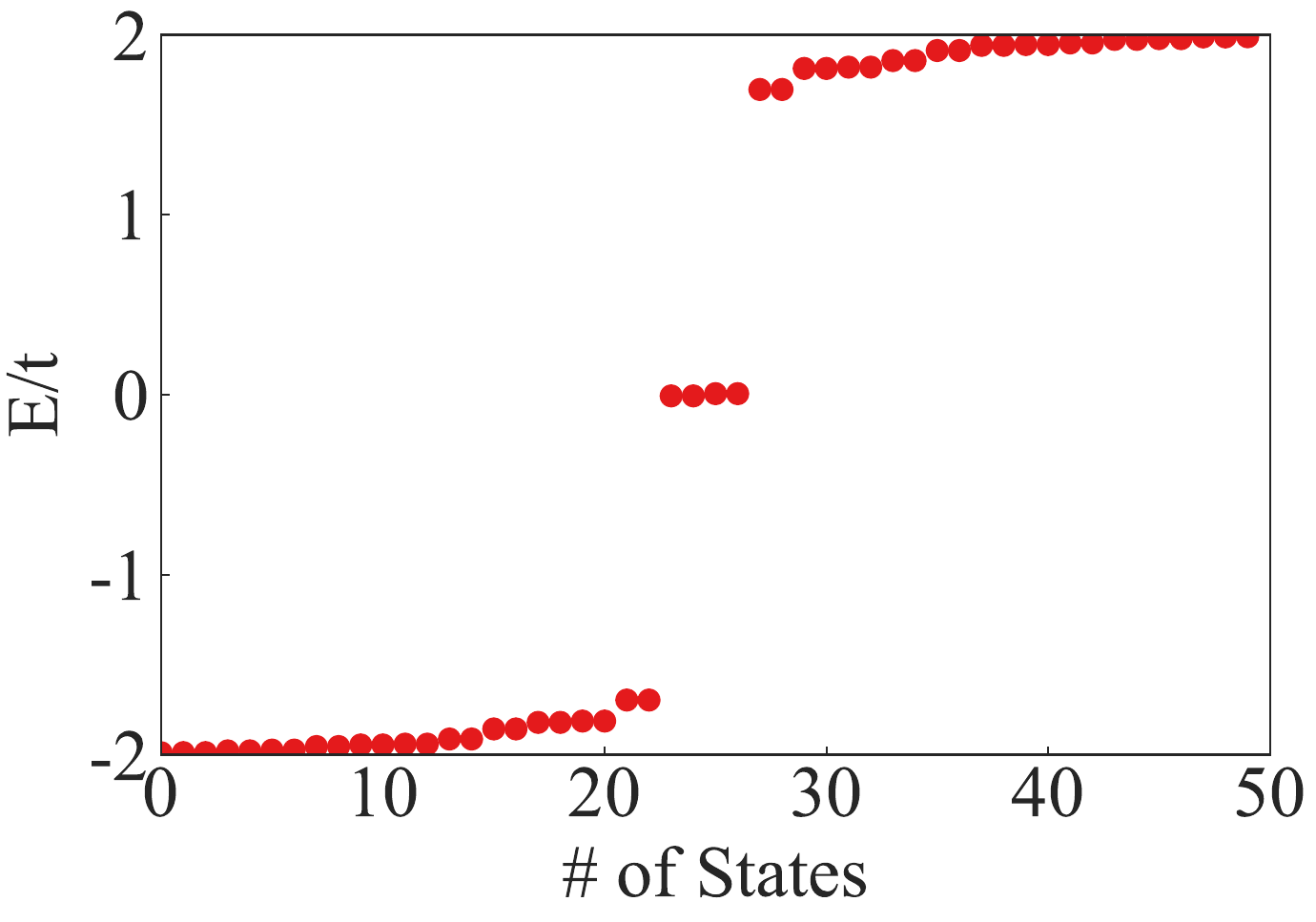}
\label{fig:OHStatelDOS}}
\caption{Diagnosis of non-trivial second homotopy class of $C_3$-symmetric planes of octupolar topological insulators with magnetic $\pi$-flux tube, oriented along $[111]$ axis. (a) The local density of states on flux tube as a function of $k_3=(k_x+k_y+k_z)/\sqrt{3}$ shows two pairs of two-fold-degenerate dispersive, mid-gap states that lead to four-fold-degenerate zero-modes for $k_3=\sqrt{3} \pi$ plane. This confirms our assertion that only $k_3=\sqrt{3} \pi$ plane carries quantized, flux of $SO(5)$ Berry connection. (b) The number of states vs, energy plot for $k_3=\sqrt{3}\pi$ plane. In contrast to this, $k_3=0$ plane only exhibits finite-energy mid-gap states, indicating trivial second homotopy class of this plane. The change of second homotopy class along $[111]$ corroborates the existence of non-trivial third homotopy class, even though mid-gap states never merge with the bulk continuum.  }
 \label{fig:OctFxInsert}
\end{figure}

What is the origin of three-dimensional winding number? In Ref.~\onlinecite{sur2022mixed}, we have shown that $n_3$ arises from $C_3$-symmetry-protected, tunneling configurations of $SO(5)$ Berry connection along $[111]$-direction. The quantized Berry flux for $C_3$-symmetric planes is carried by the matrix $\gamma_{111}=(\gamma_{12} + \gamma_{23} + \gamma_{31})/\sqrt{3}$ and the strength of tunneling is quantified by the difference of relative Chern numbers
\begin{eqnarray}
n_3=\mathfrak{C}_{111}(k_3=\sqrt{3} \pi) -\mathfrak{C}_{111}(k_3=0) \nn \\
\end{eqnarray}
with $k_{3}=(k_x + k_y +k_z)/\sqrt{3}$. For the current model, $C_R(k_3=0)=0$, and $C_{R}(k_3=\sqrt{3} \pi)=-2 \text{sgn}(t) \Theta(3-|\Delta|)$. To confirm such properties of Berry connection, 
we have probed the second homotopy class of $C_3$-symmetric planes with magnetic $\pi$-flux tube. We maintained PBC along $[111]$, $[1,-1,0]$, and $[1,1,-2]$ directions, and the flux tube was oriented along [111] direction. The results are shown in Fig.~\ref{fig:OctFxInsert}. 

\begin{figure*}
\centering
\subfigure[]{
\includegraphics[scale=0.21]{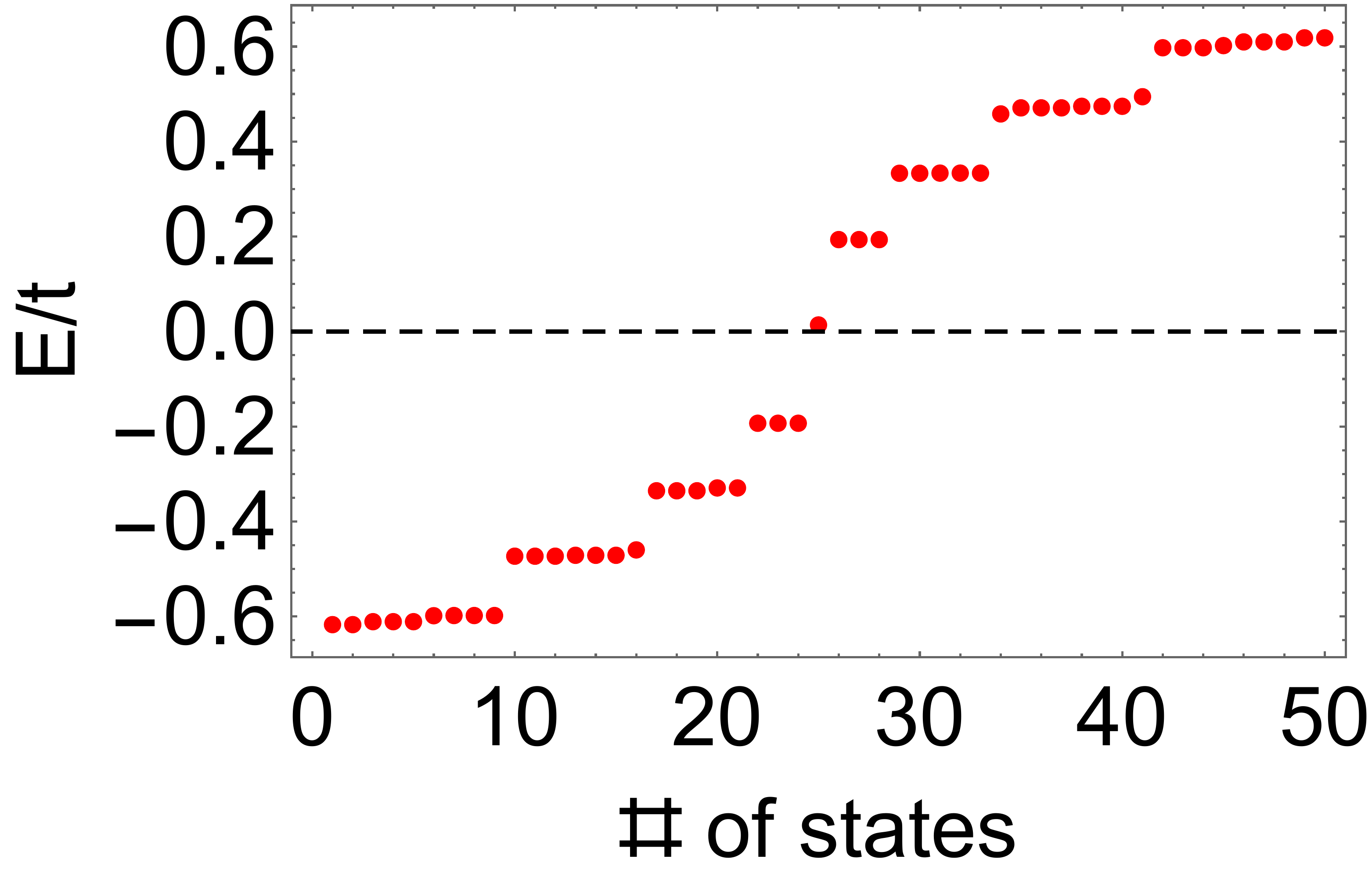}
\label{fig:7a}}
\subfigure[]{
\includegraphics[scale=0.12]{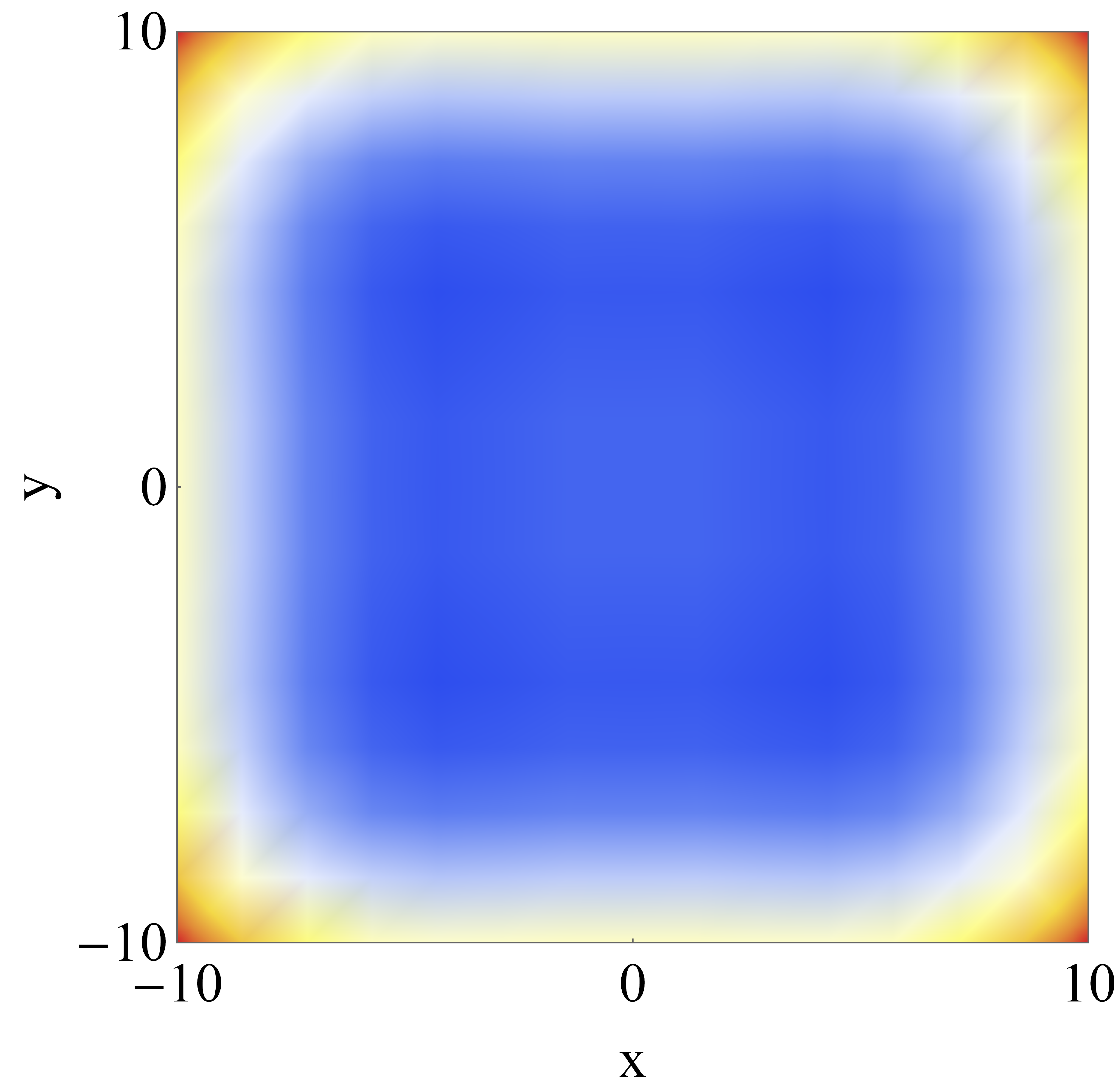}
\label{fig:7b}}
\subfigure[]{
\includegraphics[scale=0.12]{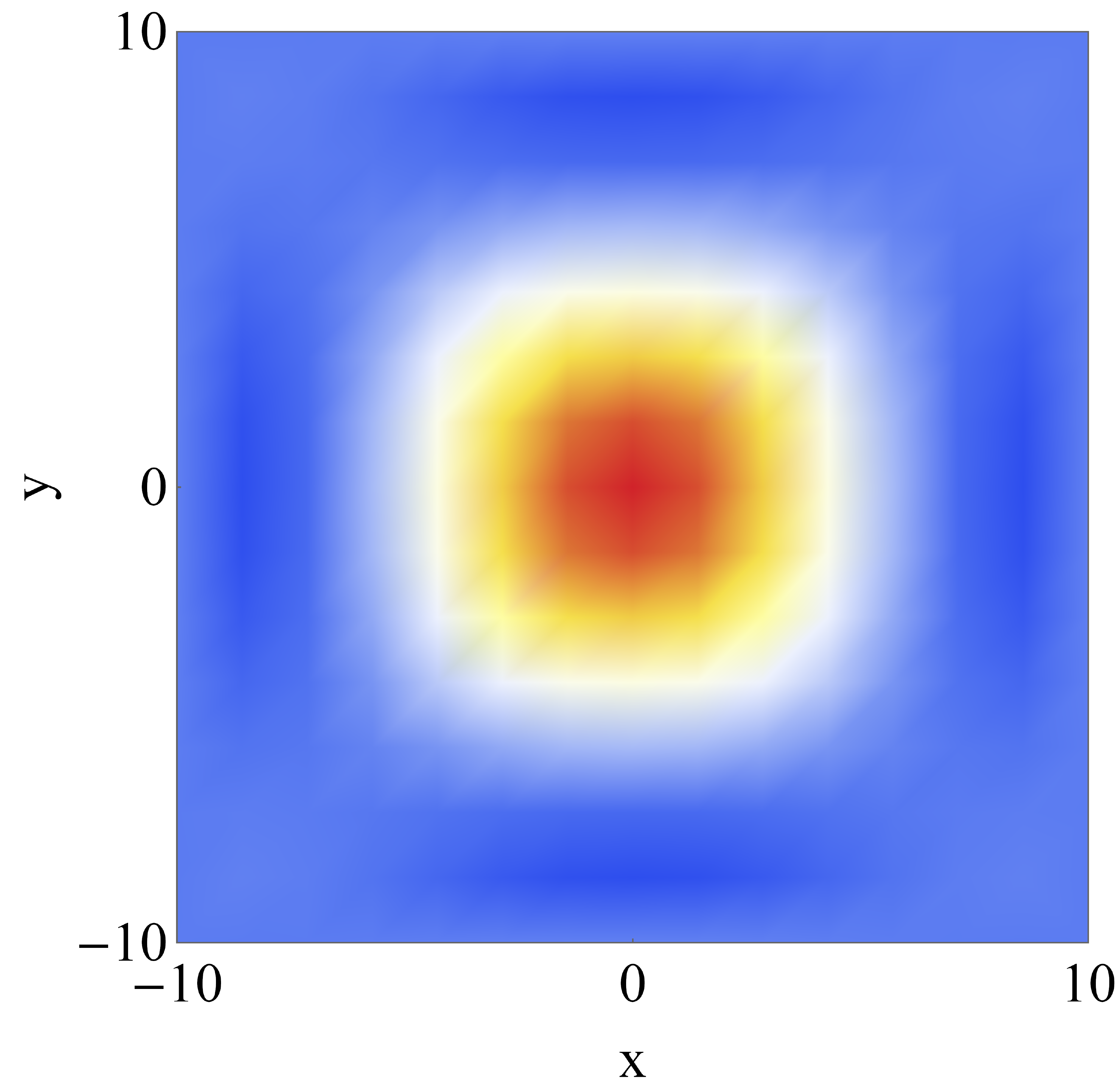}
\label{fig:7c}}
\subfigure[]{
\includegraphics[scale=0.55]{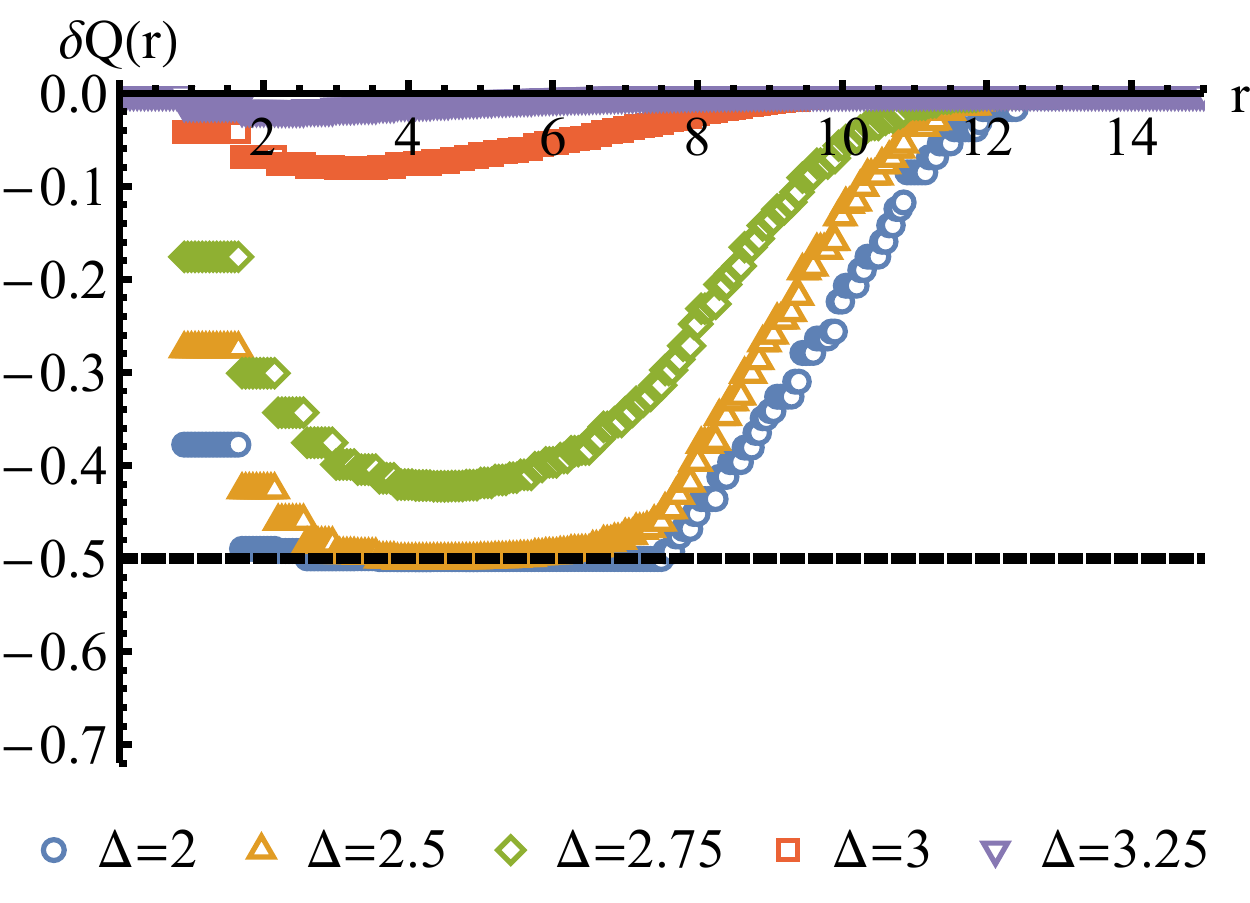}
\label{fig:7d}}
\caption{Effects of singular, pseudo-scalar, Zeeman coupling [see Eq.~\ref{Zeeman1}] on Witten effect. (a) Low-lying states in the presence of a unit strength monopole $m=+1$ for TI$_{1}$ phase (with tuning parameter $\Delta=2$), and g-factor $\tilde{g}=2$. Due to the spatial dependence of $H_Z$, the energy of the surface-bound mode is minimally shifted in energy. But the monopole-bound state is significantly shifted in energy to $E \approx +0.5 t$. The localization patterns of (b) surface-bound and (c) monopole-bound states in the $z=0$ plane. (d) Charge induced on monopole (in units of $-e$) for a half-filled system of size $L/a=10 $ as a function of $\Delta$. Thus, the observed Witten effect with pseudo-scalar Zeeman coupling confirms that TI$_1$ phase has the bulk winding number $\mathcal{N}_3=-1$. }
\label{fig:zeeman}
\end{figure*}

 \section{Conclusions}\label{conclusions}
Our comprehensive analysis of fermion spectrum, spin-charge separation with carrier doping, and Witten effect at half-filling clearly shows that monopoles can unambiguously detect $\mathbb{N}$- and $\mathbb{Z}$- classification of $\theta$. We have also showed that topological quantum critical points describe $\mathcal{C}$, $\mathcal{P}$, and $\mathcal{T}$ symmetric, Dirac semimetals, with Dirac points located at time-reversal-invariant momenta. Such symmetry-protected gapless states are associated with half-integer winding numbers and fractional axion angles $\theta_c = \pm \pi/2$. 

The analysis has been generalized to reveal non-trivial third homotopy class and quantized axion response of chiral, magnetic, and octupolar higher-order topological insulators. Therefore, we have demonstrated both first-order and higher order topological crystalline insulators can support quantized magneto-electric response with $\theta=(2n+1) \pi$ and $\theta=2n \pi$. Therefore, $\theta=2n \pi$ is not equivalent to $\theta=0$.

In our recent work~\cite{tyner2021symmetry}, we have identified $n_i$, $|n_i|$ of various Kramers-degenerate bands of elemental bismuth, respectively from a tight-binding model~\cite{schindler2018higherbismuth}, and \emph{ab initio} band structure. The analysis avoids direct calculation of Chern-Simons coefficient and identifies tunneling configurations of $SU(2)$ Berry connection of Kramers-degenerate bands by simultaneously computing in-plane Wilson loops for high-symmetry planes, and Polyakov loops along high-symmetry axes. Both tight-binding model and \emph{ab initio} band structures lead to $\theta = 2 n \pi$. Specifically, the tight-binding model~\cite{schindler2018higherbismuth} of bismuth leads to $\theta=+4\pi$. Elemental bismuth is a semimetal due to indirect band gap between conduction and valence bands. But the adiabatic deformation of band-structure of bismuth describes $\mathbb{Z}_2$-trivial higher-order topological insulators. Our detailed analysis on first- and higher- order topological insulators shows that actual bulk invariants of such phases can be identified by magneto-electric coefficients. 

So far, magneto-electric coefficients have been measured with Faraday rotation in two non-magnetic, $\mathbb{Z}_2$ topological insulators, Bi$_2$Se$_3$~\cite{wu2016quantized}, and strained HgTe~\cite{dziom2017observation}. The experiments apply uniform magnetic fields. Therefore, magnetic topological insulators of Sec.~\ref{AppA} with inversion-symmetry breaking coupling $m_{45}=0$ provides appropriate description of the experimental systems. By doping indium in Bi$_2$Se$_3$, and controlling strain in HgTe, one can drive phase transitions between $\mathbb{Z}_2$ topological insulator and $\mathbb{Z}_2$-trivial insulator (not necessarily a topologically trivial insulator). Future measurements of magneto-electric coefficient across such phase transition will help us understand topological properties of quantum critical points and $\mathbb{Z}_2$-trivial insulators.

\appendix

\section{Pseudo-scalar Zeeman coupling}\label{AppC}
Rosenberg and Franz~\cite{rosenberg2010witten} have explored the effects of Zeeman coupling 
\begin{eqnarray}\label{Zeeman1}
    H_{Z}&=&-\frac{\tilde{g}\mu_{B}}{2}\sum_i \;  \Psi^\dagger(\bs{r}_i) \mathbf{B}(\bs{r}_i) \cdot \boldsymbol \sigma \Psi (\bs{r}_i), \nn \\
    &=& -\frac{\hbar \tilde{g}\mu_{B} m }{4e} \; \sum_i \; \Psi^\dagger(\bs{r}_i) \frac{\boldsymbol \sigma \cdot \bs{r}_i} {r^3_i}  \Psi (\bs{r}_i)
    \end{eqnarray}
on Witten effect, where $\mu_{B}=e\hbar/(2m_e)$ is the Bohr magneton, $\tilde{g}$ is the $g$-factor, and $m_e$ is the mass of electron. Since $\bs{r}$ is odd under $\mathcal{P}$ and the axial vector $\Psi^\dagger \boldsymbol \sigma \Psi$ is odd (even) under $\mathcal{T}$ ($\mathcal{P}$ and $\mathcal{C}$), $H_Z$ describes a spatially varying, pseudo-scalar operator. However, $H_Z$ commutes with scalar and pseudo-scalar mass operators: $[H_Z, \Gamma_4]=[H_Z, \Gamma_5]=0$. Consequently, the spectrum of $H_{ij}+H_Z$ lacks particle-hole symmetry about $E=0$, and the bound states appear at finite energies, as shown in Figs. \ref{fig:7a}- \ref{fig:7c}. Non-zero induced electric charge at half-filling is found only in the presence of pseudo-scalar coupling [see Fig.~\ref{fig:7d}]. When $\tilde{g}=0$, the half-filled ground-state cannot exhibit Witten effect. For a thermodynamically large system ($L \to \infty$), Witten effect can be observed in $\tilde{g} \to 0$ limit. Due to the absence of particle-hole symmetry, the half-filled condition [$2 (L/a)^3$ empty and $2 (L/a)^3$ occupied states] requires $E_{F, m} \neq 0$.

\acknowledgements{
This work was supported by the National Science Foundation MRSEC program (DMR-1720139) at the Materials Research Center of Northwestern University, and the start up funds of P. G. provided by the Northwestern University. A part of this work was performed at the Aspen Center for Physics, which is supported by National Science Foundation grant PHY-1607611.}

 \bibliographystyle{apsrev4-1}
\nocite{apsrev41Control}
\bibliography{biblio.bib}


\end{document}